\documentclass[aps,rmp]{revtex4}
\usepackage{psfig}
\newcommand\lya{Ly$\alpha$}
\newcommand\aj{AJ}
\newcommand\mnras{MNRAS}
\newcommand\qjras{QJRAS}
\newcommand\pasp{PASP}
\newcommand\araa{ARAA}
\def\be#1{\begin{equation}\label{eq:#1}}
\def\be#1{\begin{equation}\label{eq:#1}}
\def\ee{\end{equation}}
\def\EC#1{(\ref{eq:#1})}
\def\bea#1{\begin{eqnarray}\label{eq:#1}}
\def\ee{\end{equation}}
\def\eea{\end{eqnarray}}
\def\la{\mathrel{\mathpalette\fun <}}
\def\ga{\mathrel{\mathpalette\fun >}}
\def\fun#1#2{\lower3.6pt\vbox{\baselineskip0pt\lineskip.9pt
        \ialign{$\mathsurround=0pt#1\hfill##\hfil$\crcr#2\crcr\sim\crcr}}}

\newcommand\cms{{\rm cm\, s^{-1}}}
\newcommand\kms{{\rm km\, s^{-1}}}
\newcommand\bfn{\mbox{\boldmath $\nabla$}}
\newcommand\bfo{\mbox{\boldmath $\omega$}}
\newcommand\bzeta{\mbox{\boldmath $\zeta$}}
\newcommand\bxi{\mbox{\boldmath $\xi$}}
\newcommand\om{\omega}
\newcommand\bemf{\mbox{\boldmath ${\cal E}$}}

\newcommand\rmt{{\,\rm rad\,m^{-2}}}

\newcommand\aod{\alpha\omega}
\newcommand\aara{AARA}

\newcommand\rrq{{\cal Q}}
\newcommand\tqcd{T_{\rm QCD}}
\newcommand\tew{T_{\rm EW}}
\newcommand\beq{B_{\rm eq}}

\begin{document}

\title{Origin of Galactic and Extragalactic Magnetic Fields}
\author{Lawrence M. Widrow}
\affiliation{Department of Physics, Queen's University,
Kingston, Ontario, Canada K7L 3N6}

\begin{abstract}

A variety of observations suggest that magnetic fields are present in
all galaxies and galaxy clusters.  These fields are characterized by a
modest strength $(10^{-7}-10^{-5}\,{\rm G})$ and huge spatial scale
$(\la 1\,{\rm Mpc})$.  It is generally assumed that magnetic fields
in spiral galaxies arise from the combined action of differential
rotation and helical turbulence, a process known as the $\aod$-dynamo.
However fundamental questions concerning the nature of the dynamo as
well as the origin of the seed fields necessary to prime it remain
unclear.  Moreover, the standard $\aod$-dynamo does not explain the
existence of magnetic fields in elliptical galaxies and clusters.  The
author summarizes what is known observationally about magnetic fields
in galaxies, clusters, superclusters, and beyond.  He then reviews the
standard dynamo paradigm, the challenges that have been leveled
against it, and several alternative scenarios.  He concludes with a
discussion of astrophysical and early Universe candidates for seed
fields.

\end{abstract}

\maketitle

\tableofcontents

\section{INTRODUCTION}
\label{sec:introduc}

The origin of galactic and extragalactic magnetic fields is one of the
most fascinating and challenging problems in modern astrophysics.
Magnetic fields are detected in galaxies of all types and in galaxy
clusters whenever the appropriate observations are made.  In addition
there is mounting evidence that they exist in galaxies at cosmological
redshifts.  It is generally assumed that the large-scale magnetic
fields observed in disk galaxies are amplified and maintained by an
$\aod$-dynamo wherein new field is regenerated continuously by the
combined action of differential rotation and helical turbulence.  By
contrast, the magnetic fields in non-rotating or slowly rotating
systems such as elliptical galaxies and clusters appear to have a
characteristic coherence scale much smaller than the size of the
system itself.  These fields may be generated by a local, turbulent
dynamo where, in the absence of rapid rotation, the field does not
organize on large scales.

In and of itself, the dynamo paradigm must be considered incomplete
since it does not explain the origin of the initial fields that act as
seeds for subsequent dynamo action.  Moreover, the timescale for field
amplification in the standard $\aod$-dynamo may be too long to explain
the fields observed in very young galaxies.

It is doubtful that magnetic fields have ever played a primary role in
shaping the large-scale properties of galaxies and clusters.  In
present-day spirals, for example, the energy in the magnetic field is
small as compared to the rotation energy in the disk.  To be sure,
magnetic fields are an important component of the interstellar medium
(ISM) having an energy density that is comparable to the energy
density in cosmic rays and in the turbulent motions of the
interstellar gas.  In addition, magnetic fields can remove angular
momentum from protostellar clouds allowing star formation to proceed.
Thus, magnetic fields can play a supporting role in the formation and
evolution of galaxies and clusters but are probably not essential to
our understanding of large-scale structure in the Universe.  

The converse is not true: An understanding of structure formation is
paramount to the problem of galactic and extragalactic magnetic
fields.  Magnetic fields can be created in active galactic nuclei
(AGN), in the first generation of stars, in the shocks that arise
during the collapse of protogalaxies, and in the early Universe.  In
each case, one must understand how the fields evolve during the epoch
of structure formation to see if they are suitable as seeds for dynamo
action.  For example, magnetic fields will be amplified during
structure formation by the stretching and compression of field lines
that occur during the gravitational collapse of protogalactic gas
clouds.  In spiral galaxies, for example, these processes occur prior
to disk formation and can amplify a primordial seed field by several
orders of magnitude.

In principle, one should be able to follow the evolution of magnetic
fields from their creation as seed fields through to the dynamo phase
characteristic of mature galaxies.  Until recently, theories of
structure formation did not possess the sophistication necessary for
such a program.  Rather, it had been common practice to treat dynamo
action and the creation of seed fields as distinct aspects of a single
problem.  Recent advances in observational and theoretical cosmology
have greatly improved our understanding of structure formation.
Ultra-deep observations, for example, have provided snapshots of disk
galaxies in an embryonic state while numerical simulations have
enabled researchers to follow an individual galaxy from linear
perturbation to a fully-evolved disk-halo system.  With these
advances, a more complete understanding of astrophysical magnetic
fields may soon be possible.

This review brings together observational and theoretical results from
the study of galactic and extragalactic magnetic fields, the pieces of
a puzzle, if you like, which, once fully assembled, will provide a
coherent picture of cosmic magnetic fields.  An outline of the review
is as follows: In Section II we summarize useful results from
magnetohydrodynamics and cosmology.  Observations of galactic and
extragalactic magnetic fields are described in Section III.  We begin
with a review of four common methods used to detect magnetic fields;
syncrotron emission, Faraday rotation, Zeeman splitting, and optical
polarization of starlight (Section III.A).  The magnetic fields in
spiral galaxies, ellipticals, and galaxy clusters are reviewed in
Sections III.B-III.D while observations of magnetic fields in objects
at cosmological redshifts are described in Section III.E.  The latter
are essential to our understanding of the origin of galactic fields
since they constrain the time available for dynamo action.  Section
III concludes with a discussion of observational limits on the
properties of cosmological magnetic fields.

Magnetic dynamos are discussed in Section IV.  We first review the
primordial field hypothesis wherein large scale magnetic fields,
created in an epoch prior to galaxy formation, are swept up by the
material that forms the galaxy and amplified by differential rotation.
The model has serious flaws but is nevertheless instructive for the
discussion that follows.  Mean-field dynamo theory is reviewed in
Section IV.B.  The equations for a disk dynamo are presented in
Section IV.C and a simple estimate for the amplification rate in
galaxies is given in Section IV.D.

The standard mean-field treatment fails to take into account
backreaction of small-scale magnetic fields on the turbulent motions
of the fluid.  Backreaction is a potentially fatal problem for the
dynamo hypothesis for if magnetic fields inhibit turbulence, the
dynamo will shut down.  These issues are discussed in Section IV.E.

Galactic magnetic fields, like galaxies themselves, display a
remarkable variety of structure and thus an understanding of galactic
dynamos has required full three-dimensional simulations.  Techniques
for performing numerical simulations are reviewed in Section IV.F and
their application to the problem of diversity in galactic magnetic
fields is discussed in Section IV.G.  In Section IV.H we turn to
alternatives to the $\aod$-dynamo.  These models were constructed to
address various difficulties with the standard scenario.  Section IV
ends with a brief discussion of the generation of magnetic fields in
elliptical galaxies and galaxy clusters.

The question of seed fields has prompted a diverse and imaginative
array of proposals.  The requirements for seed fields are derived in
Section V.A.  Section V.B describes astrophysical candidates for seed
fields while more speculative mechanisms that operate in the exotic
environment of the early Universe are discussed in Section V.C.

The literature on galactic and extragalactic magnetic fields is
extensive.  Reviews include the excellent text by Ruzmaikin, Sokoloff,
\& Shukurov (1988a) as well as articles by Rees (1987), Kronberg
(1994), and Zweibel \& Heiles (1997).  The reader interested in
magnetohydrodynamics and dynamo theory is referred to the classic
texts by Moffatt (1978), Parker (1979), and Krause \& R\"{a}dler
(1980) as well as ``The Almighty Chance'' by Zel'dovich, Ruzmaikin, \&
Sokoloff (1990).  A survey of observational results from the Galaxy to
cosmological scales can be found in Vall\'{e}e (1997).  The structure
of galactic magnetic fields and galactic dynamo models are discussed
in Sofue, Fujimoto \& Wielebinski (1986), Krause \& Wielebinski
(1991), Beck et al.\,(1996), and Beck (2000) as well as the review 
articles and texts cited above.

\section{Preliminaries}

\subsection{Magnetohydrodynamics and Plasma Physics}

Magnetohydrodynamics (MHD) and plasma physics describe the interaction
between electromagnetic fields and conducting fluids (see, for
example, Jackson 1975; Moffatt 1978; Parker 1979; Freidberg 1987;
Sturrock 1994).  MHD is an approximation that holds when charge
separation effects are negligible.  Matter is described as a single
conducting fluid characterized by a density field $\rho({\bf x},\,t)$,
velocity field ${\bf V}({\bf x},\,t)$, pressure $p({\bf x},\,t)$, and
current density ${\bf J}({\bf x},\,t)$.  The simple form of Ohm's law
is valid while the displacement current in Amp\`{e}re's Law is
ignored.  In Gaussian units, the relevant Maxwell equations take the
form

\be{Max1}
\bfn\cdot {\bf B}= 0
\ee

\be{Max2} 
\bfn\times {\bf E} + \frac{1}{c}\frac{\partial {\bf
B}}{\partial t}= 0 
\ee

\be{Max3}
\bfn\times {\bf B}= \frac{4\pi}{c}{\bf J}~\ee

\noindent and Ohm's law is given by

\be{Ohm}
{\bf J}' = \sigma{\bf E}'
\ee

\noindent where $\sigma$ is the conductivity and ``primed'' quantities
refer to the rest frame of the fluid.  Most astrophysical fluids are
electrically neutral and nonrelativistic so that ${\bf J}'={\bf J}$
and ${\bf E}'={\bf E}+\left ({\bf V}\times {\bf B}\right )/c$.
Eq.\,\EC{Ohm} becomes

\be{Ohm2}
{\bf J} = \sigma\left ({\bf E}+\frac{{\bf V}\times {\bf
B}}{c}\right )
\ee

\noindent 
which, when combined with Eqs.\,\EC{Max2} and \EC{Max3}, yields the
ideal MHD equation:

\be{MHD}
\frac{\partial {\bf B}}{\partial t} =
\bfn\times \left ({\bf V}\times {\bf B}\right )
+\eta\nabla^2{\bf B}~.
\ee

\noindent 
In deriving this equation, the molecular diffusion coefficient,
$\eta\equiv c^2/4\pi\sigma$, is assumed to be constant in space.  

In the limit of infinite conductivity, magnetic diffusion is ignored
and the MHD equation becomes

\be{MHD2}
\frac{\partial {\bf B}}{\partial t} =
\bfn\times \left ({\bf V}\times {\bf B}\right )
\ee

\noindent
or equivalently

\be{MHD3}
\frac{d{\bf B}}{dt} = 
\left (
	{\bf B} \cdot \bfn 
			\right ) {\bf V}
-{\bf B} \left (
	\bfn\cdot {\bf V}	 
			\right )
\ee

\noindent
where $d/dt=\partial/\partial t + {\bf V}\cdot\bfn$ is the convective
derivative.  The interpretation of this equation is that the flux
through any loop moving with the fluid is constant (see, for example,
Jackson 1975; Moffatt 1978; Parker 1979), i.e., magnetic field lines
are frozen into the fluid.  Using index notation we have

\bea{RHS}
\frac{dB_i}{dt} &=& B_j\frac{\partial V_i}{\partial x_j}
~-~B_i\frac{\partial V_j}{\partial x_j}\nonumber\\
&=&
B_j\left (\frac{\partial V_i}{\partial x_j}-\frac{1}{3}
\delta_{ij}\frac{\partial V_k}{\partial x_k}\right )
~-~\frac{2}{3}B_i\frac{\partial V_j}{\partial x_j}
\eea 

\noindent
where a sum over repeated indices is implied.
This equation, together with the continuity equation

\be{continuity}
\frac{d\rho}{d t}=-\rho\bfn\cdot {\bf V}~,
\ee

\noindent gives

\be{MHD4}
\frac{dB_i}{dt} = \frac{2}{3}\frac{B_i}{\rho}\frac{d\rho}{dt}
+B_j\sigma_{ij}
\ee

\noindent 
where $\sigma_{ij}=\partial_j V_i - \frac{1}{3}\delta_{ij} \partial_k
V_k$ (see, for example, Gnedin, Ferrara, \& Zweibel 2000).  

The appearance of convective derivatives in Eq.\,\EC{MHD4} suggests a
Lagrangian description in which the field strength and fluid density
are calculated along the orbits of the fluid elements.  The first term
on the right-hand side of Eq.\,\EC{MHD4} describes the adiabatic
compression or expansion of magnetic field that occurs when
$\bfn\cdot{\bf V}\ne 0$.  Consider, for example, a region of uniform
density $\rho$ and volume ${\cal V}$ that is undergoing homogeneous
collapse or expansion so that $\sigma_{ij}=0$ and $\bfn\cdot{\bf V}=C$
where $C=C(t)$ is a function of time but not position.  Eq.\,\EC{MHD4}
implies that $B\propto\rho^{2/3}\propto {\cal V}^{-2/3}$.  Thus,
magnetic fields in a system that is undergoing gravitational collapse
are amplified while cosmological fields in an expanding universe are
diluted.

The second term in Eq.\,\EC{MHD4} describes the stretching of magnetic
field lines that occurs in flows with shear and vorticity.  As an
illustrative example, consider an initial magnetic field ${\bf
B}=B_0\hat{\bf x}$ subject to a velocity field with $\partial
V_y/\partial x={\rm constant}$.  Over a time $t$, ${\bf B}$ developes
a component in the $y$-direction and its strength increases by a
factor $\left (1+\left (t\,\partial V_y/\partial x\right )^2\right
)^{1/2}$.

Combining Eq.\,\EC{MHD3} and \EC{continuity} yields the following
alternative form for the MHD equation:

\be{MHD5}
\frac{d}{dt}\left (\frac{{\bf B}}{\rho}\right )=
\left (\frac{{\bf B}}{\rho}\cdot \bfn\right ){\bf V}~.
\ee

\noindent The formal solution of this equation is

\be{Solution}
\frac{B_i\left ({\bf x},\,t\right )}
{\rho\left ({\bf x},\,t\right )}~ =~ 
\frac{B_j\left (\bxi,\,0\right )}
{\rho\left (\bxi,\,0\right )}
\frac{\partial x_i}{\partial \xi_j}
\ee

\noindent where ${\bxi}$ is the Lagrangian coordinate for the fluid:

\be{def_xi}
x_i(t) = \xi_i + \int_0^t V_i(s)ds~.
\ee

\noindent
It follows that if a ``material curve'' coincides with a magnetic field
line at some initial time then, in the limit $\eta=0$, it will
coincide with the same field line for all subsequent times.  Thus, the
evolution of a magnetic field line can be determined by following
the motion of a material curve (in practice, traced out by test
particles) as it is carried along by the fluid.

The equation of motion for the fluid is given by

\be{euler}
\frac{\partial{\bf V}}{\partial t}+
\left ({\bf V}\cdot \bfn\right ){\bf V}=
-\frac{1}{\rho}\bfn p -\bfn\Phi +
\frac{1}{c\rho}\left ({\bf J}\times {\bf B}\right )
+\nu\nabla^2 {\bf V}
\ee

\noindent 
where $\nu$ is the viscosity coefficient and $\Phi$ the gravitational
potential.  In many situations, the fields are weak and the Lorentz
term in Eq.\,\EC{euler} can be ignored.  This is the kinematic regime.
In the limit that the pressure term is also negligible, the vorticity
$\bzeta\equiv \bfn\times {\bf V}$ obeys an equation that is similar,
in form, to Eq.\,\EC{MHD}:

\be{vorticityI}
\frac{\partial\bzeta}{\partial t} =
\bfn\times \left ({\bf V}\times {\bzeta}\right )
+\nu\nabla^2{\bzeta}~.
\ee

\noindent 
Moreover, if viscosity is negligible, then $\bzeta$ satisfies the
Cauchy equation (Moffatt 1978):

\be{cauchy}
\frac{\zeta_i\left ({\bf x},\,t\right )}
{\rho\left ({\bf x},\,t\right )}~ =~ 
\frac{\zeta_j\left (\bxi,\,0\right )}
{\rho\left (\bxi,\,0\right )}
\frac{\partial x_i}{\partial \xi_j}~.
\ee

\noindent 
However, Eq.\,\EC{cauchy} is not a solution of the vorticity equation
so much as a restatement of Eq.\,\EC{vorticityI} since $\partial
x/\partial\xi$ is determined from the velocity field which, in turn,
depends on ${\bf x}$.  By contrast, in the kinematic regime and in the
absence of magnetic diffusion, Eq.\,\EC{Solution} provides an explicit
solution of Eq.\,\EC{MHD4}.

The magnetic energy density associated with a field of strength $B$ is
$\epsilon_B=B^2/8\pi$.  For reference, we note that the energy density
of a $1\,{\rm G}$ field is $\simeq 0.040\,{\rm erg\,cm^{-3}}$.  A
magnetic field that is in equipartition with a fluid of density $\rho$
and rms velocity $v$ has a field strength $B\simeq \beq\equiv \left
(4\pi\rho v^2\right )^{1/2}$.  In a fluid in which magnetic and
kinetic energies are comparable, hydromagnetic waves propagate at
speeds close to the so-called Alfv\'{e}n speed, $v_A\equiv \left
(B^2/4\pi\rho\right )^{1/2}$.

It is often useful to isolate the contribution to the magnetic field
associated with a particular length scale $L$.  Following Rees \& 
Reinhardt (1972) we write

\be{field_decomp}
\left\langle\frac{B^2}{8\pi}\right\rangle 
= \int_{\cal V}\frac{B(L)^2}{8\pi}\frac{dL}{L}
\ee

\noindent 
where $\langle B^2/8\pi\rangle$ is the magnetic field energy density
averaged over some large volume $\cal{V}$.  $B(L)$ is roughly the
component of the field with characteristic scale between $L$ and $2L$.
Formally, ${\bf B}(L) = \left ( k^3/2\pi^2{\cal V}\right )^{1/2}{\bf
B}_k$ where ${\bf B}_k\equiv \int d^3x \exp{\left (i{\bf k}\cdot {\bf
x}\right )} {\bf B}({\bf x})$ is the Fourier component of ${\bf B}$
associated with the wavenumber $k=2\pi/L$.

In the MHD limit, magnetic fields are distorted and amplified (or
diluted) but no net flux is created.  A corollary of this statement is
that if at any time ${\bf B}$ is zero everywhere, it must be zero at
all times.  This conclusion follows directly from the assumption that
charge separation effects are negligible.  When this assumption breaks
down, currents driven by non-electromagnetic forces can create
magnetic fields even if ${\bf B}$ is initially zero.

\subsection{Cosmology}

Occasionally, we will make reference to specific cosmological models.
A common assumption of these models is that on large scales, the
Universe is approximately homogeneous and isotropic.  Spacetime
can then be described by the Robertson-Walker metric:

\be{RW_metric}
ds^2 = c^2dt^2 - a^2(t)dr^2
\ee

\noindent 
where $a(t)$ is the scale factor and $dr$ is the three-dimensional
line element which encodes the spatial curvature of the model (flat,
open, or closed).  For convenience, we set $a(t_0)=1$ where $t_0$ is
the present age of the Universe.  The evolution of $a$ is described by
the Friedmann equation (see, for example, Kolb \& Turner 1990)

\bea{friedmann}
H(t)^2 & = & 
\left (\frac{1}{a}\frac{da}{dt}\right )^2 \nonumber \\
	& = &\frac{8\pi G}{3}
		\left (\epsilon_r +
		\epsilon_m + 
		\epsilon_\Lambda\right )
			-\frac{k}{a^2}
\eea

\noindent
where $\epsilon_r$, $\epsilon_m$, and $\epsilon_\Lambda$ are the
energy densities in relativistic particles, nonrelativistic particles,
and vacuum energy respectively, $k=0,\,\pm 1$ parametrizes the spatial
curvature, and $H$ is the Hubble parameter.  Eq.\,\EC{friedmann} can
be recast as

\be{friedmann2} 
\frac{1}{a}\frac{da}{dt} \\ = H_0\left
(\frac{\Omega_r}{ a^{4}} + \frac{\Omega_m}{ a^{3}} + \Omega_\Lambda +
\frac{\left (1-\Omega_r-\Omega_m-\Omega_\Lambda\right )}{a^2} \right
)^{1/2}  
\ee

\noindent 
where $H_0\equiv H(t_0)$ is the Hubble constant and $\Omega$ is the
present-day energy density in units of the critical density
$\epsilon_{\rm c}\equiv 3H^2/8\pi G$, i.e., $\Omega_m\equiv
\epsilon_m/\epsilon_{\rm c}$, etc..

Recent measurements of the angular anisotropy spectrum of the CMB
indicate that the Universe is spatially flat or very nearly so (Balbi et al.
2000; Melchiorvi et al. 2000; Pryke et al. 2001).  If these results are combined
with dynamical estimates of the density of clustering matter (i.e.,
dark matter plus baryonic matter) and with data on Type Ia supernova,
a picture emerges of a universe with zero spatial curvature,
$\Omega_m\simeq 0.15-0.4$, and $\Omega_\Lambda = 1-\Omega_m$ (see, for
example, Bahcall et al.\,1999).  In addition, the Hubble constant has
now been determined to an accuracy of $\sim 10\%$: The published value
from the Hubble Space Telescope Key Project is $71\pm 6\,{\rm
km\,s^{-1}}\,{\rm Mpc}^{-1}$ (Mould, et al.~2000).

\section{Observations of Cosmic Magnetic Fields}

Observations of galactic and extragalactic magnetic fields can be
summarized as follows:

\begin{itemize}

\item Magnetic fields with strength $\sim 10\mu{\rm G}$ are found in
spiral galaxies whenever the pertinent observations are made.  These
fields invariably include a large-scale component whose coherence
length is comparable to the size of the visible disk.  There are also
small-scale tangled fields with energy densities approximately equal
to that of the coherent component.

\item The magnetic field of a spiral galaxy often exhibits patterns or
symmetries with respect to both the galaxy's spin axis and equatorial
plane.

\item Magnetic fields are ubiquitous in elliptical galaxies, though in
contrast with the fields found in spirals, they appear to be random
with a coherence length much smaller than the galactic scale.
Magnetic fields have also been observed in barred and irregular
galaxies.

\item Microgauss magnetic fields have been observed in the
intracluster medium of a number of rich clusters.  The coherence
length of these fields is comparable to the scale of the cluster
galaxies.

\item There is compelling evidence for galactic-scale magnetic fields
in a redshift $z\simeq 0.4$ spiral.  In addition, microgauss fields
have been detected in radio galaxies at $z\ga 2$.  Magnetic fields may
also exist in damped \lya~systems at cosmological redshifts.

\item There are no detections of purely cosmological fields (i.e.,
fields not associated with gravitationally bound or collapsing
structures).  Constraints on cosmological magnetic fields have been
derived by considering their effect on big bang nucleosynthesis, the
cosmic microwave background, and polarized radiation from
extragalactic sources.

\end{itemize}

These points will be discussed in detail.  Before doing so, we
describe the four most common methods used to study astrophysical
magnetic fields.  A more thorough discussion of observational
techniques can be found in various references including Ruzmaikin,
Shukurov, and Sokoloff (1988a).

\subsection{Observational Methods}

\subsubsection{Synchrotron Emission}

Synchroton emission, the radiation produced by relativistic electrons
spiralling along magnetic field lines, is used to study magnetic
fields in astrophysical sources ranging from pulsars to superclusters.
The total synchrotron emission from a source provides one of the two
primary estimates for the strength of magnetic fields in galaxies and
clusters while the degree of polarization is an important indicator of
the field's uniformity and structure.

For a single electron in a magnetic field ${\bf B}$, the emissivity
as a function of frequency $\nu$ and electron energy $E$ is

\be{single_e}
J(\nu,E)~\propto~ B_\perp\left (\frac{\nu}{\nu_c}\right)^{1/3}
f\left (\frac{\nu}{\nu_c}\right )
\ee

\noindent 
where $B_\perp$ is the component of the magnetic field perpendicular
to the line of sight, $\nu_c\equiv \nu_L \left (E/mc^2\right )^2$ is
the so-called critical frequency, $\nu_L =\left (eB_\perp/2\pi
mc\right )$ is the Larmor frequency, and $f(x)$ is a cut-off function
which approaches unity for $x\to 0$ and vanishes rapidly for $x\gg 1$.

The total synchrotron emission from a given source depends on the energy
distribution of electrons, $n_e(E)$.  A commonly used class of models is
based on a power-law distribution

\be{electron_df}
n_e(E) dE = n_{e0}\left (\frac{E}{E_0}\right )^{-\gamma}dE
\ee

\noindent 
assumed to be valid over some range in energy.  The exponent 
$\gamma$ is called the spectral index while the constant
$n_{e0}\equiv n_e(E_0)$ sets the normalization of the distribution.  A
spectral index $\gamma\simeq 2.6-3.0$ is typical for spiral galaxies.

The synchrotron emissivity is $j_\nu \equiv \int J(\nu,\,E)n_e(E)dE$.
Eq.\,\EC{single_e} shows that synchrotron emission at frequency $\nu$
is dominated by electrons with energy $E \simeq m_ec^2\left
(\nu/\nu_L\right )^{1/2}$, i.e., $\nu\simeq \nu_c$, so that to a good
approximation, we can write $J(\nu,\,E)\propto B_\perp\nu_c\delta
\left (\nu-\nu_c\right )$.  For the power-law distribution
Eq.\,\EC{electron_df} we find

\be{j_nu}
j_\nu
~\propto~ n_{e0}\nu^{(1-\gamma)/2}B_\perp^{(1+\gamma)/2}~.
\ee

\noindent
Alternatively, we can write the distribution of electrons as a
function of $\nu_c$: $n(\nu_c)\equiv n_e(E)dE/d\nu_c\propto
j_{\nu_c}/\nu_c B_\perp$.  (See Leahy 1991 for a more detailed
discussion).
 
The energy density in relativistic electrons is $\epsilon_{re}
=\int n(E) E dE$.  Thus, the synchrotron emission spectrum can be
related to the energy density in relativistic electrons
$\epsilon_{re}$ and the strength of the magnetic field (Burbidge 1956;
Pacholczyk 1970; Leahy 1991).  It is standard practice to write the
total kinetic energy in particles as $\epsilon_k = \left (1+k\right
)\epsilon_{re}$ where $k\sim 100$ is a constant (see, for example,
Ginzberg \& Syrovatskii 1964; Cesarsky 1980).  The total energy
(kinetic plus field) is therefore $\epsilon_{\rm tot} = \left
(1+k\right )\epsilon_{re} + \epsilon_B$.  One can estimate the
magnetic field strength either by assuming equipartition ($\left
(1+k\right )\epsilon_{re}=\epsilon_B$) or by minimizing $\epsilon_{\rm
tot}$ with respect to $B$.

The standard calculation of $\epsilon_{re}$ uses a fixed integration
interval in frequency, $\nu_L\le\nu\le\nu_U$:

\be{relativistic_electrons}
\epsilon_{re} ~=~ \int_{\nu_L}^{\nu_U} En(\nu_c)d\nu_c
~\propto~ B^{-3/2}\Theta^2 S_\nu(\nu_0)
\ee

\noindent 
where $S_\nu$ is the total flux density, $\Theta$ is the angular size
of the source, and $\nu_0$ is a characteristic frequency between
$\nu_L$ and $\nu_U$.  Assuming either equipartition or minimum energy,
this expression leads to an estimate for $B$ of the form $B_{\rm
eq}\propto S_\nu^{2/7}\Theta^{-4/7}$.  However Beck et al.~(1996) and
Beck (2000) pointed out that a fixed frequency range corresponds to
different ranges in energy for different values of the magnetic field
(see also Leahy (1991) and references therein).  From Eq.\,\EC{j_nu}
we have $n_{e0}\propto j(\nu)\nu^{(\gamma-1)/2}
B_\perp^{-(\gamma+1)/2}$.  Integrating over a fixed energy interval
gives $\epsilon_{re}\propto \Theta^2 S_\nu B^{-(\gamma+1)/2}$ which
leads to a minimum energy estimate for $B$ of the form $B_{\rm
eq}\propto S_\nu^{2/(\gamma+5)}\Theta^{-4/(\gamma+5)}$.

Interactions between cosmic rays, supernova shock fronts, and magnetic
fields can redistribute energy and therefore, at some level, the
minimum energy condition will be violated.  For this reason, the
equipartition/minimum energy method for estimating the magnetic field
strength is under continous debate.  Duric (1990) argued that
discrepancies of more than a factor of $10$ between the derived and
true values for the magnetic field require rather extreme conditions.
Essentially, $B/\beq$ sets the scale for the thickness of radio
synchrotron halos.  A field as small as $0.1\beq$ requires higher
particle energies to explain the synchrotron emission data.  However,
high energies imply large propagation lengths and hence an extended
radio halo (scale height $\sim 30\,{\rm kpc}$) in conflict with
observations of typical spiral galaxies.  Conversely, a field as large
as $10\beq$ would confine particles to a thin disk $(\sim 300\,{\rm
pc})$ again in conflict with observations.

In the Galaxy, the validity of the equipartition assumption can be
tested because we have direct measurements of the local cosmic-ray
electron energy density and independent estimates of the local
cosmic-ray proton density from diffuse continuum $\gamma$-rays.  A
combination of the radio synchrotron emission measurements with these
results yields a field strength in excellent agreement with the
results of equipartition arguments (Beck 2002).

While synchroton radiation from a single electron is elliptically
polarized, the emission from an ensemble of electrons is only
partially polarized.  The polarization degree $p$ is defined as the
ratio of the intensity of linearly polarized radiation to the total
intensity.  For a regular magnetic field and power-law electron
distribution (Eq.\,\EC{electron_df}) $p$ is fixed by the spectral
index $\gamma$.  In particular, if the source is optically thin with
respect to synchrotron emission (a good assumption for galaxies and
clusters),

\be{partial_p}
p = p_H \equiv \frac{\gamma + 1}{\gamma + 7/3}
\ee

\noindent 
(Ginzburg \& Syrovatskii 1964; Ruzmaikin, Shukurov, \& Sokoloff 1988a).
For values of $\gamma$ appropriate to spiral galaxies, this implies a
polarization degree in the range $p=0.72-0.74$.  The observed values
--- $p=0.1-0.2$ for the typical spiral --- are much smaller.

There are various effects which can lead to the depolarization of the
synchrotron emission observed in spiral galaxies.  These effects
include the presence of a fluctuating component to the magnetic field,
inhomogeneities in the magneto-ionic medium and relativistic electron
density, Faraday depolarization (see below) and beam-smearing (see,
for example, Sokoloff et al.\,1998).  Heuristic arguments by Burn
(1966) suggest that for the first of these effects, the polarization
degree is reduced by a factor equal to the ratio of the energy density
of the regular field $\overline{B}$ to the energy density of the
total field:

\be{p_reduced}
p = p_H\frac{\overline{B}^2}{B^2}~.
\ee

\noindent
(This expression is useful only in a statistical sense since one does
not know {\it a priori} the direction of the regular field.)  Thus,
perhaps only $\sim 25\%$ of the total magnetic field energy in a
typical spiral is associated with the large-scale component.  Of
course, the ratio $\overline{B}/B$ would be higher if other
depolarization effects were important.

\subsubsection{Faraday rotation}  

Electromagnetic waves, propagating through a region of both magnetic
field and free electrons, experience Faraday rotation wherein left and
right-circular polarization states travel with different phase
velocities.  For linearly polarized radiation, this results in a
rotation with time (or equivalently path length) of the electric field
vector by an angle

\be{rotation_angle}
\varphi = \frac{e^3\lambda^2}{2\pi m_e^2 c^4}\int_{0}^{l_s} 
n_e(l) B_\parallel(l) dl + \varphi_0
\ee

\noindent 
where $m_e$ is the mass of the electron, $\lambda$ is the wavelength
of the radiation, $\varphi_0$ is the initial polarization angle, and
$B_\parallel$ is the line-of-sight component of the magnetic field.
Here, $n_e(l)$ is the density of thermal electrons along the line of
sight from the source ($l=l_s$) to the observer ($l=0$).  $\varphi$ is
usually written in terms of the rotation measure, RM:

\be{rotation_measure}
\varphi = \left (RM\right )\lambda^2 + \varphi_0
\ee

\noindent where

\bea{RM_def}
RM &\equiv &\frac{e^3}{2\pi m_e^2 c^4}
\int_{0}^{l_s} 
n_e(l) B_\parallel(l) dl\nonumber \\
&\simeq & 
810\,\frac{{\rm rad}}{m^2}\int_{0}^{l_s}\, 
\left (\frac{n_e}{{\rm cm}^{-3}}\right )
\left (\frac{B_\parallel}{\mu{\rm G}}\right )
\left (\frac{dl}{{\rm kpc}}\right )
\eea

\noindent 
In general, the polarization angle must be measured at three or more
wavelengths in order to determine RM accurately and remove the
$\varphi\equiv\varphi\pm n\pi$ degeneracy.

By convention, RM is positive (negative) for a magnetic field directed
toward (away from) the observer.  The Faraday rotation 
angle includes contributions from all magnetized regions along
the line of sight to the source.  Following Kronberg \& Perry (1982)
we decompose RM into three basic components:

\be{RM_decompose}
RM = RM_g + RM_s + RM_{ig}
\ee

\noindent
where $RM_g,~RM_s,$ and $RM_{ig}$ are respectively the contributions
to the rotation measure due to the Galaxy, the source itself, and the
intergalactic medium.

Faraday rotation from an extended source leads to a decrease in the
polarization: The combined signal from waves originating in different
regions of the source will experience different amounts of Faraday
rotation thus leading to a spread in polarization directions.  Faraday
depolarization can, in fact, be a useful measure of magnetic field in
the foreground of a source of polarized synchrotron emission.

\subsubsection{Zeeman Splitting}  

In vacuum, the electronic energy levels of an atom are independent of
the direction of its angular momentum vector.  A magnetic field lifts
this degeneracy by picking out a particular direction in space.  If
the total angular momentum of an atom is ${\bf J}$ ($=$ spin ${\bf S}$
plus orbital angular momentum ${\bf L}$) there will be $2j+1$ levels
where $j$ is the quantum number associated with ${\bf J}$.  The
splitting between neighboring levels is $\Delta E = g\mu B$ where $g$
is the Lande factor which relates the angular momentum of an atom to
its magnetic moment and $\mu = e\hbar/2m_e c= 9.3\times 10^{-21}\,{\rm
erg\,G}^{-1}$ is the Bohr magneton.  This effect, known as Zeeman
splitting, is of historical importance as it was used by Hale (1908)
to discover magnetic fields in sunspots, providing the first known
example of extraterrestrial magnetic fields.

Zeeman splitting provides the most direct method available for
observing astrophysical magnetic fields.  Once $\Delta E$ is measured,
$\overline{B}$ can be determined without additional assumptions.
Moreover, Zeeman splitting is sensitive to the regular magnetic field
at the source.  By contrast, synchrotron emission and Faraday rotation
probe the line-of-sight magnetic field.

Unfortunately, the Zeeman effect is extremely difficult to observe.
The line shift associated with the energy splitting is

\be{zeeman}
\frac{\Delta\nu}{\nu} = 1.4 g\left (\frac{B}{\mu G}\right )
\left (\frac{{\rm Hz}}{\nu}\right )~.
\ee

\noindent
For the two most common spectral lines in Zeeman-effect observations
--- the 21 cm line for neutral hydrogen and the 18 cm OH line for
molecular clouds --- $\Delta \nu/\nu\simeq 10^{-9}g\left (B/\mu
G\right )$.  A shift of this amplitude is to be compared with Doppler
broadening, $\Delta\nu/\nu \simeq v_T/c\simeq 6\times 10^{-7}\left
(T/100\,K\right )^{1/2}$ where $v_T$ and $T$ are the mean thermal
velocity and temperature of the atoms respectively.  Therefore Zeeman
splitting is more aptly described as abnormal broadening, i.e., a
change in shape of a thermally broadened line.  Positive detections
have been restricted to regions of low temperature and high magnetic
field.
 
Within the Galaxy, Zeeman effect measurements have provided
information on the magnetic field in star forming regions and near the
Galactic center.  Of particular interest are studies of Zeeman
splitting in water and $OH$ masers.  Reid \& Silverstein (1990), for
example, used observations of 17 $OH$ masers to map the large-scale
magnetic field of the Galaxy.  Their results are consistent with those
found in radio observations and, as they stress, provide {\it in situ}
measurements of the magnetic field as opposed to the integrated field
along the line-of-sight.  Measurements of Zeeman splitting of the
$21\,{\rm cm}$ line have been carried out for a variety of objects.
Kaz\`{e}s, Troland, \& Crutcher (1991), for example, report positive
detections in High Velocity HI clouds as well as the active galaxy NGC
1275 in the Perseus cluster.  However, Verschuur (1995) has challenged
these results, suggesting that the claimed detections are spurious
signals, the result of confusion between the main beam of the
telescope and its sidelobes.  Thus, at present, there are no confirmed
detections of Zeeman splitting in systems beyond the Galaxy.

\subsubsection{Polarization of Optical Starlight}

Polarized light from stars can reveal the presence of large-scale
magnetic fields in our Galaxy and those nearby.  The first
observations of polarized starlight were made by Hiltner (1949a,b) and
Hall (1949).  Hiltner was attempting to observe polarized radiation
produced in the atmosphere of stars by studying eclipsing binary
systems.  He expected to find time-variable polarization levels of
1-2\%.  Instead, he found polarization levels as high as 10\% for some
stars but not others.  While the polarization degree for individual
stars did not show the expected time-variability, polarization levels
appeared to correlate with position in the sky.  This observation led
to the conjecture that a new property of the interstellar medium (ISM)
had been discovered.  Coincidentally, it was just at this time that
Alfv\'{e}n (1949) and Fermi (1949) were proposing the existence of a
galactic magnetic field as a means of confining cosmic rays (See
Trimble 1990 for a further discussion of the early history of this
subject).  A connection between polarized starlight and a galactic
magnetic field was made by Davis and Greenstein (1951) who suggested
that elongated dust grains would have a preferred orientation in a
magnetic field: for prolate grains, one of the short axes would
coincide with the direction of the magnetic field.  The grains, in
turn, preferentially absorb light polarized along the long axis of the
grain, i.e., perpendicular to the field.  The net result is that the
transmitted radiation has a polarization direction parallel to the
magnetic field.

Polarization of optical starlight has limited value as a probe of
extragalactic magnetic fields for three reasons.  First, there is at
least one other effect that can lead to polarization of starlight,
namely anisotropic scattering in the ISM.  Second, the starlight
polarization effect is self-obscuring since it depends on extinction.
There is approximately one magnitude of visual extinction for each 3\%
of polarization (see, for example, Scarrott, Ward-Thompson, \&
Warren-Smith 1987).  In other words, a 10\% polarization effect must
go hand in hand with a factor of 20 reduction in luminosity.  Finally,
the precise mechanism by which dust grains are oriented in a magnetic
field is not well understood (see, for example, the review by
Lazarian, Goodman, \& Myers 1997).

Polarized starlight does provide information that is complementary to
what can be obtained from radio observations.  The classic
polarization study by Mathewson \& Ford (1970) of 1800 stars in the
Galaxy provides a vivid picture of a field that is primarily in the
Galactic plane, but with several prominent features rising above and
below the plane.  In addition, there are examples of galaxies where a
spiral pattern of polarized optical radiation has been observed
including NGC 6946 (Fendt, Beck, \& Neininger 1998), M51 (Scarrott,
Ward-Thompson, \& Warren-Smith 1987), and NGC 1068 (Scarrott et
al.\,1991).  The optical polarization map of M51, for example,
suggests that its magnetic field takes the form of an open spiral
which extends from within $200\,{\rm pc}$ of the galactic center out
to at least $5\,{\rm kpc}$.  Radio polarization data also indicates a
spiral structure for the magnetic field for this galaxy providing
information on the magnetic configuration from $3\,{\rm kpc}$ to
$15\,{\rm kpc}$ (Berkhuijsen et al.\,1997).  Nevertheless, it is
sometimes difficult to reconcile the optical and radio data.  Over
much of the M51 disk, the data indicates that the same magnetic field
gives rise to radio synchrotron emission and to the alignment of dust
grains (Davis-Greenstein mechanism).  However in one quadrant of the
galaxy, the direction of the derived field lines differ by $\sim
60^\circ$ suggesting that either the magnetic fields responsible for
the radio and optical polarization reside in different layers of the
ISM or that the optical polarization is produced by a mechanism other
than the alignment of dust grains by the magnetic field.

\subsection{Spiral Galaxies}

Spiral galaxies are a favorite laboratory for the study of cosmic
magnetic fields.  There now exist estimates for the magnetic field
strength in well over $100$ spirals and, for a sizable subset of those
galaxies, detailed studies of their magnetic structure and morphology.

\subsubsection{Field Strength}

The magnetic field of the Galaxy has been studied through synchrotron
emission, Faraday rotation, optical polarization, and Zeeman
splitting.  The latter provides a direct determination of the {\it in
situ} magnetic field at specific sites in the Galaxy.  Measurements of
the $21$-cm Zeeman effect in Galactic HI regions reveal regular
magnetic fields with $\overline{B}\simeq 2-10\,\mu{\rm G}$, the higher
values being found in dark clouds and HI shells (Heiles 1990 and
references therein).  Similar values for the Galactic field have been
obtained from Faraday rotation surveys of galactic and extragalactic
sources (i.e., estimates of $RM_{g}$).  Manchester (1974) has compiled
RM data for 38 nearby pulsars and was able to extract the Galactic
contribution.  He concluded that the coherent component of the local
magnetic field is primarily toroidal with a strength
$\overline{B}\simeq 2.2\pm 0.4\,\mu{\rm G}$.  Subsequent RM studies
confirmed this result and provided information on the global structure
of the Galactic magnetic field (see for example Rand \& Lyne 1994 and
also Frick, Stepanov, Shukurov, \& Sokoloff (2001) who describe a new
method for analysing RM data based on wavelets).

Early estimates of the strength of the magnetic field from synchrotron
data were derived by Phillipps et al.\,(1981).  Their analysis was
based on a model for Galactic synchrotron emission in which the
magnetic field in the Galaxy is decomposed into regular and tangled
components.  An excellent fit to the data was obtained when each
component was assumed to have a value of $3\,{\rm \mu G}$.  More recent
estimates give $\sim 4\mu{\rm G}$ for the regular and $\sim 6\mu{\rm
G}$ for the total local field strength (Beck 2002).

Magnetic fields in other galaxies are studied primarily through
synchrotron and Faraday rotation observations.  An interesting case is
provided by M31.  Polarized radio emission in this galaxy is confined
to a prominent ring $\sim 10\,{\rm kpc}$ from the galaxy's center.
The equipartition field strength in the ring is found to be $\sim
4\,\mu{\rm G}$ for both regular and random components.

Fitt \& Alexander (1993) applied the minimum energy method to a sample
of 146 late-type galaxies.  The distribution of field strengths across
the sample was found to be relatively narrow with an average value of
$\langle\beq\rangle \simeq 11\pm 4\, \mu{\rm G}$ (using $k=100$), in
agreement with earlier work by Hummel et al.\,(1988).  The
magnetic field strength does not appear to depend strongly on galaxy
type although early-type galaxies have a slightly higher mean.

A few galaxies have anomalously strong magnetic fields.  A favorite
example is M82 where the field strength, derived from radio continuum
observations, is $\simeq 50\,\mu{\rm G}$ (Klein, Wielebinski, \& Morsi
1988).  This galaxy is characterized by an extraordinarily high star
formation rate.

\subsubsection{Global Structure of the Magnetic Field in Spirals}

Analysis of RM data as well as polarization maps of synchrotron
emission can be used to determine the structure of magnetic fields in
galaxies.  It is common practice to classify the magnetic field
configurations in disk galaxies according to their symmetry properties
under rotations about the spin axis of the galaxy.  The simplest
examples are the axisymmetric and bisymmetric spiral patterns shown in
Figure 1.  In principle, an RM map can distinguish between the
\begin{figure}
\begin{center}
\begin{tabular}{|c|}\hline
\psfig{file=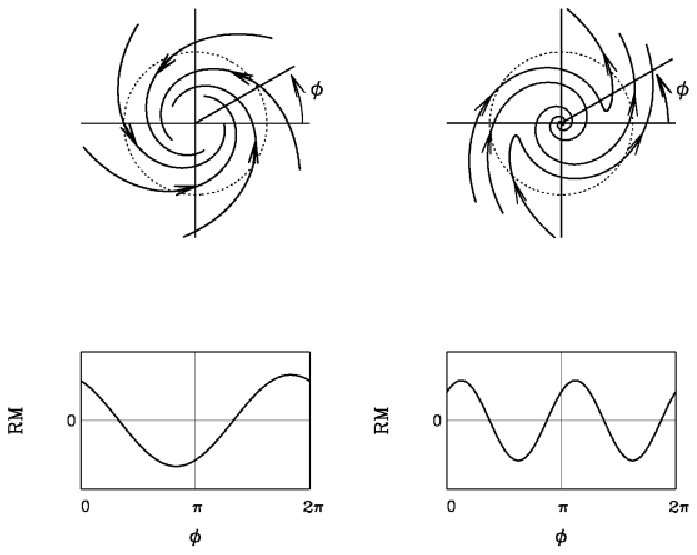,width=10cm,height=9cm} \\ \hline
\end{tabular}
\end{center}

\caption{Axisymmetric and bisymmetric field configurations for disk
systems along with the corresponding RM vs. $\phi$ plots.  Top panels
show toroidal field lines near the equatorial plane.  
Lower panels show RM as a function of azimuthal angle $\phi$ for 
observations at the circle (dotted line) in the corresponding 
top panel.  Note that the pitch angle has the opposite sign
for the two cases shown.}

\end{figure}
different possibilities (Tosa \& Fujimoto 1978; Sofue, Fujimoto, \&
Wielebinski 1986).  For example, one can plot RM as a function of the
azimuthal angle $\phi$ at fixed physical distance from the galactic
center.  The result will be a single (double) periodic distribution
for a pure axisymmetric (bisymmetric) field configuration.  The
RM-$\phi$ method has a number of weaknesses as outlined in Ruzmaikin,
Sokoloff, Shukurov, \& Beck (1990) and Sokoloff, Shukurov, \& Krause
(1992).  In particular, the method has difficulty disentangling a
magnetic field configuration that consists of a superposition of
different modes.  In addition, determination of the RM is plagued by
the ``$n\pi$ degeneracy'' and therefore observations at a number of
wavelengths is required.  An alternative is to consider the
polarization angle $\psi$ as a function of $\phi$.  and model
$\psi(\phi)$ as a Fourier series: $\psi(\phi) = \sum_n
a_n\cos{(n\phi)}+ b_n\sin{(n\phi)}$.  The coefficients $a_n$ and $b_n$
then provide a picture of the azimuthal structure of the field.  Of
course, if an estimate of the field strength is desired,
multiwavelength observations are again required (Ruzmaikin, Sokoloff,
Shukurov, \& Beck 1990; Sokoloff, Shukurov, \& Krause 1992).

In M31, both $RM(\phi)$ and $\psi(\phi)$ methods suggest strongly that
the regular magnetic field in the outer parts of the galaxy (outside
the synchroton emission ring) is described well by an axisymmetric
field.  Inside the ring, the field is more complicated and appears to
have a significant admixture of either $m=1$ or $m=2$ modes
(Ruzmaikin, Sokoloff, Shukurov, \& Beck 1990).  These higher harmonics
may be an indication that the dynamo is modulated by the two-arm
spiral structure observed in this region of the galaxy.  The polarized
synchrotron emissivity along the ring may provide a further clue as to
the structure of the magnetic field.  The emissivity is highly
asymmetric --- in general much stronger along the minor axis of the
galaxy.  Urbanik, Otmianowska-Mazur, \& Beck (1994) suggested that
this pattern in emission are better explained by a superposition of
helical flux tubes that wind along the axis of the ring rather than a
pure azimuthal field.  (For a further discussion of helical flux tubes
in the context of the $\aod$-dynamo see Donner \& Brandenburg 1990).

Field configurations in disk galaxies can also be classified according
to their symmetry properties with respect to reflections about the
central plane of the galaxy.  Symmetric, or even parity field
configurations are labeled Sm where, as before, $m$ is the azimuthal
mode number.  Antisymmetric or odd parity solutions are labelled $Am$.
Thus as S0 field configuration is axisymmetric (about the spin axis)
and symmetric about the equatorial plane.  An A0 configuration is also
axisymmetric but is antisymmetric with respect to the equatorial plane.
As shown in Figure 2, the poloidal component of a symmetric
(antisymmetric) field configuration has a quadruple (dipole)
structure.


\begin{figure}
\centerline{\psfig{file=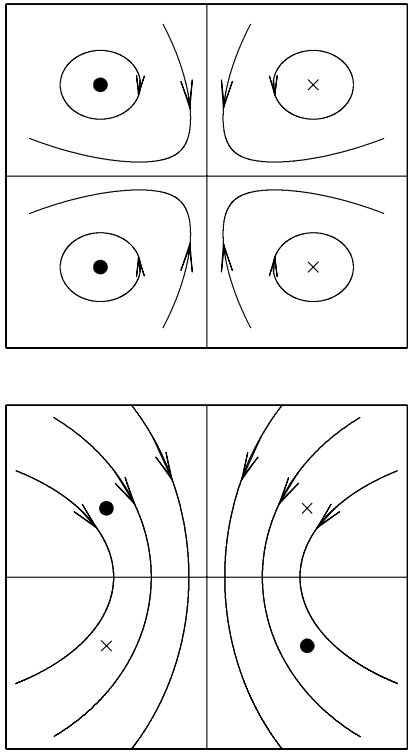,width=12.5cm}}

\caption{Field lines for even (top panel) and odd (bottom panel)
configurations.  Shown are cross-sections perpendicular to the equatorial
plane and containing the symmetry axis of the galaxy (i.e., poloidal
planes).  The toroidal field is indicated by an `x' (field out of the page)
or `dot' (field into the page).}

\label{parity}
\end{figure}


The parity of a field configuration in a spiral galaxy is extremely
difficult to determine.  Indeed, evidence in favor of one or the other
type of symmetry has been weak at best and generally inconclusive
(Krause \& Beck 1998).  One carefully studied galaxy is the Milky Way
where the magnetic field has been mapped from the RMs of galactic and
extragalactic radio sources.  An analysis by Han, Manchester,
Berkhuijsen, \& Beck (1997) of over 500 extragalactic objects suggests
that the field configuration in the inner regions of the Galaxy is
antisymmetric about its midplane.  On the other hand the analysis by
Frick et al.\,(2001) indicates that the field in the solar
neighborhood is symmetric.  Evidently, the parity of the field
configuration can change from one part of a galaxy to another.  A
second well-studied case is M31 where an analysis of the RM across its
disk suggests that the magnetic field is symmetric about the
equatorial plane, i.e., an even-parity axisymmetric (S0) configuration
(Han, Beck, \& Berkhuijsen 1998).

Among S0-type galaxies, there is an additional question as to the
direction of the magnetic field, namely whether the field is oriented
inward toward the center of the galaxy or outward (Krause \& Beck
1998).  The two possibilities can be distinguished by comparing the
sign of the RM (as a function of position on the disk) with velocity
field data.  Krause \& Beck (1998) point out that in four of five
galaxies where the field is believed to be axisymmetric, those fields
appear to be directed inward.  This result is somewhat surprising
given that a magnetic dynamo shows no preference for one type of
orientation over the other.  It would be premature to draw conclusions
based on such a small sample.  Nevertheless, if, as new data becomes
available, a preference is found for inward over outward directed
fields (or more realistically, a preference for galaxies that are in
the same region of space to have the same orientation), it would
reveal a preference in initial conditions and therefore speak directly
to the question of seed fields.

\subsubsection{Connection with Spiral Structure}

Often, the spiral magnetic structures detected in disk galaxies appear
to be closely associated with the material spiral arms.  A possible
connection between magnetic and optical spiral structure was first
noticed in observations of M83 (Sukumar \& Allen 1989), IC 342 and M81
(Krause, Hummel, \& Beck 1989a, 1989b).  A particularly striking
example of magnetic spiral structure is found in the galaxy NGC 6946,
as shown in Figure 3 (Beck \& Hoernes 1995; Frick et al.\,2000).  In
each case, the map of linearly polarized synchrotron emission shows
clear evidence for spiral magnetic structures across the galactic
disk.  The magnetic field in IC 342 appears to be an inwardly-directed
axisymmetric spiral while the field in M81 is more suggestive of a
bisymmetric configuration (Sofue, Takano, \& Fujimoto 1980; Krause,
Hummel, \& Beck 1989a, 1989b; Krause 1990).  In many cases, magnetic
spiral arms are strongest in the regions between the optical spiral
arms but otherwise share the properties (e.g., pitch angle) of their
optical counterparts.  These observations suggest that either the
dynamo is more efficient in the interarm regions or that magnetic
fields are disrupted in the material arms.  For example, Mestel \&
Subramanian (1991, 1993) proposed that the $\alpha$-effect of the
standard dynamo contains a non-axisymmetric contribution whose
configuration is similar to that of the material spiral arms.  The
justification comes from one version of spiral arm theory in which the
material arm generates a spiral shock in the interstellar gas.  The
jump in vorticity in the shock may yield an enhanced $\alpha$-effect
with a spiral structure.  Further theoretical ideas along these lines
were developed by Shukurov (1998) and a variety of numerical
simulations which purport to include nonaxisymmetric turbulence have
been able to reproduce the magnetic spiral structures found in disk
galaxies (Rohde \& Elstner 1998; Rohde, Beck, \& Elstner 1999;
Elstner, Otmianowska-Mazur, von Linden, \& Urbanik 2000).  Along
somewhat different lines, Fan \& Lou (1996) attempted to explain
spiral magnetic arms in terms of both slow and fast
magnetohydrodynamic waves.


\begin{figure}
\centerline{
\psfig{file=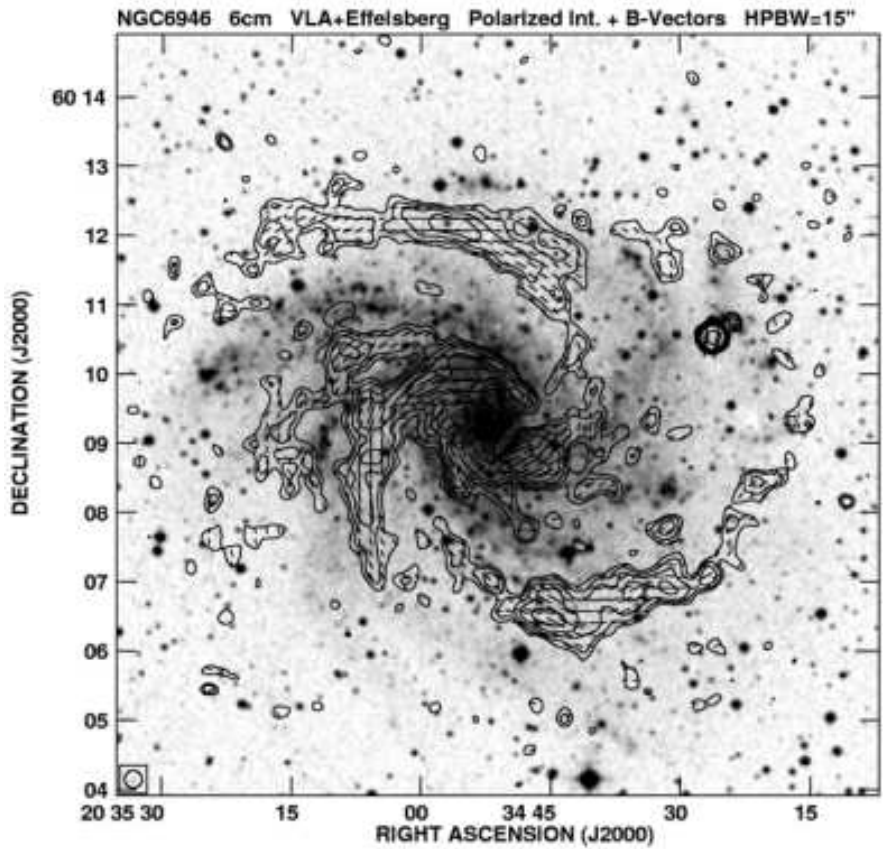,width=12cm,height=12cm}
}

\caption{Polarized synchrotron intensity (contours) and magnetic field
orientation of NGC 6946 (obtained by rotating $E$-vectors by
$90^\circ$) observed at $\lambda 6.2\,{\rm cm}$ with the VLA (12.5
arcsec synthesized beam) and combined with extended emission observed
with the Effelsberg $100-{\rm m}$ telescope (2.5 arcmin resolution).
The lengths of the vectors are proportional to the degree of
polarization.  (From Beck \& Hoernes 1996.)}

\label{NGC6946}
\end{figure}


Recently, Beck et al.\,(1999) discovered magnetic fields in the barred
galaxy NGC 1097.  Models of barred galaxies predict that gas in the
region of the bar is channeled by shocks along highly non-circular
orbits.  The magnetic field in the bar region appears to be aligned
with theoretical streamlines suggesting that the field is mostly
frozen into the gas flow in contrast with what is expected for a
dynamo-generated field.  The implication is that a dynamo is required
to generate new field but that inside the bar simple stretching by the
gas flow is the dominant process (see Moss et al.~2001).

\subsubsection{Halo Fields}

Radio observations of magnetic fields in edge-on spiral galaxies
suggest that in most cases the dominant component of the magnetic
field is parallel to the disk plane (Dumke, Krause, Wielebinski, \&
Klein 1995).  However, for at least some galaxies, magnetic fields are
found to extend well away from the disk plane and have strong vertical
components.  Hummel, Beck, and Dahlem (1991) mapped two such galaxies,
NGC 4631 and NGC 891, in linearly polarized radio emission and found
fields with strength $\sim 5$ and $\sim 8\,\mu{\rm G}$ respectively
with scale heights $\sim 5-10\,{\rm kpc}$.  The fields in these two
galaxies have rather different characteristics: In NGC 4631 (Figure
4), numerous prominent radio spurs are found throughout the halo.  In
all cases where the magnetic field can be determined, the field
follows these spurs (Golla \& Hummel 1994).  (Recent observations by
T\"{u}llmann et al.\,(2000) revealed similar structures in the edge-on
spiral NGC 5775.)  Moreover, the large-scale structure of the field is
consistent with that of a dipole configuration (anti-symmetric about
the galactic plane) as in the bottom panel of Figure 2.  The field in
NGC 891 is more disorganized, that is, only ordered in small regions
with no global structure evident.

\begin{figure}
\centerline{\psfig{file=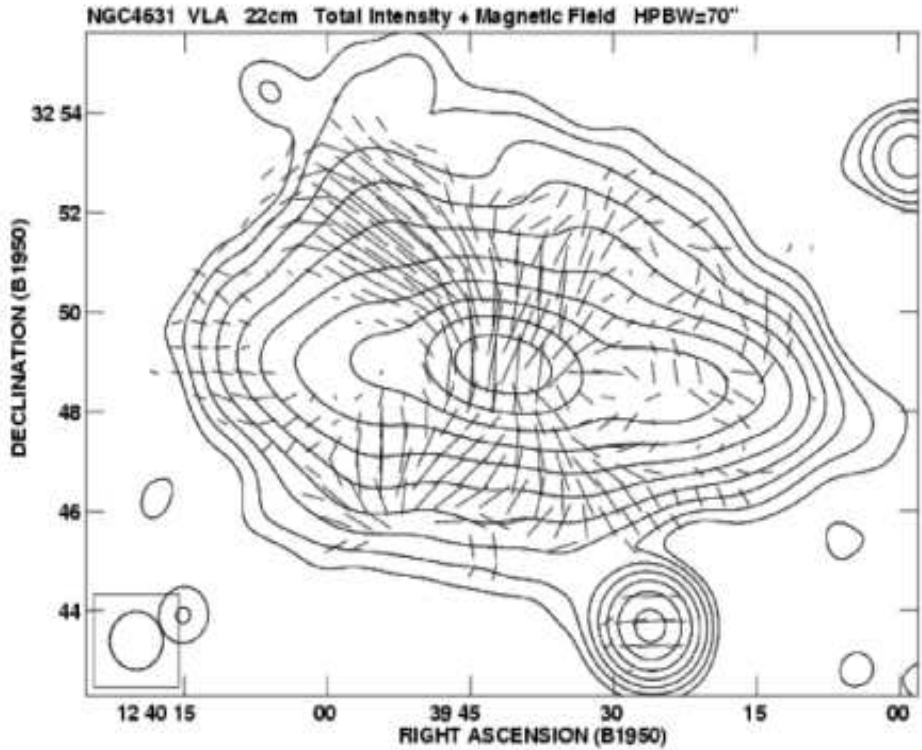,width=12.5cm,height=12.5cm,angle=-90}}

\caption{Total radio emission and $B$ vectors of polarized emission of
NGC 4631 at $\lambda 22\,{\rm cm}$ (VLA, $70''$ synthesized beam).
The $B$ vectors have been corrected for Faraday rotation; their length
is proportional to the polarized intensity (From Krause \& Beck,
unpublished)}

\label{NGC4631}
\end{figure}


Magnetic fields are but one component of the ISM found in the halos of
spiral galaxies.  Gas (which exists in many different phases), stars,
cosmic rays, and interstellar dust, are also present.  Moreover, the
disk and halo couple as material flows out from the disk and into the
halo only to eventually fall back completing a complex circulation of
matter (see, for example, Dahlem 1997).  At present, it is not clear
whether halo fields are the result of dynamo action in the halo or
alternatively, fields produced in the disk and carried into the halo
by galactic winds or magnetic buoyancy (see Section IV.G). 

\subsubsection{Far Infrared-Radio Continuum Correlation}

An observation that may shed light on the origin and evolution of
galactic magnetic fields is the correlation between galactic far
infrared (FIR) emission and radio continuum emission.  This
correlation was first discussed by Dickey \& Salpeter (1984) and de
Jong, Klein, Wielebinski, \& Wunderlich (1985).  It is valid for
various types of galaxies including spirals, irregulars, and cluster
galaxies and has been established for over four orders of magnitude in
luminosity (see Niklas \& Beck (1997) and references therein).  The
correlation is intriguing because the FIR and radio continuum
emissions are so different.  The former is thermal and presumably
related to the star formation rate (SFR).  The latter is mostly
nonthermal and produced by relativistic electrons in a magnetic field.
Various proposals to explain this correlation have been proposed (For
a review, see Niklas \& Beck (1997).)  Perhaps the most appealing
explanation is that both the magnetic field strength and the star
formation rate depend strongly on the volume density of cool gas
(Niklas \& Beck 1997).  Magnetic field lines are anchored in gas
clouds (Parker 1966) and therefore a high number density of clouds
implies a high density of magnetic field lines.  Likewise, there are
strong arguments in favor of a correlation between gas density and the
SFR of the form SFR$\propto \rho^n$ (Schmidt 1959).  With an index
$n=1.4\pm 0.3$, taken from survey data of thermal radio emission
(assumed to be an indicator of the SFR), where able to provide a
self-consistent picture of the FIR and radio continuum correlation.

\subsection{Elliptical and Irregular Galaxies}

Magnetic fields are ubiquitous in elliptical galaxies though they are
difficult to observe because of the paucity of relativistic electrons.
Nevertheless, their presence is revealed through observations of
synchrotron emission.  In addition, Faraday rotation has been observed
in the polarized radio emission of background objects.  One example is
that of a gravitationally lensed quasar where the two quasar images
have rotation measures that differ by $100\,{\rm rad\,m^{-2}}$
(Greenfield, Roberts, \& Burke 1985).  The conjecture is that light
for one of the images passes through a giant cD elliptical galaxy
whose magnetic field is responsible for the observed Faraday rotation.
A more detailed review of the observational literature can be found in
Moss \& Shukurov (1996).  These authors stress that while the evidence
for microgauss fields in ellipticals is strong, there are no positive
detections of polarized synchrotron emission or any other
manifestation of a regular magnetic field.  Thus, while the inferred
field strengths are comparable to those found in spiral galaxies, the
coherence scale for these fields is much smaller than the scale of the
galaxy itself.

Recently, magnetic fields were observed in the dwarf irregular galaxy
NGC 4449.  The mass of this galaxy is an order of magnitude lower than
that of the typical spiral and shows only weak signs of global
rotation.  Nevertheless, the regular magnetic field is measured to be
$6-8\,\mu\,{\rm G}$, comparable to that found in spirals (Chyzy et
al.\,2000).  Large domains of non-zero Faraday rotation indicate that
the regular field is coherent on the scale of the galaxy.  This field
appears to be composed of two distinct components.  First, there is a
magnetized ring $2.2\,{\rm kpc}$ in radius in which clear evidence for
a regular spiral magnetic field is found.  This structure is
reminiscent of the one found in M31.  Second, there are radial
``fans'' -- coherent magnetic structures that extend outward from the
central star forming region.  Both of these components may be
explained by dynamo action though the latter may also be due to
outflows from the galactic center which can stretch magnetic field
lines.

\subsection{Galaxy Clusters}

Galaxy clusters are the largest non-linear systems in the Universe.
X-ray observations indicate that they are filled with a tenuous hot
plasma while radio emission and RM data reveal the presence of
magnetic fields.  Clusters are therefore an ideal laboratory to test
theories for the origin of extragalactic magnetic fields (see, for
example, Kim, Tribble, \& Kronberg 1991; Tribble 1993).

Data from the Einstein, ROSAT, Chandra, and XMM-Newton observatories
provide a detailed picture of rich galaxy clusters.  The intracluster
medium is filled with a plasma of temperature $T\simeq 10^7-10^8\,K$
that emits X-rays with energies $\sim 1-10\,{\rm keV}$.  Rich clusters
appear to be in approximate hydrostatic equilibrium with virial
velocities $\sim 1000\,{\rm km\,s^{-1}}$ (see, for example, Sarazin
1986).  In some cluster cores, the cooling time for the plasma due to
the observed X-ray emission is short relative to the dynamical time.
As the gas cools, it is compressed and flows inward under the combined
action of gravity and the thermal pressure of the hot outer gas
(Fabian, Nulsen, \& Canizares 1984).  These cooling flows are found in
elliptical galaxies and groups as well as clusters.  The primary
evidence for cooling flows comes from X-ray observations.  In
particular, a sharp peak in the X-ray surface brightness distribution
is taken as evidence for a cooling flow since it implies that the gas
density is rising steeply towards the cluster center (see, for
example, Fabian 1994).

A small fraction of rich clusters have observable radio halos.
Hanisch (1982) examined data from four well-documented examples and
found that radio-halo clusters share a number of properties ---
principally, a large homogeneous hot intracluster medium and the
absence of a central dominant (cD) galaxy.  He concluded that radio
halos are short-lived phenomena, symptoms of a transient state in the
lifetime of a cluster.

Magnetic fields appear to exist in galaxy clusters regardless of
whether there is evidence of cooling flows or extended radio emission.
Taylor, Barton, \& Ge (1994), working from the all-sky X-ray sample of
galaxies of Edge et al.\,(1992), concluded that over half of all
cooling flow clusters have $RM>800\rmt$ and a significant number have
$RM>2000\,\rmt$.  Furthermore, they found a direct correlation between
the cooling flow rate and the observed RM.  Estimates of the
regular magnetic field strength for clusters in their sample range from
$0.2-3\,\mu{\rm G}$.

Evidence for magnetic fields in radio-halo clusters is equally strong.
Kim et al.\,(1990) determined the RM for 18 sources behind the Coma
cluster and derived an intracluster field strength of $B\sim 2.5\left
(L/10\,{\rm kpc}\right )^{-1/2}\,\mu{\rm G}$ where $L$, a model
parameter, is the typical scale over which the field reverses
direction.  Unfortunately, for most clusters, there are no more than a
few radio sources strong enough to yield RM measurements.  To
circumvent this problem, several authors, beginning with Lawler \&
Dennison (1982), employed a statistical approach by combining data
from numerous clusters.  For example, Kim, Tribble, \& Kronberg (1991)
used data from $\sim 50$ clusters (including radio-halo and
cooling-flow clusters), to plot the RM of background radio sources as
a function of their impact parameter from the respective cluster
center.  The dispersion in RMs rises from a background level of
$15\,\rmt$ to $\sim 200\,\rmt$ near the cluster center revealing the
presence of magnetic fields in most, if not all, of the clusters in
the sample.  Recently, Clarke, Kronberg, \& B\"{o}hringer (2001)
completed a similar study of 16 ``normal'' Abell clusters selected to
be free of widespread cooling flows and strong radio halos.  Once
again, the dispersion in RM is found to increase dramatically at low
impact parameters indicating strong ($0.1-1\,\mu G$) magnetic fields
on scales of order $10{\rm} \,{\rm kpc}$.

Radio emission is of course produced by relativistic electrons
spiralling along magnetic field lines.  These same electrons can
Compton scatter CMB photons producing a non-thermal spectrum of X-rays
and $\gamma$-rays.  At high energies, these Compton photons can
dominate the thermal X-ray emission of the cluster (e.g., Rephaeli
1979).  In contrast to synchrotron emission, the flux of the Compton
X-rays is a decreasing function of $B$ --- $j_c(\nu)\propto B^{-\left
(1+\gamma\right )/2}$ for the power-law electron distribution in
Eq.\,\EC{electron_df} --- so that an upper limit on the non-thermal
X-ray flux translates to a lower limit on the magnetic field strength
in the cluster.  Using this method, Rephaeli \& Gruber (1988) found a
lower limit $\sim 10^{-7}\,\mu{\rm G}$ for several Abell clusters
in agreement with the positive detections described above.

\subsection{Extracluster Fields}

There are hints that magnetic fields exist on supercluster scales.
Kim et al.\,(1989) detected faint radio emission in the region between
the Coma cluster and the cluster Abell 1367.  These two clusters are
$40\,{\rm Mpc}$ apart and define the plane of the Coma supercluster.
Kim et al.\,(1989) observed a portion of the supercluster plane using
the Westerbork Synthesis Radio Telescope.  They subtracted emission
from discrete sources such as extended radio galaxies and found, in
the residual map, evidence of a `bridge' in radio emission (Figure 5).
The size of the `bridge' was estimated to be $1.5h_{75}^{-1}\,{\rm
Mpc}$ in projection where $h_{75}$ is the Hubble constant $H_0$ in
units of $75\,{\rm km\,s^{-1}\,Mpc^{-1}}$.  They concluded that the
`bridge' was a feature of the magnetic field of the Coma-Abell 1367
supercluster with a strength, derived from minimum energy arguments,
of $0.2-0.6\,\mu{\rm G}$.


\begin{figure}
\centerline{\psfig{file=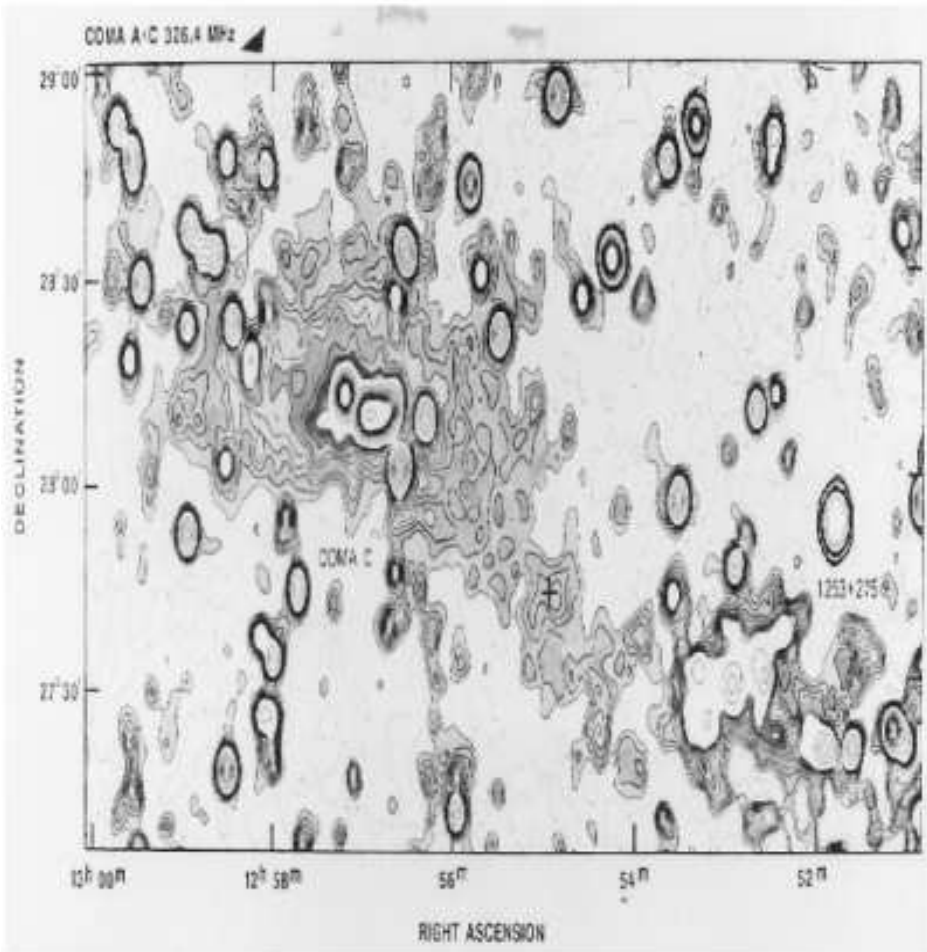,width=12.5cm}}

\caption{WSRT map of the Coma cluster of galaxies at 326\,{\rm MHz}
from 6 12-h observing sessions.  The projected linear of the bridge
is $\sim 1.5h_{75}^{-1}\,{\rm Mpc}$.  The + symbol marks the location
of NGC 4839.  The peak surface brightness is 9.1\,mJy per beam
and contours are shown at -1,1,2,...,9,10,20,...100,200,400 times
4 mJy per peam.}

\label{coma}
\end{figure}

Indirect evidence of extracluster magnetic fields may exist in radio
observations by Ensslin et al.\,(2001) of the giant radio galaxy NGC
315.  New images reveal significant asymmetries and peculiarities in
this galaxy.  These features can be attributed to the motion of the
galaxy through a cosmological shock wave $10-100$ times the dimension
of a typical cluster.  Polarization of the radio emission suggests the
presence of a very-large scale magnetic field associated with the
shock.

\subsection{Galactic Magnetic Fields at Intermediate Redshifts}

Evidence of magnetic fields in galaxies at even moderate redshifts
poses a serious challenge to the galactic dynamo hypothesis since it
would imply that there is limited time available for field
amplification.  At present, the most convincing observations of
galactic magnetic fields at intermediate redshifts come from RM
studies of radio galaxies and quasars.  Kronberg, Perry, \& Zukowski
(1992) obtained an RM map of the radio jet associated with the quasar
PKS 1229-121.  This quasar is known to have a prominent absorption
feature presumably due to an intervening object at $z\simeq 0.395$.
(The intervener has not been imaged optically).  Observations indicate
that the RM changes sign along the ``ridge line'' of the jet in a
quasi-oscillatory manner.  One plausible explanation is that the
intervener is a spiral galaxy with a bisymmetric magnetic field as
illustrated in Figure 6.  Alternatively, the field in the intervening
galaxy might be axisymmetric with reversals along the radial
direction.  (A configuration of this type has been suggested for the  
Milky Way (see, for example, Poedz, Shukurov, \& Sokoloff 1993).)


\begin{figure}
\centerline{\psfig{file=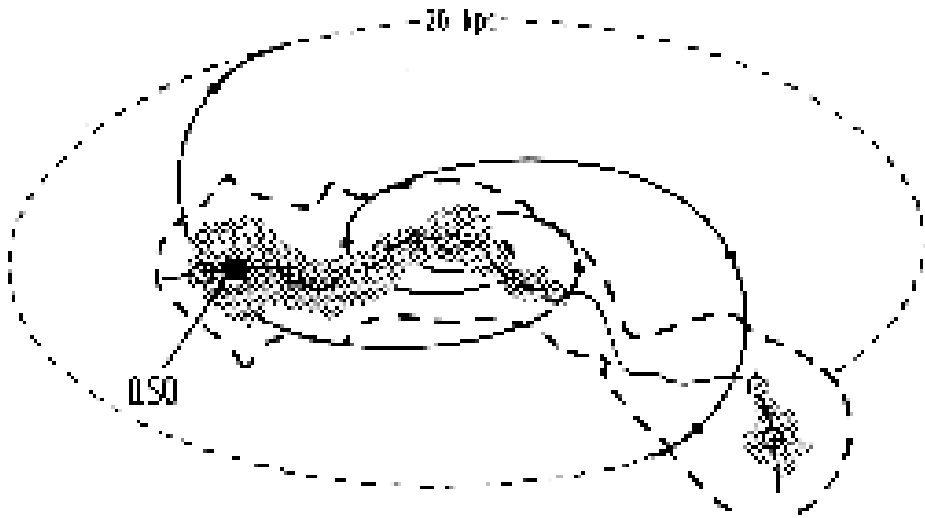,width=12.5cm}}

\caption{Map of the M81 bisymmetric spiral magnetic field (Krause
1990) projected to have the same linear scale at $z=0.395$ as that of
the jet of PKS 1229-021 at the same $z$.  It is shown superposed on
the rotation measure distribution of the jet.  The shaded area shows
the region where RM data have been collected and the dashed outline
shows the approximate region of the total radiation zone.  The ridge
line of the jet is shown, and the positions of the maxima and minima
in RM are shown by circled plus and minus signs indicating magnetic
field directions toward and away from the observer
(from Kronberg, Perry, \& Zukowski 1992).}

\label{pks_rm}
\end{figure}


Athreya et al.\,(1998) studied 15 high redshift ($z\ga 2$) radio
galaxies at multiple frequencies in polarized radio emission and
found significant RMs in almost all of them with several of the
objects in the sample having $RM\ga 1000\,\rmt$.  The highest
RM in the sample is $6000\,\rmt$ for the $z\simeq 2.17$ galaxy
1138-262.  RMs of this magnitude require microgauss fields that are
coherent over several ${\rm kpc}$.

\subsection{Cosmological Magnetic Fields}

A truly cosmological magnetic field is one that cannot be associated
with collapsing or virialized structures.  Cosmological magnetic
fields can include those that exist prior to the epoch of galaxy
formation as well as those that are coherent on scales greater than
the scale of the largest known structures in the Universe, i.e., $\ga
50\,{\rm Mpc}$.  In the extreme, one can imagine a field that is
essentially uniform across our Hubble volume.  At present, we do not
know whether cosmological magnetic fields exist.

Observations of magnetic fields in the Coma supercluster and in
redshift $z\simeq 2$ radio galaxies hint at the existence of
widespread cosmological fields and lend credence to the hypothesis
that primordial fields, amplified by the collapse of a protogalaxy
(but not necessarily by dynamo action), become the microgauss fields
observed in present-day galaxies and clusters.  (An even bolder
proposal is that magnetic fields play an essential role in galaxy
formation.  See, for example, Wasserman 1978; Kim, Olinto, \& Rosner
1996.)  The structure associated with the magnetic `bridge' in the
Coma supercluster (Kim et al.\,1989) is dynamically young so that
there has been little time for dynamo processes to operate.  The
observations by Athreya et al.~(1998) and, to a lesser extent,
Kronberg, Perry, \& Zukowski (1992), imply that a similar problem
exists on galactic scales.  Interest in the primordial field
hypothesis has also been fueled by challenges to the standard dynamo
scenario.

A detection of sufficiently strong cosmological fields would provide
tremendous support to the primordial field hypothesis and at the same
time open a new observational window to the early Universe.  Moreover,
since very weak cosmological fields can act as seeds for the galactic
dynamo the discovery of even the tiniest cosmological field would help
complete the dynamo paradigm.

For the time being, we must settle for limits on the strength of
cosmological fields.  Constraints have been derived from Faraday
rotation studies of high-redshift sources, anisotropy measurements of
the CMB, and predictions of light element abundances from big bang
nucleosynthesis (BBN).

\subsubsection{Faraday Rotation due to a Cosmological Field}

Faraday rotation of radio emission from high redshift sources can be
used to study cosmological magnetic fields.  For a source at a
cosmological distance $l_s$, the rotation measure is given by the
generalization of Eq.\,\EC{RM_def} appropriate to an expanding
Universe:

\be{cosmic_RM}
\frac{RM}{{\rm rad\,m^{-2}}}
~\simeq~ 8.1\times 10^5\int_{0}^{l_s} 
\left (\frac{n_e(l)}{{\rm cm}^{-3}}\right )
\left (\frac{B_\parallel(l)}{\mu{\rm G}}\right )
\left (1+z\right )^{-2}
\frac{dl}{{\rm Mpc}}~.
\ee

\noindent 
The factor of $(1+z)^{-2}$ accounts for the redshift of the
electromagnetic waves as they propagate from source to observer.  We
consider the contribution to this integral from cosmological magnetic
fields.  If the magnetic field and electron density are homogeneous
across our Hubble volume, an all-sky RM map will have a dipole
component (Sofue, Fujimoto, Kawabata 1968; Brecher \& Blumenthal 1970;
Vall\'{e}e 1975; Kronberg 1977; Kronberg \& Simard-Normandin 1976;
Vall\'{e}e 1990).  The amplitude of this effect depends on the
evolution of $B$ and $n_e$.  The simplest assumption is that the
comoving magnetic flux and comoving electron number density are
constant, i.e., $B(z)=B_0\left (1+z\right )^2$ and $n_e(z) =
n_{e0}\left (1+z\right )^3$.  The cosmological component of the RM is
then

\be{RM_open}
\frac{RM_{ig}}{{\rm rad\,m^{-2}}}~\simeq~
3.2\times 10^4 h^{-1}_{75}\cos{\theta}\left (
\frac{n_{e0}}{10^{-5}\,{\rm cm}^{-3}}\right )\,\left (
\frac{B_0}{\mu\,{\rm G}}\right )\,F
\left (\Omega_m,\,\Omega_\Lambda;\,z\right )
\ee

\noindent 
where $\theta$ is the angle between the source and the magnetic field,

\be{def_F}
F(z) = \frac{H_0}{c}
\int_0^{z_s}dz
\left (1+z\right )^3
\frac{dl}{dz}
\ee

\noindent and

\be{dldz}
\frac{H_0}{c}
\frac{dl}{dz} 
= \left (1+z\right )^{-1}
\left (\Omega_m\left (1+z\right )^3+
\left (1-\Omega_m-\Omega_\Lambda\right )\left (1+z\right
)^2+\Omega_\Lambda\right )^{-1/2}~.
\ee

\noindent 
In Figure 7, we plot $F$ and $RM_{ig}$ as a function of $z_s$ for
selected cosmological models.  The path length to a source and hence
the cosmological contribution to the RM are increasing functions of
$z_s$, as is evident in Figure 7.  In addition, for fixed $z_s$, the
$RM_{ig}$ is greater in low-$\Omega_m$ models than in the Einstein-de
Sitter model, a reflection of the fact that the path length per unit
redshift interval is greater in those models.


\begin{figure}
\centerline{\psfig{file=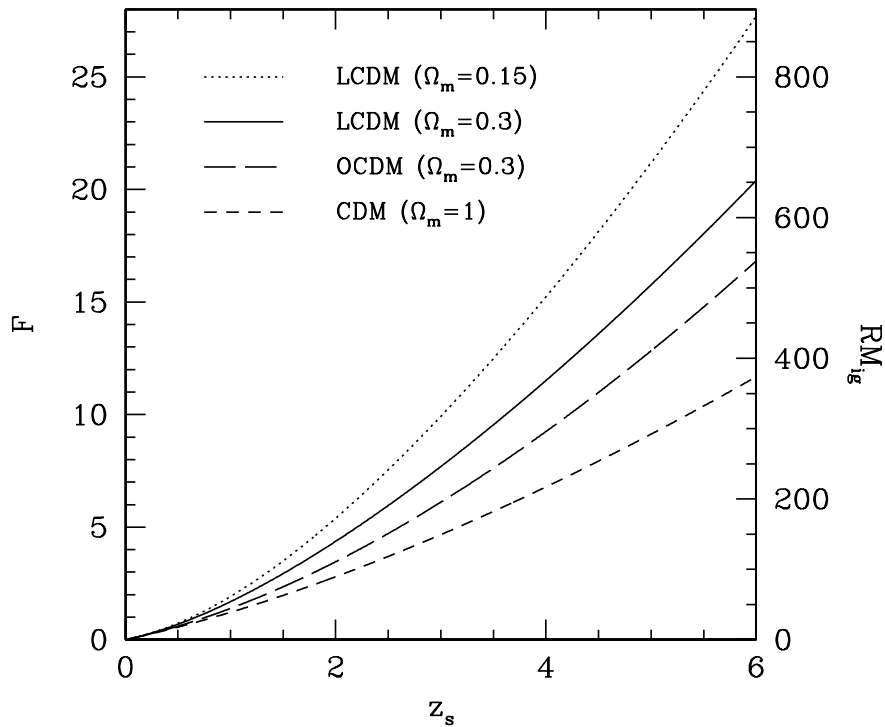,width=12.5cm}}

\caption{Intergalactic contribution to the rotation measure
($RM_{ig}$) as a function of source redshift $z_s$.  The left-hand
vertical axis gives the function $F(z_s)$ as defined in
Eq.\,\EC{def_F}.  The right-hand axis gives $RM_{ig}$ assuming a
$1\,\mu{\rm G}$ field, $h_{75}=1,~n_{e0}=10^{-5}\,{\rm cm}^{-3},$ and
$\theta=0$.  Curves shown are for the standard cold dark matter (CDM) model
($\Omega_m=1$), two LCDM models (CDM with a cosmological constant)
($\Omega_m=0.15; \Omega_\Lambda=0.85$ and and $\Omega_m=0.3;
\Omega_\Lambda=0.7$) and an open CDM model ($\Omega_m=0.3;
\Omega_\Lambda=0$).}

\label{cosmic_rm}
\end{figure}


Eqs.\,\EC{RM_open}-\EC{dldz}, together with RM data for high-redshift
galaxies and quasars, can be used to constrain the strength of
Hubble-scale magnetic fields.  The difficulty is that the source and
Galaxy contributions to the RM are unknown.  (Indeed, Sofue, Fujimoto,
and Kawabata (1968) reported a positive detection of a $10^{-9}\,{\rm
G}$ cosmological field, a result which was refuted by subsequent
studies.)  By and large, the Galactic contribution is $RM_{\rm g}\la
200\,\rmt$ and in general decreases with increasing angle relative to
the Galactic plane.  However, Kronberg and Simard-Normandin (1976)
found that even at high Galactic latitude, some objects have $RM\ga
200\rmt$.  In particular, the high Galactic latitude subsample that
they considered was evidently composed of two distinct populations,
one with $\langle RM^2\rangle^{1/2} \simeq 50\rmt$ and another with
$\langle RM^2\rangle^{1/2} \simeq 200\rmt$.  (A similar decomposition
is not possible at low Galactic latitudes where the contribution to
the RM from the Galactic magnetic field is stronger.  However, the RM
sky at low Galactic latitudes does reveal a wealth of structure in the
Galactic magnetic field (see, for example, Duncan, Reich, Reich, \&
F\"{u}rst (1999) and Gaensler et al.\,(2001).)  This observation
suggests that the best opportunity to constrain Hubble-scale magnetic
fields comes from the high Galactic latitude, low-RM subsample.  For
example, Vall\'{e}e (1990) tested for an RM dipole in a sample of 309
galaxies and quasars.  The galaxies in this sample extended to
$z\simeq 3.6$ though most of the objects were at $z\la 2$.  Vall\'{e}e
derived an upper limit to $RM_{ig}$ of about $2\,\rmt$, corresponding
to an upper limit of $6\times 10^{-12}\,{\rm G} \left
(n_{e0}/10^{-5}\,{\rm cm}^{-3}\right )^{-1}$ on the strength of the
uniform component of a cosmological magnetic field.

If either the electron density or magnetic field vary on scales less
than the Hubble distance, the pattern of the cosmological contribution
to the RM across the sky will be more complicated than a simple
dipole.  Indeed, if variations in $n_e$ and $B$ occur on scales much
less than $c/H_0$, a typical photon from an extragalactic source will
pass through numerous ``Faraday screens''.  In this case, the average
cosmological RM over the sky will be zero.  However, $\sigma_{\rm
RM}^2$, the variance of RM, will increase with $z$.  Kronberg and
Perry (1982) considered a simple model in which clouds of uniform
electron density and magnetic field are scattered at random throughout
the Universe.  The rotation measure associated with a single cloud at
a redshift $z$ is given by

\be{RM_single}
\frac{RM_{c}(z)}{\rm rad\, m^{-2}}~\simeq~
 2.6\left (1+z\right )^{-2}
\left (\frac{B_\parallel}{\mu\,{\rm G}}\right )
\left (\frac{N_e}{10^{19}{\rm cm}^{-2}}\right )
\ee

\noindent 
where $r_c(z)\ll c/H$ and $N_e\equiv \int n_e(z) dl\simeq
2n_e(z)r_c(z)$ are the cloud radius and electron column density
respectively.  For simplicity, all clouds at a given redshift are
assumed to share the same physical characteristics.  The contribution
to $\sigma_{RM}^2$ from clouds between $z$ and $z+dz$ can be written

\be{differential_var}
d\sigma_{RM}^2(z) = RM_{c}(z)^2\frac{d{\cal N}}{dz}dz
\ee

\noindent 
where 

\be{def_dndz}
\frac{d{\cal N}}{dz}=\pi r_c^2(z)n_c(z)\,\frac{dl}{dz}
\ee

\noindent
and $n_c(z)$ is the number density of clouds in the Universe.
Kronberg \& Perry (1982) estimated $\sigma_{RM}^2$ under the
assumption that the comoving number density of clouds, as well as
their comoving size, electron density, and magnetic flux are constant,
i.e., $n_c(z)=n_{c0}\left (1+z\right )^{3},~ r_c(z)=r_{c0}\left
(1+z\right )^{-1},~ n_e(z)=n_{e0}\left (1+z\right )^{3},~$ and
$B_\parallel(z)=B_{0}\left (1+z\right )^{2}$.  The expression for
$\sigma_{RM}^2$ then becomes

\be{sigRM2}
\frac{\sigma_{RM}(z)^2}{{\rm rad^2\,m^{-4}}}~=~
3.2\times 10^4
\left (\frac{n_{e0}}{{\rm cm^{-3}}}\right )^2
\left (\frac{r_{c0}}{{\rm kpc}}\right )^4
\left (\frac{n_{c0}}{{\rm Mpc}^{-3}}\right )
\left (\frac{B_0}{\mu{\rm G}}\right )F'(z)
\ee

\noindent
where $F'$ is an integral similar to the one in Eq.\,\EC{def_F}.

The Kronberg-Perry model was motivated by spectroscopic observations
of QSOs which reveal countless hydrogen absorption lines spread out in
frequency by the expansion of the Universe.  This dense series of
lines, known as the \lya-forest, implies that there are a large number
of neutral hydrogen clouds at cosmological distances.  Kronberg \&
Perry (1982) selected model parameters motivated by the \lya-cloud
observations of Sargent et al.\,(1980), specifically, $n_{e0}\simeq
2.5\times 10^{-6}\,{\rm cm}^{-3}$, $n_{c0}\simeq 5\,{\rm Mpc}^{-3}$,
and $r_{c0}\simeq 60\,{\rm kpc}$.  For these parameters, the estimate
for $\sigma_{RM}^2$ is disappointingly small.  For example, in a
spatially flat, $\Omega_m=0.3$ model, $\sigma_{RM}^2(z=3) \simeq
700\left (B_0/\mu G\right )^2\,{\rm rad^2 m^{-4}}$.  Detection above
the ``noise'' of the galactic contribution requires $\sigma_{\rm
RM}\ga 4\,\rmt$ or equivalently $B_0\ga 0.1\,\mu{\rm G}$.  A magnetic
field of this strength would have been well above the equipartition
strength for the clouds.

Clearly, the limits derived from RM data depend on the model one
assumes for the \lya-clouds.  Recently, Blasi, Burles, \& Olinto
(1999) suggested that the conclusions of Kronberg \& Perry (1982) were
overly pessimistic.  Their analysis was motivated by a model for the
intergalactic medium by Bi \& Davidsen (1997) and Coles \& Jones
(1991) in which the Universe is divided into cells of uniform electron
density while the magnetic field is parametrized by its coherence
length and mean field strength.  Random lines of sight are simulated
for various model universes.  The results suggest that a detectable
variance in RM is possible for magnetic fields as low as $B_0\simeq
6\times 10^{-9}\,{\rm G}$.  The enhanced sensitivity relative to the
Kronberg \& Perry (1982) observations is primarily due to the larger
filling factor assumed for the clouds: The clouds in Kronberg \&
Perry (1982) have a filling factor of order $10^{-3}$ while those of
Blasi, Burles, and Olinto (1999) have a filling factor of $\sim 1$.

Further improvements in the use of Faraday rotation to probe
cosmological magnetic fields may be achieved by looking for
correlations in RM (Kolatt, 1998).  The correlation function for the
RM from sources with an angular separation that is small compared to
the angular size of the clouds increases as $N_l^2$, i.e., $\langle
RM_1 RM_2\rangle\simeq N_l^2\,RM_c^2$, where $N_l$ is the average
number of clouds along the line of sight (Kolatt 1998).  By contrast,
$\sigma_{\rm RM}^2$ increases linearly with $N_l$, i.e., $\langle
RM^2\rangle\simeq N_l RM_c^2$.  Thus, the signal in a correlation map
can be enhanced over the $\sigma_{\rm RM}$ signal by an order of
magnitude or more.  Moreover, in a correlation map, the ``noise'' from
the Galaxy is reduced.  Finally, the correlation method can provide
information about the power spectrum of cosmological fields and the
statistical properties of the clouds.

A different approach was taken by Kronberg \& Perry (1982) who argued
that since the $RM_{ig}$ due to \lya-clouds is small, a large observed
$RM_{ig}$ must be due to either gas intrinsic to the QSO or to a few
rare gas clouds (e.g., a gaseous galactic halo) along the QSO line of
sight.  The implication is that the distribution of $RM_{ig}$ in a
sample of QSOs will be highly nongaussian (i.e., large for a subset of
QSOs but small for many if not most of the others) and correlated
statistically with redshift and with the presence of damped
\lya-systems.  Kronberg \& Perry (1982), Welter, Perry, \& Kronberg
(1984) and Wolfe, Lanzetta, \& Oren (1992) all reported evidence for
these trends in RM data from QSO surveys.  Wolfe, Lanzetta, \& Oren
(1992), for example, found that in a sample of 116 QSOs, the 5 with
known damped \lya-systems had large $RM_{ig}$ as compared with $35$
(i.e., $\simeq 30\%$) of those in the rest of the sample.  However
Perry, Watson, \& Kronberg (1993) argued that the case for strong
magnetic fields in damped \lya-systems is unproven, the most serious
problems in the Wolfe, Lanzetta, \& Oren (1992) analysis arising from
the sparsity and heterogeneous nature of the data.  In particular,
since electron densities can vary by at least an order of magnitude, a
case by case analysis is required.  Moreover, a subsequent study by
Oren \& Wolfe (1995) of an even larger data set found no evidence for
magnetic fields in damped \lya-systems.

\subsubsection{Evolution of Magnetic Fields in the Early Universe}

Limits on cosmological magnetic fields from CMB observations
and BBN constraints are discussed in the two subsections that follow.
These limits are relevant to models in which magnetic fields arise
in the very early Universe (see Section V.C).  In this subsection,
we discuss briefly the pre-recombination evolution of magnetic
fields.  

During most of the radiation-dominated era, magnetic fields are frozen
into the cosmic plasma.  So long as this is the case, a magnetic
field, coherent on a scale $L$ at a time $t_1$, will evolve, by a
later time $t_2$ according to the relation

\be{b_evolution}
B\left (\frac{a(t_2)}{a(t_1)}L,\,t_2\right ) =
\left (\frac{a(t_1)}{a(t_2)}\right )^2
B\left (L,\,t_1\right )~.
\ee

\noindent
Jedamzik, Katalinic, \& Olinto (1998) pointed out that at certain
epochs in the early Universe, magnetic field energy is converted 
into heat in a process analogous to Silk
damping (Silk 1968).  In particular, at recombination, the photon 
mean free path and hence radiation diffusion length scale becomes
large and magnetic field energy is dissipated.  The damping of 
different MHD modes --- Alfv\'{e}n, fast magnetosonic, and slow 
magnetosonic --- is a complex problem and the interested reader is
referred to Jedamzik, Katalinic, \& Olinto (1998).  In short,
modes whose wavelength is larger than the Silk damping scale at
decoupling (comoving length $\lambda_{\rm silk}\simeq 50\,{\rm Mpc}$)
are unaffected while damping below the Silk scale depends on the 
type of mode and the strength of the magnetic field.

\subsubsection{Limits from CMB Anisotropy Measurements}

A magnetic field, present at decoupling ($z_d\simeq 1100$) and
homogeneous on scales larger than the horizon at that time, causes the
Universe to expand at different rates in different directions.  Since
anisotropic expansion of this type distorts the CMB, measurements of
the CMB angular power spectrum imply limits on the cosmological
magnetic fields (Zel'dovich \& Novikov 1983; Madsen 1989; Barrow,
Ferreira, \& Silk 1997).

The influence of large-scale magnetic fields on the CMB is easy to
understand (see, for example, Madsen 1989).  Consider a universe that
is homogeneous and anisotropic where the isotropy is broken by a
magnetic field that is unidirectional but spatially homogeneous.
Expansion of the spacetime along the direction of the field stretches
the field lines and must therefore do work against magnetic tension.
Conversely, expansion orthogonal to the direction of the field is
aided by magnetic pressure.  Thus, the Universe expands more slowly
along the direction of the field and hence the cosmological redshift
of an object in this direction is reduced relative to what it would be
in a universe in which ${\bf B}=0$.

Zel'dovich \& Novikov (1983) and Madsen (1989) considered a spatially
flat model universe that contains a homogeneous magnetic field.  The
spacetime of this model is Bianchi type I, the simplest of the nine
homogeneous and anisotropic three-dimensional metrics known
collectively as the Bianchi spacetimes.  The analysis is easily
extended to open homogeneous anisotropic spacetimes (i.e., universes
with negative spatial curvature), known as Bianchi type V (Barrow,
Ferreira, \& Silk 1997).  In the spatially flat case, the model is
described by two dimensionless functions of time.  The first is the
ratio of the energy density in the field relative to the energy
density in matter:

\be{def_q}
\rrq\equiv \frac{B^2}{8\pi\epsilon_m}~.
\ee

\noindent The second function is the difference of expansion rates
orthogonal to and along the field divided by the Hubble parameter.
The fact that the angular anisotropy of the CMB is small on all
angular scales implies that the both of these functions are small.
Madsen (1989) found that to a good approximation

\be{deltaT}
\frac{\Delta T}{T} > 4\rrq_d
\ee

\noindent 
where the subscript `d' refers to the decoupling epoch.  If we assume
that the field is frozen into the plasma, then $B\propto (1+z)^2$ and
${\cal Q}\propto (1+z)$.  By definition $\epsilon_m/\Omega = 3H^2/8\pi G$.
The constraint implied by Eq.\,\EC{deltaT} can therefore be written

\bea{CMB1}
B_{\rm cosmic} &\la & 2\times 10^{-4}\Omega^{1/2} h_{75}
\left (\frac{\Delta T/T}{1+z_d}\right )^{1/2}{\rm G}\nonumber\\
&\la & 3\times 10^{-8}{\rm G}~.
\eea

\noindent 
While this expression was derived assuming a pure Bianchi type I model
(i.e., homogeneous on scales larger than the present day horizon) the
main contribution to the limit comes from the expansion rate at
decoupling and therefore the result should be valid for scales as
small as the scale of the horizon at decoupling.

Barrow, Ferreira, \& Silk (1997) carried out a more sophisticated
statistical analysis based on the 4-year Cosmic Background Explorer
(COBE) data for angular anisotropy and derived the following limit
for primordial fields that are coherent on scales larger than the
present horizon:

\be{CMB2}
B_{\rm cosmic}\la 5\times 10^{-9}h_{75}\Omega^{1/2}{\rm G}~.
\ee

Measurements of the CMB angular anisotropy spectrum now extend to
scales $\sim 500$ times smaller than the present-day horizon (Balbi et
al.  2000; Melchiorri et al. 2000; Pryke et al. 2001).  These
measurements imply that for fields with a comoving coherence length
$\ga 10\,{\rm Mpc}$ their strength, when scaled via
Eq.\,\EC{b_evolution} to the present epoch, must be $\la 10^{-8}\,{\rm
G}$ (Durrer, Kahniashvili, \& Yates 1998; Subramanian \& Barrow 1998).
Future observations should be able to detect or limit magnetic fields
on even smaller scales.  However, the interpretation of any limit
placed on magnetic fields below the Silk scale is complicated by the
fact that such fields are damped by photon diffusion (Jedamzik,
Katalinic, \& Olinto 1998; Subramanian \& Barrow 1998).  If a
particular MHD mode is efficiently damped prior to decoupling, then
any limit on its amplitude is essentially useless.

Jedamzik, Katalinic, \& Olinto (2000) pointed out that as MHD modes
are damped, they heat the baryon-photon fluid.  Since this process
occurs close to the decoupling epoch, it leads to a distortion of the
CMB spectrum.  Using data from the COBE/FIRAS experiment (Fixen 1996),
they derived a limit on the magnetic field strength of $B\la 3\times
10^{-8}\,{\rm G}$ (scaled to the present epoch) between comoving
scales $\simeq 400\,{\rm pc}$ and $0.6\,{\rm Mpc}$.  

The existence of a magnetic field at decoupling may induce a
measurable Faraday rotation in the polarization signal of the CMB.
Kosowsky \& Loeb (1996) showed that a primordial field with strength
corresponding to a present-day value of $10^{-9}\,{\rm G}$ induces a
$1^\circ$ rotation at $30\,{\rm GHz}$ and a strategy to measure this
effect in future CMB experiments was suggested.

\subsubsection{Constraints from Big Bang Nucleosynthesis}

Big bang nucleosynthesis (BBN) provides the earliest quantitative test
of the standard cosmological model (see, for example, Schramm \&
Turner 1998; Olive, Steigman, \& Walker 2000).  BBN took place between
$10^{-2}$ and $1\,{\rm s}$ after the Big Bang and is responsible for
most of the $^4$He, $^3$He, D, and $^7$Li in the Universe.  Numerical
calculations yield detailed predictions of the abundances of these
elements which can be compared to observational data.  Over the years,
discrepancies between theory and observation have come and gone.
Nevertheless, at present, BBN must be counted as an unqualified
success of the Big Bang paradigm.

Magnetic fields can alter the predictions of BBN.  Thus, the success
of BBN --- specifically, the agreement between theoretical predictions
and observations of the light element abundances --- imply limits on
the strength of primordial fields.  Limits of this type were first
proposed by Greenstein (1969) and O'Connell \& Matese (1970).  They
identified the two primary effects of magnetic fields on BBN: (i)
nuclear reaction rates change in the presence of strong magnetic
fields, and (ii) the magnetic energy density leads to an increase
cosmological expansion rate.  During the 1990's, detailed calculations
were carried out by numerous groups including Cheng, Schramm, \&
Truran (1994), Grasso \& Rubenstein (1996), Kernan, Starkman, \&
Vachaspati (1996), and Cheng, Olinto, Schramm, \& Truran (1996).
Though the results from these groups did not always agree (see, for
example, Kernen, Starkman, \& Vachaspati (1997).  The general
concensus is that the dominant effect comes from the change in 
the expansion rate due to magnetic field energy.

The effects of a magnetic field on BBN can be understood in terms of
the change they induce in the neutron fraction.  For example, a magnetic
field affects the neutron fraction by altering the electron density of
states.  In a uniform magnetic field, the motion of an electron can be
decomposed into linear motion along the direction of the field and
circular motion in the plane perpendicular to the field.  According to
the principles of quantum mechanics, the energy associated with the
circular motion is quantized and the total energy of the particle can
be written

\be{landau}
E = \left (p_z^2 c^2 + m_e^2 c^4 + 2 e B \hbar c n_s\right )^{1/2}
\ee

\noindent
where $n_s = 1, 2, \dots$ is the quantum number for the different
energy eigenstates known as Landau levels (Landau \& Lifshitz ) which
are important for field strengths $B\ga B_c \equiv m_e^2 c^3/e\hbar =
4.4\times 10^{13}\,{\rm G}$.  The (partial) quantization of the
electron energy implied by Eq.\,\EC{landau} therefore changes the
density of states of the electrons which in turn affects processes
such as neutron decay where it leads to an increase in the decay rate.

If the only effect of the magnetic field was to increase nuclear
reaction rates, it would lead to a decrease in the number of neutrons
at the time of BBN and hence a decrease in the $^4He$ abundance.
However, a magnetic field also contributes to the energy density of
the Universe and therefore alters the time-temperature relationship.
If the correlation length of the field is greater than the horizon
scale, the field causes the Universe to expand anisotropically.  On
the other hand, a field whose correlation scale is much smaller than
the horizon can be treated as a homogeneous and isotropic component of
the total energy density of the Universe.  In either case, the
magnetic field increases the overall expansion rate thus decreasing
the time over which nucleosynthesis can occur and in particular the
time over which neutrons can decay.  The net result is an increase in
the $^4{\rm He}$ abundance.  The helium abundance is fixed when the
age of the Universe is $t\simeq 1\,{\rm s}$ and the temperature is
$kT\simeq 1\,{\rm MeV}$.  At this time, the energy density of the
Universe is $2\times 10^{25}\,{\rm erg\,cm^{-3}}$ which is comparable
to the energy density in a $6\times 10^{12}\,{\rm G}$ magnetic field.
The magnetic field must be somewhat less than this value so as not to
spoil the predictions of BBN.  If one assumes that the magnetic field
scales according to Eq.\EC{b_evolution} then this leads to the
following constraint on the magnetic field at the present epoch
$B<10^{-6}\,{\rm G}$.

\subsubsection{Intergalactic Magnetic Fields and High Energy Cosmic
Rays}

Cosmic rays are relativistic particles (primarily electrons and
protons with a small admixture of light nuclei and antiprotons) that
propagate through the Galaxy with energies ranging from $10^9 -
10^{20}\,{\rm eV}$ (Hillas 1998).  Their energy spectrum is
characterized by a power law up to the `knee' ($E\simeq 10^{15}\,{\rm
eV}$), a slightly steeper power law between the knee and the `ankle',
($E\simeq 10^{19}\,{\rm eV}$), and a flattened distribution (the
ultrahigh energy cosmic rays or UHECRs) above the ankle.  The origin
of the UHECRs is a mystery.  Circumstantial evidence suggests that
these particles are created outside the Galaxy but within $50-100
\,{\rm Mpc}$.  Since the gyrosynchrotron radius for a particle in the
Galactic magnetic field with $E\ga 10^{19}\,{\rm eV}$ is larger than
the Galaxy, if UHECRs originated in the Galaxy, the arrival direction
would point back to the source.  However, to date, no sources have
been identified.  On the other hand, protons with energies above
$5\times 10^{19}\,{\rm eV}$ interact with CMB photons producing pions
over a mean free path of order $50-100 \,{\rm Mpc}$.  Therefore if
most UHECRs originated at cosmological distances, their energy
spectrum would show a distinct drop known as the GZK cut-off (Greisen,
1966; Zatsepin \& V. A. Kuzmin 1966).  The absence of such a cut-off
implies that UHECRs are produced within $100\,{\rm Mpc}$.

A number of authors have considered the fate of UHECRs that are
produced in the local supercluster (LSC) under the assumption that the
LSC is magnetized (Lemoine, et al. 1997; Blasi \& Olinto 1999 and
references therein).  This assumption is reasonable given the
detection of magnetic fields in the Coma supercluster (Kim et
al.\,1989).  Blasi \& Olinto (1999) found that for a LSC field of
$10^{-7}\,{\rm G}$ cosmic rays with energies below $10^{19}\,{\rm eV}$
execute a random walk as they travel from source to observer while
those above $10^{20}\,{\rm eV}$ follow a relatively straight path.
They argued that the break in the energy spectrum at the ankle is a
consequence of the transition from random walk to free-stream
propagation.  A corollary of this result is that the detection of
source counterparts for the $10^{19}\,{\rm eV}$ particles (or
alternatively clustering in the source distribution) would imply a
limit on the strength of the magnetic field in the LSC.

The effects of large-scale magnetic fields on UHECRs were also
considered by Waxman \& Miralda-Escud\'{e} (1996).  Following Waxman
(1995), Vietri (1995) and Milgrom \& Usov (1995) they assumed that the
same astrophysical objects responsible for $\gamma$-ray bursts also
produce UHECRs.  The expected rate for $\gamma$-ray bursts within
$100\,{\rm Mpc}$ is only $1$ per $50\,{\rm yr}$.  Since Takeda et
al.\,(1998) observed 7 events above $10^{20}\,{\rm eV}$ over 8 years
a dispersion in arrival times of $\ga 50\,{\rm yrs}$ for cosmic
rays produced in a single burst needs to be invoked.  Waxman \&
Miralda-Escud\'{e} (1996) proposed that such a dispersion is due to
the deflections of UHECRs by a large-scale magnetic field.  The
induced time delay is estimated to be $\tau\sim 50\,{\rm yr} \left
(L/10\,{\rm Mpc}\right )\left (B/5\times 10^{-10}\right )^2$ for a
cosmic ray energy of $10^{20}\,{\rm eV}$ and source at $100\,{\rm
Mpc}$.  Magnetic fields have a predictable effect on the angular
position and time of flight for cosmic rays of a given energy.  Hence,
future cosmic ray experiments should be able to determine not only
whether or not UHECRs are indeed produced by $\gamma$-ray bursters but
whether they are deflected by a large-scale magnetic field on route
from source to observer.

Along rather different lines, Plaga (1995) proposed a technique for
detecting cosmological magnetic fields at extremely low levels.  The
idea is to look at the arrival times of $\gamma$-ray photons from
cosmological sources (e.g., $\gamma$-ray bursts, flare events in AGN).
High energy photons suffer collisions in diffuse extragalactic
radiation fields.  At $\gamma$-ray energies, the dominant process is
electron-positron pair production.  The electrons and positrons can
then inverse Compton scatter off CMB photons producing high energy
photons.  An intergalactic magnetic field will deflect the electrons
and protons and therefore delay the secondary pulse.  In principle,
this technique could be able to detect fields as weak as
$10^{-24}\,{\rm G}$!

\section{Galactic and Extragalactic Dynamos}

A magnetic dynamo consists of electrically conducting matter moving in
a magnetic field in such a way that the induced currents amplify and
maintain the original field.  The dynamo principle was known in the
1800s though it was Larmor (1919) who first suggested that dynamo
processes might be responsible for astrophysical magnetic fields such
as those found in the Sun and Earth.  Steenbeck, Krause, \& R\"{a}dler
(1966) recognized the importance of helical turbulence for dynamos in
stars and planets.  Their ideas were soon applied to the problem of
galactic magnetic fields by Parker (1971) and Vainshtein \& Ruzmaikin
(1971, 1972).

Over the years, a standard galactic dynamo model known as the
$\aod$-dynamo has emerged whose essential features are as follows:
Turbulent motions in the ISM driven, for example, by stellar winds,
supernova explosions, and hydromagnetic instabilities, carry loops of
toroidal magnetic field out of the plane of the disk.  These loops are
twisted into the poloidal plane by the Coriolis effect while toriodal
field is regenerated from the poloidal field by differential rotation.
The $\aod$-dynamo can operate in any differentially rotating,
turbulent medium and is widely accepted as the primary mechanism for
the maintenance of magnetic fields in the Sun (Krause \& R\"{a}dler
1980; Zel'dovich, Ruzmaikin, \& Sokoloff 1983; Ruzmaikin, Sokoloff, \&
Shukurov 1988a).  Its applicability to galaxies has been more
controversial and numerous variants and alternative models have been
proposed.  Nevertheless, the general idea that galactic fields are
maintained by differential rotation and small-scale velocity
fluctuations is compelling.  More speculative is the conjecture that
magnetic dynamos act on supergalactic scales.  On the other hand, if
structure formation proceeds hierarchically, it is plausible that
dynamo processes operate sequentially from subgalactic to galactic
scales.

An essential feature of a dynamo is its ability to continuously
regenerate large-scale magnetic fields.  For galaxies, the alternative
is that magnetic fields are relics of the early Universe.  A magnetic
field that permeates the protogalactic medium will be amplified by
compression as a galaxy forms and by differential rotation once the
disk is fully developed (Hoyle 1958; Piddington 1964, 1972; Kulsrud
1990).  The relic field hypothesis has been challenged vigorously by
Parker (1973b) and others on the grounds that turbulent diffusion
destroys a primordial field on a relatively short timescale.  Since
these arguments provide a good introduction to the $\aod$-dynamo, we
repeat them below.  NB: The dynamo hypothesis does not preclude
primordial magnetic fields.  To the contrary, while a dynamo can
amplify existing fields, the first fields might have been primordial,
i.e., created in the very early Universe.
  
After a brief discussion of the primordial field hypothesis, we review
the essentials of mean-field dynamo theory and the standard
$\aod$-dynamo.  Particular attention is paid to the assumptions
necessary for the development of this model.  Some of these
assumptions have been challenged as being demonstrably false while
others are seen as simply too restrictive.  These concerns have led to
alternative models for galactic magnetic fields which are discussed at
the end of the Section.

\subsection{Primordial Field Hypothesis}

Soon after the discovery of galactic magnetic fields, Hoyle (1958)
began to contemplate their origin.  An astrophysical battery seemed an
implausible explanation for galactic fields since the voltage required
to drive the requisite currents, $V\sim 3\times 10^{13}\,{\rm Volts}$,
is so enormous.  Instead, Hoyle considered a scenario in which
magnetic fields are present {\it ab initio} in the material that
collapses to form a galaxy.  Piddington (1964, 1972) championed the
primordial field hypothesis and developed models for the structure and
evolution of an initially homogeneous field (presumed to be of
primordial origin) in a rotating disk galaxy.  The primordial
field hypothesis hypothesis has been studied recently by Howard
\& Kulsrud (1997).

The following idealized example illustrates the difficulties with the
primordial field hypothesis (Parker 1973b).  Consider a differentially
rotating disk with angular velocity $\omega = \omega(R)$.  (We are
using cylindrical ($R,\,\phi,\,z$) coordinates.)  Suppose that at
$t=0$, the magnetic field is uniform and lies in the disk plane.
Without loss of generality, we can orient the x-axis to be along the
initial direction of the field, i.e., ${\bf B}({\bf
x},\,0)=B_0\hat{\bf x}$.  So long as magnetic diffusion and 
backreaction effects are negligible, the field at time $t>0$ will be
given by

\be{spiral}
{\bf B}\left (R,\,\phi,\,t\right ) = 
B_0\left (\hat{\bf b} + t\frac{d\omega}{d\ln{R}}
\left (\cos\left (\omega + \phi t\right )\right )
\hat{\mbox{\boldmath ${\phi}$}}\right )
\ee

\noindent
(see Eq.\,\EC{Solution}) where $\hat{\bf b}=\hat{\bf b}(R)=\cos{\omega
t}\,\hat{\bf x}+\sin{\omega t}\,\hat {\bf y}$.  Field lines for an
illustrative example are shown in Figure 8.  Due to differential
rotation, the azimuthal component of the field grows linearly with $t$
while gradients in the field grow as $t^2$.  Note that both the
initial and final field configurations are bisymmetric.
(Alternatively, if the magnetic field is initially oriented along the
spin axis of the disk, the field will be axisymmetric but with odd
parity about the equatorial plane of the disk, that is, an A0
configuration (see, for example, Ruzmaikin, Shukurov, \& Sokoloff
1988a.)  Differential rotation, with an axisymmetric velocity field
(i.e., $\omega$ independent of $\phi$) does not alter the symmetry
properties of the field.  Roughly speaking, we have $\nabla B\sim
\left (\omega t \right )^2/L$ where $L$ is the disk scale length and
$d\omega/dr\sim \omega/L$.  Eventually, magnetic diffusion becomes
important with the diffusion timescale, $\tau_d\equiv
B/\eta\nabla^2B\simeq L^2/\eta\omega^2 t^2$, decreasing with $t$.
Linear growth lasts until $t\simeq \tau_d$ or equivalently $t\simeq
\left (L^2/\omega^2\eta\right )^{1/3}$ after which the field decays
rapidly.  In galactic disks, $L\simeq 3\,{\rm kpc}$, $\omega\simeq
10^{-15}\,{\rm s}^{-1}$, and $\eta\simeq 10^{26}\,{\rm cm}^2 s^{-1}$
implying a decay time of $t\simeq 3\times 10^8\,{\rm yrs}$ which is
much shorter than the age of a galaxy.  (NB: The timescale for the
field to decay depends on the structure of the field.  If, for
example, the field is concentrated in intermittent rope-like
structures, the decay time will be much longer than indicated by the
estimate above (Subramanian 1998).)  Equally problematic is the
observation that galactic magnetic fields form, by and large, a
loosely wound spiral as in Figure 3 rather than the tightly wound
spiral suggested by Figure 8.  The implication is that galactic
magnetic fields are generated continuously.  This argument is similar
to the one given in support of the hypothesis that spiral arms are
(continuously-generated) density waves propagating through the
galactic disk.

These conclusions were challenged recently by Howard \& Kulsrud (1997)
who investigated the primordial field hypothesis within the context of
a simple model for galaxy formation.  In particular, they considered a
rotating spherical protogalaxy threaded by a constant magnetic field
which then collapses to form a disk galaxy.  This process leads to
amplification of the initial field by several orders of magnitude (see
also Lesch \& Chiba 1995).  The field is then wound up by differential
rotation as described above.  Howard \& Kulsrud (1997) point out that
the magnetic field observed in spiral galaxies is an average of the
true detailed field.  Their contention is that the fine scale
structure of the field {\it is} that of Figure 8 and only appears as a
toroidal azimuthal field because of inadequate resolution.


\begin{figure}
\centerline{\psfig{file=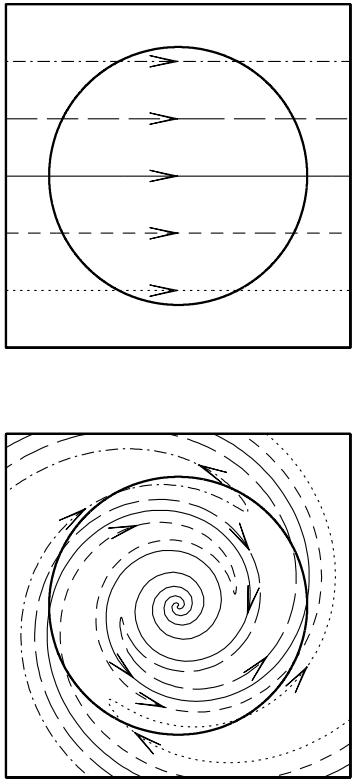,width=12.5cm}}

\caption{Distortion of magnetic field lines under the action of
differential rotation.  The upper panel shows the initial homogeneous
magnetic field configuration.  The different line types are for
visualization purposes.  The lower panel shows the same field lines
after they have been distorted by differential rotation.}

\label{diff_rot}
\end{figure}

\subsection{Mean-Field Dynamo Theory}

Most discussions of astrophysical dynamos make use of a mean-field
approximation to describe the effects of turbulence.  In addition,
backreaction of the field on the fluid is typically ignored so that
the evolution of the field reduces to a purely kinematic problem.
Detailed treatments of mean-field dynamo theory can be found in
numerous references including Steenbeck, Krause, \& R\"adler (1966),
Moffatt (1978), Parker (1979), Krause \& R\"{a}dler (1980),
Zel'dovich, Ruzmaikin, \& Sokoloff 1983, Ruzmaikin, Shukurov, \&
Sokoloff (1988a), Krause \& Wielebinski (1991), Beck et al.\,(1996)
and Kulsrud (1999).

The evolution of a magnetic field in the MHD limit is given by
Eq.\EC{MHD}.  In a mean-field analysis, we write

\be{meanb}
{\bf B}={\bf \overline{B}}+{\bf b}
~~~~~~~~~~~~
{\bf V}={\bf \overline{V}}+{\bf v}
\ee

\noindent where $\overline{\bf B}$ and $\overline{\bf V}$ represent
ensemble averages of the magnetic and velocity fields and ${\bf b}$
and ${\bf v}$ are the corresponding small-scale tangled components.
The ensemble average of Eq.\,\EC{MHD2} is

\be{mean_mhd}
\frac{\partial{\bf \overline{B}}}{\partial t} = 
\bfn\times\left ({\bf\overline{V}}\times{\bf\overline{B}}\right )+
\bfn\times\left (\overline{{\bf v}\times{\bf b}}\right )
\ee

\noindent with the residual equation

\be{fluc_mhd}
\frac{\partial{\bf b}}{\partial t} = 
\bfn\times\left ({\bf v}\times{\bf\overline{B}}+
\overline{{\bf V}}\times{\bf b}+
{\bf v}\times{\bf b}-
\overline{{\bf v}\times{\bf b}}\right )~.
\ee

\noindent 
(We have assumed that molecular diffusion can be neglected.)

The $\overline{{\bf V}}\times{\bf b}$ term in Eq.\,\EC{fluc_mhd} can
be eliminated by transforming to the rest frame of the fluid while the
${\bf v}\times{\bf b}$ terms in this equation are usually ignored.
Eq.\,\EC{fluc_mhd} is then used to eliminate ${\bf b}$ in favor of
${\bf v}$ and ${\bf B}$ in Eq.\,\EC{mean_mhd}, though this step
requires additional assumptions about the statistical properties of
the turbulence.  The result, which can be derived in a variety of ways
(see Ruzmaikin, Shukurov, and Sokoloff 1988a and references therein),
is

\be{dynamo}
\frac{\partial{\bf{\overline{B}}}}{\partial t} = 
\bfn\times\left (\overline{\bf V}\times\overline{\bf B}\right )
+\bfn\times\bemf~.
\ee

\noindent 
$\bemf$, the effective electromotive force due to turbulent motions of
the magnetic field as it is carried around by the fluid, is often written
in terms of two tensors, $\alpha$ and $\beta$:

\be{emf}
{\cal E}_i = \alpha_{ij}\overline{B}_j + \beta_{ijk}
\frac{\partial \overline{B}_j}{\partial x_k}~.
\ee
 
\noindent
Explicit expressions for the $\alpha$ and $\beta$ tensors can be found
in various sources including Ruzmaikin, Shukurov, \& Sokoloff (1988a).

The classic example of the $\alpha$-effect is the distortion of a
magnetic field line by a localized helical disturbance or cyclonic
event (Parker 1970).  In the context of a galactic dynamo, we can
think of a cyclonic event as a plume of gas rising above the disk and
acted upon by differential rotation and the Coriolis effect.  Consider
a magnetic field configuration that is initially purely toroidal and
focus on a single field line at a radius $R=R_0$.  The velocity field
of the plume is assumed to be constant during the ``event'' ($t_i < t
< t_f$) and given by the expression

\be{plume}
{\bf v}_p({\bf x},\,t)=
\left (\frac{xz}{b^2},\,\frac{yz}{b^2},\,1-\frac{x^2+y^2}{a^2}\right )
e^{-\left (\frac{x^2+y^2}{a^2}+\frac{z^2}{b^2}\right )}
\ee

\noindent
where we have introduced a local Cartesian coordinate system with
$\hat{\bf x}=\hat{\mbox{\boldmath $\phi$}}$ and $\hat{\bf y}=\hat{\bf
R}$.  By design, $\bfn\cdot{\bf v}_p=0$.  The field line of interest
is initially ${\bf B}(R=R_0)=B_0\hat{\bf x}$.  As before, it is traced
by particles that are carried along with the fluid.  In the absence of
rotation, these test particles obey the equation of motion $d{\bf
u}/dt = \left ({\bf v}_p\cdot \bfn\right ){\bf v}_p$ where ${\bf
u}={\bf u}(t)$ is the velocity of a test particle as distinct from the
fluid velocity field of the plume ${\bf v}_p$.  The field line after a
cyclonic event is shown in Figure 9(a).  In a rotating system, it is
easiest to follow the evolution of a field line (i.e., the motion of
the tracer particles) in a frame rotating with angular velocity
$\bfo_0=\bfo(R_0)$.  The equation of motion for the particles is then

\be{eofm}
\frac{d{\bf u}}{dt} = \left ({\bf
v_p}\cdot \bfn\right ){\bf v_p}-2\bfo_0\times {\bf u}+\dots
\ee


\begin{figure}
\begin{center}
\begin{tabular}{|c|}\hline
\psfig{file=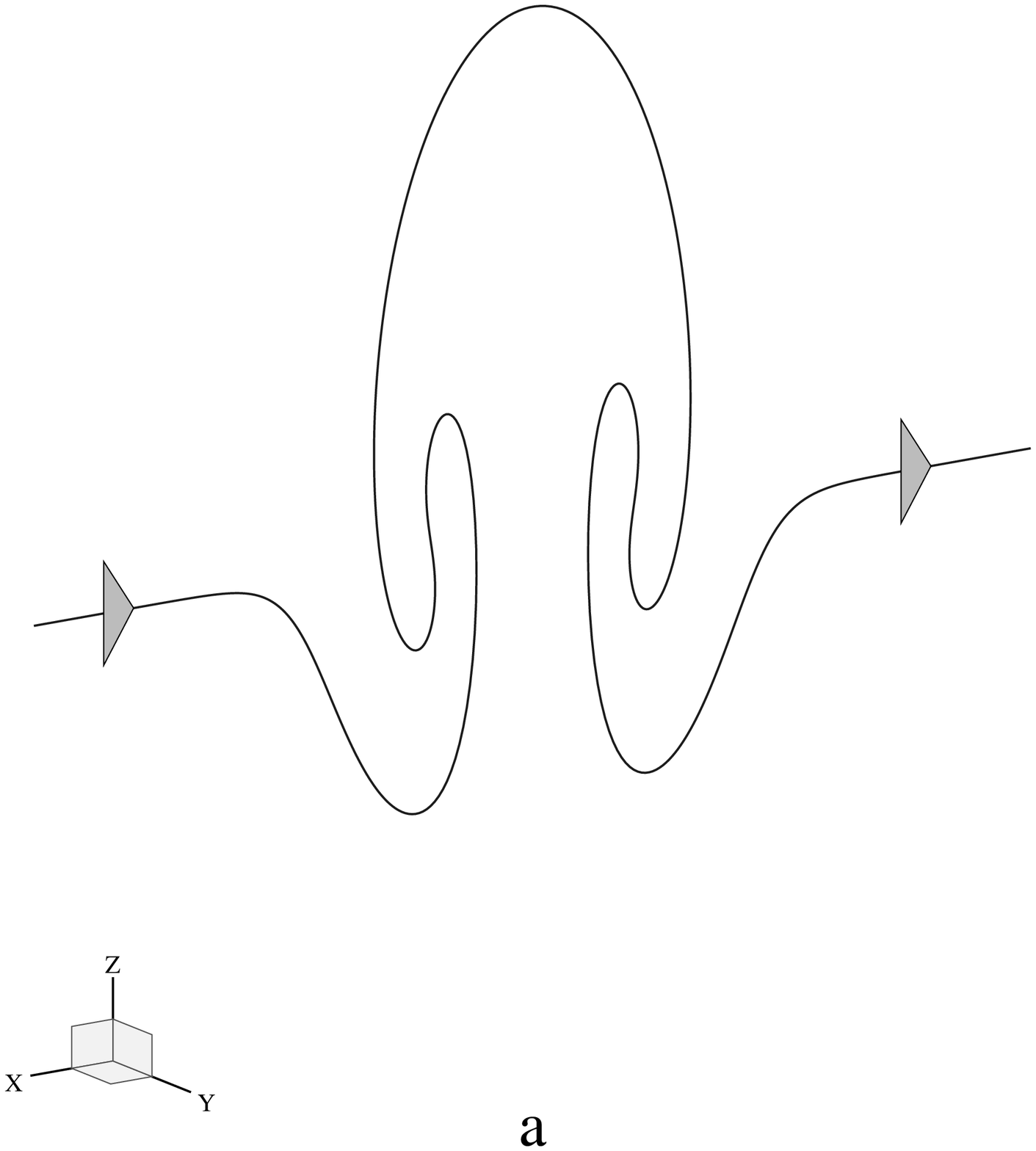,width=10cm,height=10cm} \\ \hline
\psfig{file=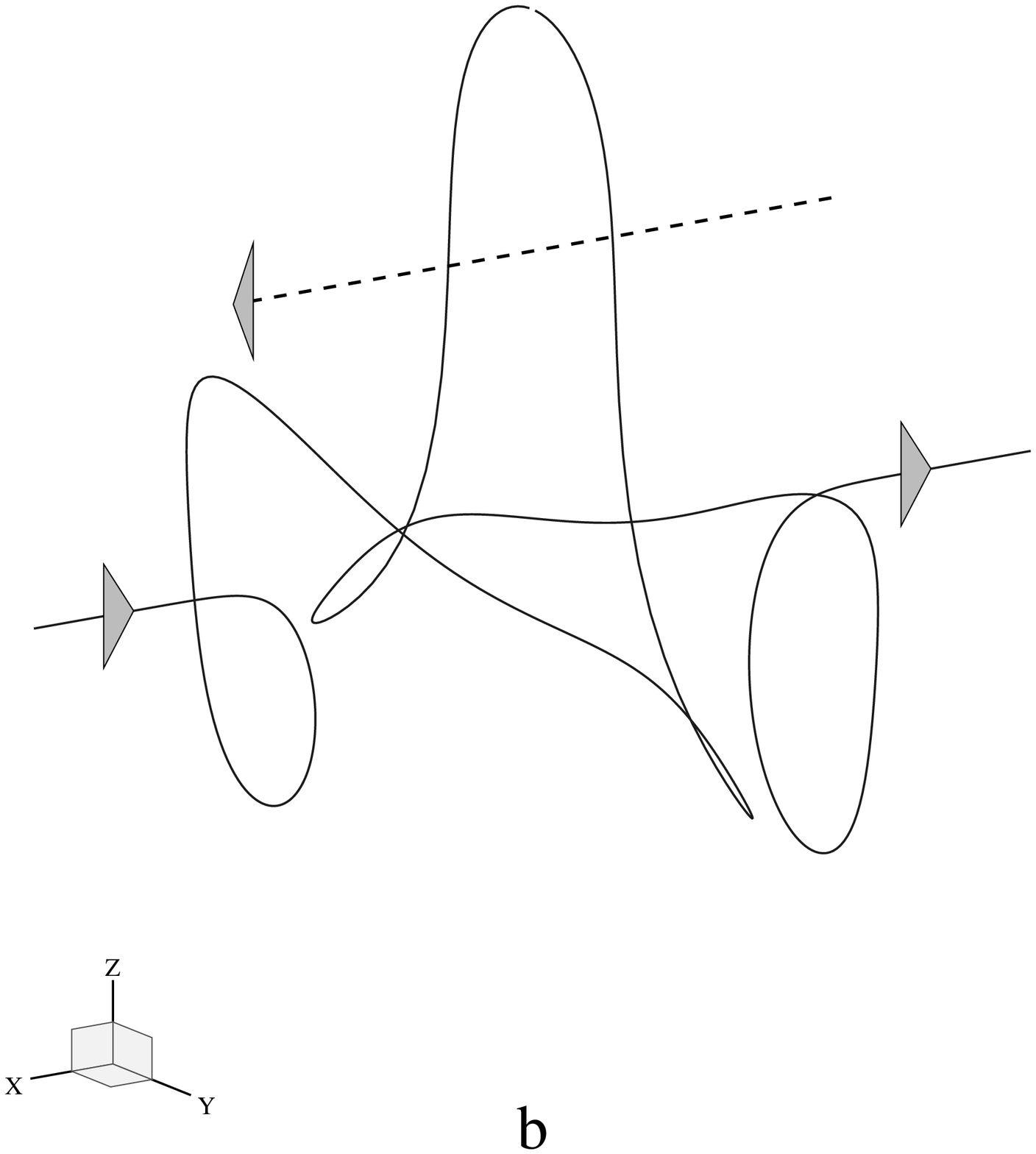,width=10cm,height=10cm} \\ \hline
\end{tabular}
\end{center}
\caption{Cyclonic event as an illustration of the $\alpha$-effect.  We
consider a single magnetic field line initially oriented along the
x-axis.  In (a) we show field line after it has been distorted by the
plume velocity field given in Eq.\,\EC{plume}.  In (b) we include the
Coriolis effect.  The dashed line and arrow represent the electric
current associated with the loop of magnetic field in the $yz$-plane.}
\label{alpha}
\end{figure}


\noindent
where the differential rotation term is not shown explicitly and is
ignored in this illustrative example.  The second term on the
right-hand side of Eq.\,\EC{eofm} describes the Coriolis effect which
twists the magnetic plume into the poloidal plane (i.e., the yz-plane)
as shown in Figure 9(b).  Note that the current associated with this
loop is anti-parallel to the initial magnetic field line.

Given a specific model for the small-scale velocity field the $\alpha$
and $\beta$ tensors can be calculated from first principles (see, for
example, Ferri\`{e}re (1992, 1993, 1998)).  However, most calculations
make use of {\it ad hoc} phenomenological forms for these functions.  A
simplifying though questionable assumption is that helical turbulence
in galactic disks is isotropic.  In this case, the $\alpha$ and
$\beta$ tensors take the form

\be{alpha}
\alpha_{ij} = \alpha\delta_{ij}~~~~~~~~~~
\alpha = -\frac{\tau}{3}\langle{\bf v}\cdot\left (\bfn
\times{\bf v}\right )\rangle~,
\ee

\noindent and

\be{beta}
\beta_{ijk}=\beta\epsilon_{ijk}~~~~~~~~~~
\beta = \frac{\tau}{3}\langle v^2\rangle~.
\ee

\noindent 
where $\delta_{ij}$ and $\epsilon_{ijk}$ are the unit tensor and
three-dimensional permutation tensor respectively and $\tau$ is the
correlation time of the turbulence.  With these expressions for $\alpha$
and $\beta$ the dynamo equation becomes

\be{dynamo_eq}
\frac{\partial{\bf B}}{\partial t} = 
\bfn\times\left ({\bf V}\times{\bf B}\right )+
\bfn\times\left (\alpha{\bf B}-\beta\bfn\times {\bf B}\right )~.
\ee

The cyclonic event described above illustrates the dubious nature of
the isotropy assumption: Instabilities tend to develop in the
direction perpendicular to the disk while vorticity is generated along
the spin axis of the galaxy.  Moreover, the largest turbulent eddies
in the Galaxy are of order $100\,{\rm pc}$ in size which is not much
smaller than the scale height of the disk.
Ferri\`{e}re (1992, 1993, 1998) has calculated the $\alpha$ and
$\beta$ tensors for a model in which turbulence is driven by
correlated supernova explosions.  The results serve as an explicit
example of a case where turbulence is decidedly anisotropic.
In this model, the $\alpha$-tensor takes the form

\be{SN_alpha}
\alpha =
 \left (
\begin{array}{ccc}
\alpha_R & -V_{\rm esc} & 0\\
V_{\rm esc} & \alpha_\phi & 0\\
0 & 0 & \alpha_z
\end{array} \right )~.
\ee

\noindent
The off-diagonal terms describe the convection of magnetic field lines
away from the equatorial plane (a ``$\bfn\times\left ({\bf V}\times {\bf
B }\right )$''-term in the induction equation) while the diagonal
elements characterize the strength of the $\alpha$-effect along the
three coordinate axes.  However, while the calculations of
Ferri\`{e}re account for the stretching of magnetic field lines by the
expanding supernova shells and concomitant twisting by the Coriolis
effect they do not take into account for the $\alpha$-effect due to
turbulence that is undoubtedly generated by the expanding shells.
Thus, her analysis probably underestimates the strength of the
$\alpha$-effect associated with supernovae.

Eq.\,\EC{dynamo} is the starting point for most discussions of
astrophysical dynamos.  A self-consistent treatment of the fluid
requires an equation of motion for $\overline{\bf V}$ which includes a
Lorentz force term describing the backreaction of the field on the
fluid.  In addition, backreaction may affect the fluctuation fields
${\bf v}$ and ${\bf b}$ thus modifying $\alpha$ and $\beta$.  These
effects will be discussed below.  Here, we assume that backreaction is
negligible so that $\overline{\bf V}$, $\alpha$, and $\beta$ can be
specified as model inputs that are independent of $\overline{\bf B}$.
If we further suppose that they are time-independent, the solutions to
Eq.\,\EC{dynamo} will be of the form $\overline{\bf B}\propto
\exp{\Gamma t}$ where the eigenvalue $\Gamma$ depends on boundary
conditions for the field.  The solutions of interest are, of course,
ones where $\Gamma$ is real and positive.

\subsection{Disk Dynamos}

We consider an axisymmetric differentially-rotating disk galaxy with a
large-scale velocity field $\overline{\bf V}(R)=\bfo\times {\bf R}$
where ${\bfo}=\om(R){\bf\hat {z}}$.  In general $\alpha$ and $\beta$
are functions of $R$ and $z$ with the proviso that $\alpha(-z)=
-\alpha(z)$.  Rotational invariance of the dynamo equations suggests
the following Fourier decomposition:

\be{fourier}
{\bf B}\left ({\bf r},\,t\right )
=\sum_m {\bf B}_m\left (r,\,z,\,t\right )e^{im\phi}~.
\ee

\noindent 
One might guess, and detailed calculations confirm, that the growth
rate decreases with increasing $m$ so that the fastest growing mode
has azimuthal symmetry ($m=0$) (Ruzmaikin, Shukurov, and Sokoloff
1988a).  The $m=1$ bisymmetric mode can also be important, though in
general this mode is difficult to excite without introducing
nonaxisymmetric forms for $\alpha$, $\beta$, and/or $\omega$.  We will
return to this point below and proceed with a discussion of axisymmetric
solutions.

In component notation, Eq.\,\EC{dynamo} becomes

\be{dynamo_r}
\frac{\partial B_R}{\partial t} = 
-\frac{\partial}{\partial z}\left (\alpha B_\phi\right )+
\beta\left (\nabla^2B\right )_R
\ee

\be{dynamo_p}
\frac{\partial B_\phi}{\partial t} = 
\frac{d\omega}{d\ln R} B_R
+\frac{\partial}{\partial z}\left (\alpha B_R\right )
-\frac{\partial}{\partial R}\left (\alpha B_z\right )
+\beta\left (\nabla^2B\right )_\phi
\ee

\be{dynamo_z}
\frac{\partial B_z}{\partial t} = 
\frac{1}{R}\frac{\partial}{\partial R}
\left (R\alpha B_\phi\right )+
\beta\left (\nabla^2B\right )_z
\ee

\noindent 
where the overbar and subscript $m (=0)$ are omitted for the sake of
clarity.  It is generally assumed that the $\alpha$-terms in the
equation for $B_\phi$ are small compared to the $\omega$-term (i.e.,
the toroidal field is generated through the action of differential
rotation rather than turbulence).  Eqs.\,\EC{dynamo_r}-\EC{dynamo_z}
are symmetric under the transformation $z\to -z$.  Therefore, the
solutions will have definite parity with both even and odd parity
solutions (${\bf B}^+$ and ${\bf B}^-$ respectively) possible:

\be{pparity}
B^\pm_R(-z)=\pm B^\pm R(z)~~~~~~~~~~~
B^\pm_\phi(-z)=\pm B^\pm\phi(z)~~~~~~~~~~~
B^\pm_z(-z)=\mp B^\pm_z(z)
\ee

\noindent 
For $m=0$, ${\bf B}^+$ and ${\bf B}^-$ correspond respectively to the
even parity ($S0$) and odd parity ($A0$) configurations described in
Section III.B.2 and shown in Figure 2.

Since the dynamo equations are linear in the fields they admit
solutions of the form $\overline{\bf B}\propto \exp{\Gamma t}$.
Furthermore, for disk-like geometries, variations in the field with
respect to $z$ will be much greater than those with respect to $R$ ---
in our own Galaxy, the scale height of the disk is a factor of 10
smaller than the scale radius of the disk.  This situation suggests a
quasi-separation of variables of the form

\be{ansatz}
{\bf B}\left (R,\,z,\,t\right ) = 
Q_n(R)\tilde{{\bf B}}(R,z)e^{\Gamma_n t}
\ee

\noindent 
(see, for example, Ruzmaikin, Shukurov, and Sokoloff 1988a) where $n$
labels the different eigenfunctions $Q_n$ and eigenvalues $\Gamma_n$.
(For a more rigorous discussion of the thin-disk limit in which the
Schr\"{o}dinger-type radial equation (see below) appears as an
approximation to the full (integro-differential) see Priklonsky et al. 2000.)
$\tilde{\bf B}$ depends on $R$ parametrically, i.e., no derivatives
with respect to $R$ appear in the equation for $\tilde{{\bf B}}$.
With this ansatz, the dynamo equations become

\be{dynamo_rr}
\gamma(R) \tilde{B}_R = 
-\frac{\partial}{\partial z}\left (\alpha \tilde{B}_\phi\right )+
\beta\frac{\partial^2\tilde{B}_R}{\partial z^2}~,
\ee

\be{dynamo_pp}
\gamma(R) \tilde{B}_\phi = 
\frac{d\omega}{d\ln R} \tilde{B}_R+
\beta\frac{\partial^2\tilde{B}_\phi}{\partial z^2}
\ee

\noindent and

\be{Q_n}
\frac{\beta}{R^2}\frac{d}{dR}\left (
\frac{1}{R}\frac{d}{dR}\left (RQ_n\right )\right )
+\left (\gamma(R)-\Gamma_n\right )Q_n = 0~.
\ee

\noindent 
(Since $\tilde{B_z}$ does not appear in the equations for
$\tilde{B_r}$ and $\tilde{B_\phi}$ we may ignore it at this time.
Once $\tilde{B_r}$ and $\tilde{B_\phi}$ have been determined,
$\tilde{B_z}$ may be found from the condition $\bfn\cdot {\bf B}=0$ or
alternatively by solving the $\tilde{B_z}$ equation.)

Eqs.\,\EC{dynamo_rr}-\EC{Q_n} can be solved once a complete set of
boundary conditions are specified.  For simplicity, we assume that the
disk is defined by sharp boundaries at $z=\pm h$ and $R=R_d$ with
force-free fields (i.e., $\bfn\times{\bf B}=0$) outside the disk.
These conditions imply that $B_\phi(z=h)=0$ and $B_R(z=h)\simeq 0$.
($B_R(z=h)$ is identically zero in the limit $h/R_D\to 0$ but may be
small for a finite disk.)

Eqs.\,\EC{dynamo_rr} and \EC{dynamo_pp} make clear the essential
features of the standard $\aod$-dynamo.  Turbulence via the
$\alpha$-effect generates $B_R$ (and $B_z$) from $B_\phi$ while
differential rotation regenerates $B_\phi$.  The sequence of steps for
an A0 dynamo is shown schematically in Figure 10.  We begin in Figure
10(a) with a pure poloidal dipole-like field.  Differential rotation
stretches field lines creating toroidal field.  This is illustrated in
Figure 10(b) where a single field line from Figure 10(a) is shown after
it has been acted upon by differential rotation.  The field in the
equatorial plane is characterized by strong gradients and high
magnetic tension.  This tension can be relieved either by turbulent
diffusion, via the $\beta$-effect, or by some other process (e.g.,
magnetic reconnection, see below).  The net result is to decouple the
toroidal field in the upper and lower hemispheres, as shown in Figure
10(c).  Next we assume that cyclonic events occur throughout the disk.
The toroidal field is distorted in the vertical direction as in Figure
10(d).  The loops of vertical field are then twisted into the poloidal
plane by the Coriolis effect (Figure 10(e), see, also Figure 9(b)).
Once again, some form of diffusion or dissipation is needed to
eliminate magnetic field near the equatorial plane.  Provided this
occurs, poloidal loops in upper and lower hemispheres can combine to
yield a dipole-like field which reinforces the original field (Figure
10(f)).

This example illustrates the importance of diffusion for
the dynamo.  The $\alpha$ and $\omega$ effects twist, shear, and
stretch magnetic field lines but do not create new ones.  While they
can increase the magnetic field energy in the system, they cannot
change the net flux through a surface that encloses it.  Diffusion
eliminates unwanted flux.  In the odd parity dynamo of Figure 10, this
process occurs in the equatorial region.  In an even parity dynamo, 
diffusion allows flux of the wrong sign to escape by moving to high
galactic latitudes.


\begin{figure}
\begin{center}
\begin{tabular}{|c|c|}\hline
\psfig{file=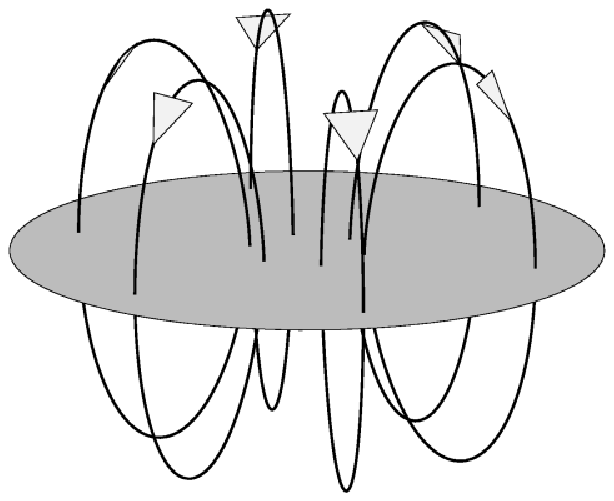,width=6cm,height=6cm} &
\psfig{file=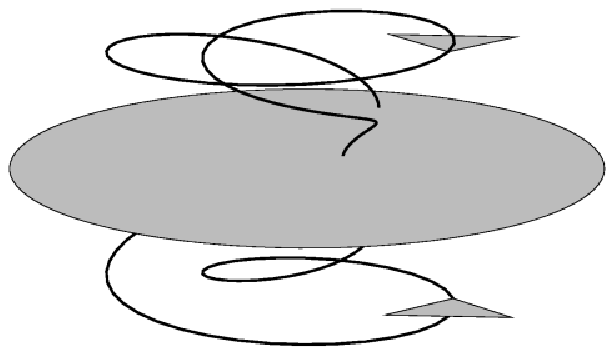,width=6cm,height=6cm} \\ \hline
\psfig{file=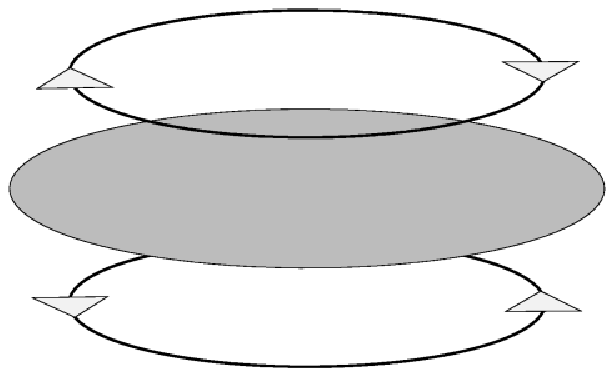,width=6cm,height=6cm} &
\psfig{file=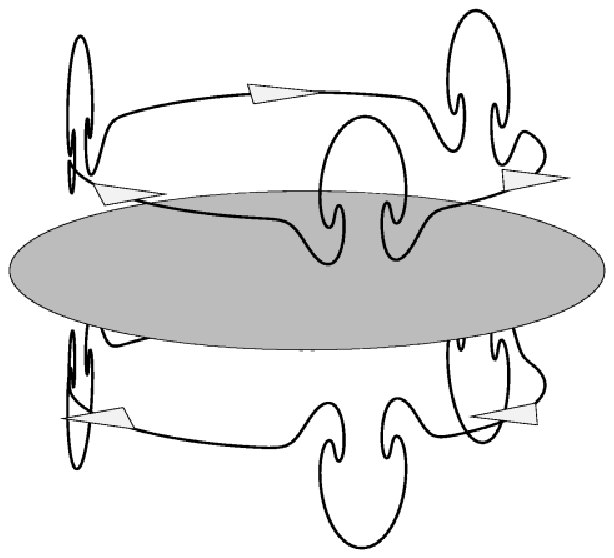,width=6cm,height=6cm}\\ \hline
\psfig{file=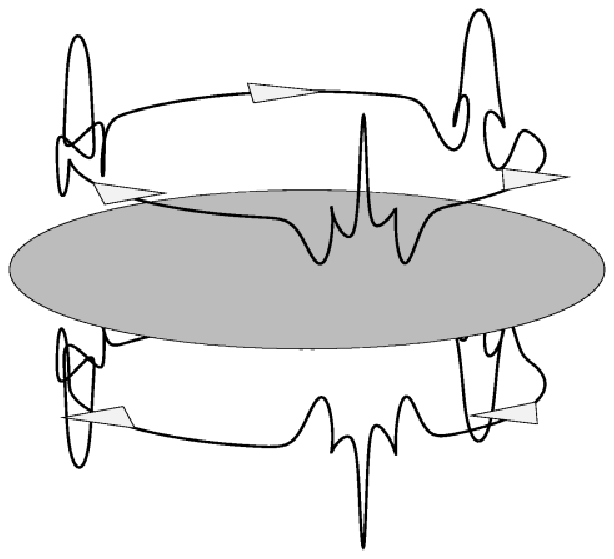,width=6cm,height=6cm} &
\psfig{file=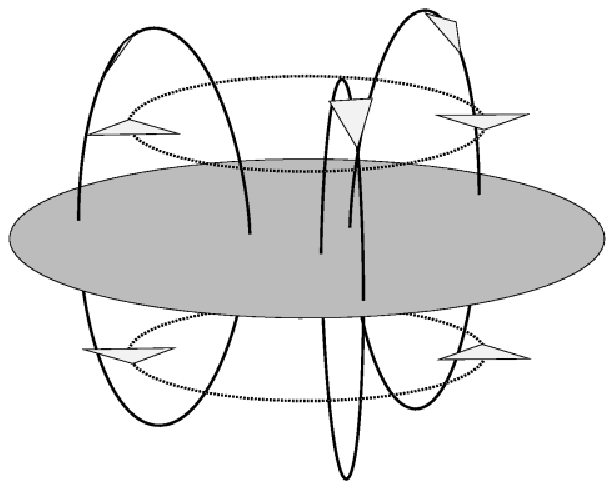,width=6cm,height=6cm}\\ \hline
\end{tabular}
\end{center}

\caption{Sequence of events illustrating an axisymmetric odd-parity
(A0) $\aod$-dynamo.  (a) Dipole-like poloidal (A0) field.  (b) Single
field line from (a) after it has been stretched by differential
rotation.  (c) Toroidal component from (b) assuming turbulent
diffusion and/or reconnection have decoupled field in upper and lower
hemispheres.  (d) Plumes created from the toroidal field by cyclonic
events similar to the one illustrated in Figure 9.  (e) Field loops
from (d) twisted by the Coriolis effect. (f) Poloidal loops created
from the those shown in (e).}
\label{dynamo}
\end{figure}


The importance of diffusion can be illustrated by considering $F_R$,
the radial flux through the surface at $R=R_D$ and $F_\phi$, the
azimuthal flux through a surface at fixed $\phi$ (see, for example,
Zel'dovich, Ruzmaikin, \& Sokoloff 1983):

\be{radial_flux}
F_R = \int_0^{2\pi} R_Dd\phi\int_0^h dz\, B_R
~~~~~~~~~~~~~~
F_\phi = \int_0^{R_D} dR\int_0^h dz\, B_\phi~.
\ee

\noindent 
From Eq.\,\EC{dynamo_r} we find

\be{drfluxdt}
\frac{dF_R}{dt}=\left. 2\pi R_D
\beta\frac{\partial B_R}{\partial z}\right |_0^h
\ee

\noindent 
where contributions from the $\alpha$-term in Eq.\,\EC{dynamo_r}
vanish because $\alpha(z=0)=0$ and $B_\phi(z=h)=0$.  With $\beta=0$,
$F_R$ is constant, i.e., growing mode solutions require $\beta\ne 0$.

The equation for $F_\phi$ is also interesting (Zel'dovich, Ruzmaikin,
\& Sokoloff 1983).  From Eq.\,\EC{dynamo_p} we find

\be{dpfluxdt}
\frac{dF_\phi}{dt}=\int_0^{R_D} dR\int_0^h dz \,
\frac{d\omega}{dR}RB_R+\left.
\beta\frac{\partial B_\phi}{\partial z}\right |_0^h~.
\ee

\noindent 
For the first term, we integrate by parts and use the condition
$\bfn\cdot {\bf B}=0$ to find

\be{first_term}
\int_0^{R_D} dR\int_0^h dz \,
\frac{d\omega}{dR}RB_R =
\int_0^{R_D}dR\,\omega \, B_z \big |_0^h~.
\ee

\noindent 
Thus, with $\beta=0$, $F_\phi$ grows linearly with time, but only if
field lines exit the disk through the surface at $z=h$.  If the field
lines are confined to the disk, than $F_\phi$ remains constant even
though the energy in the field is increasing due to differential
rotation.

Thus, magnetic flux must be expelled from a galaxy inorder for the net
large-scale field to grow.  Flux expulsion can occur by a number of
mechanisms including magnetic buoyance and supernovae or superbubble
explosions.  The latter was considered by Rafikov \& Kulsrud (2000)
who concluded that the gravitational field of the disk would severely
limit flux expulsion.  Clearly, this issue deserves further attention,
perhaps through three-dimensional numerical simulations, which can
model a galaxy and its immediate environment.

Eqs.\,\EC{dynamo_rr} and \EC{dynamo_pp} can be simplified by
introducing the dimensionless quantities $\tilde{t}= t/\tau_D$,
$\tilde{\gamma}=\gamma\tau_D$, $\tilde{z}= z/h$,
$\tilde{\alpha}=\alpha/\alpha_0$ and $\tilde{\om}=-\left
(d\om/d\ln{R}\right )/\om_0$ where $\tau_D=h^2/\beta$ is the
(turbulent) diffusion timescale for the large-scale field.  We have

\be{dynamo_rrr}
\left (\tilde{\gamma} - 
\frac{\partial^2}{\partial \tilde{z}^2}\right )B_R
= -R_\alpha\frac{\partial}{\partial \tilde{z}}
\left (\tilde{\alpha} B_\phi\right )
\ee

\be{dynamo_ppp}
\left (\tilde{\gamma} - 
\frac{\partial^2}{\partial \tilde{z}^2}\right )B_\phi
= -R_\om B_R
\ee

\noindent 
where we have introduced the dimensionless dynamo coefficients
$R_\alpha = \alpha_0 \tau_D/h$ and $R_\om = \om_0 \tau_D$.  Upon
rescaling the ratio $B_R/B_\phi$ by $R_\alpha$, we see that
Eqs.\,\EC{dynamo_rrr} and \EC{dynamo_ppp} can be characterized by the
single dimensionless parameter, $D=R_\alpha R_\om = \om_0
\alpha_0/\tau_D^2 h$, known as the dynamo number.

\subsection{Growth Rate for the Galactic Magnetic Field}

Eqs.\,\EC{dynamo_rrr} and \EC{dynamo_ppp} can be solved using standard
techniques.  In the thin-disk limit, the fastest growing solution has
positive parity (i.e., the quadrupole (S0) mode).  Ruzmaikin,
Shukurov, Sokoloff (1988a) have determined the growth rate (i.e.,
eigenvalue $\gamma$) in the thin-disk limit as a function of the
dynamo number $D$ for various forms of $\alpha(z)$ (e.g.,
$\tilde{\alpha} = \sin{\pi z},~z,$ and $\Theta(z)-\Theta(-z)$ where
$\Theta$ is the Heaviside function).  Growing mode solutions are found
for values of $D>D_{\rm cr}$ where $D_{\rm cr}$, the critical dynamo
number, varies from $6$ to $11$ for the different forms of $\alpha$
mentioned above.  For $D\gg D_{\rm cr}$, $\gamma\propto D^{1/2}$.  For
example, in the case $\tilde{\alpha} = \sin{\pi z}$, the growth rate
can be approximated by the fitting formula $\tilde{\gamma}\simeq 0.78
D^{1/2}-0.23$ (Field 1995).

The actual growth rate depends on the parameters $\om_0$, $\alpha_0$,
and $t_D$.  It is relatively straightforward to determine $\om_0$.
For example, at the position of the Sun in the Galaxy,
$\om(R=R_s)=\om_0=2A \simeq 29\,{\rm km\,s^{-1}\,kpc^{-1}}$ where $A$
is the first Oort constant (Binney and Merrifield, 1998).  $\alpha_0$
and $\tau_D$ are more difficult to estimate.  However, if we write
$\gamma$ in units of $\om_0$,

\bea{growth_rate}
\frac{\gamma}{\om_0}
&\simeq& 0.78\left (\frac{\alpha_0}{h\om_0}\right )^{1/2}
-0.23\left (\frac{1}{\om_0 \tau_D}\right )^{1/2}\nonumber \\
&\la &0.78\left (\frac{R_\alpha}{R_\omega}\right )^{1/2}~,
\eea

\noindent 
we see that the dependence on $\tau_D$ is relatively weak, and in any
case one can ignore the second term and derive an upper limit for the
growth rate or a lower bound on the e-folding time for the field.
Given an observed value for the field at a time $t_f$ (e.g., in the
Galaxy, $B_f=3\mu {\rm G}$ with $t_f = t_0$) and an estimate for an
initial time $t_i$ when the dynamo begins to operate, one can derive a
lower bound for the requisite seed field.

\subsection{Criticisms of Mean-Field Dynamo Theory}

The validity of mean-field dynamo theory was questioned almost as soon
as it was proposed.  The most serious criticisms revolve around the
assertion that backreaction of the magnetic field on the fluid is
unimportant, an assumption that has been challenged on the following
grounds: In a highly conducting turbulent fluid, the magnetic field on
small scales builds up rapidly via small-scale dynamo action and also
the tangling of weak large-scale fields if they are present.  In a
mean-field dynamo, amplification of the regular field (length scale
much larger than the outer scale of the turbulent velocity field)
takes place on a timescale much longer than the eddy diffusion time
associated with the turbulence.  The difficulty is that Lorentz forces
on small scales can react back on the fluid altering the turbulent
motions (Piddington 1964, 1972; Cattaneo \& Vainshtein 1991;
Vainshtein \& Cattaneo 1992; Kulsrud \& Anderson 1992).  If turbulent
motions are suppressed, then so too will be turbulent diffusion and
the $\alpha$-effect effectively shutting off the dynamo.

The crux of the putative problem lies in the high magnetic Reynolds
number (or alternatively, high electric conductivity) found in
astrophysical fluids.  In a highly conducting turbulent fluid,
magnetic flux tubes are continuously stretched into intricate ribbons
thus leading to a build up of energy to small scales until
backreaction effects become important.  Various authors have suggested
that the suppression of the $\alpha$ effect takes the form $\alpha\sim
\alpha_T/\left (1 + \left (\overline B/B_{\rm eq}\right )^2R_M^p\right
)$ where $p$ is a constant of order unity and $\alpha_T\simeq v_T l$
is the standard result calculated in the absence of backreaction,
$v_T$ and $l$ are the characteristic velocity and length scale
associated with the turbulence, and $ B_{\rm eq}^2=4\pi\rho v_T^2$ is
the corresponding equipartition field strength (see, for example,
Cattaneo \& Vainshtein 1991; Vainshtein \& Cattaneo 1992; Gruzinov \&
Diamond (1994)).  Thus, if $p\simeq 1$, as has often been suggested,
dynamo action in galaxies is strongly suppressed.

The form for $\alpha$-suppression given above can be traced to
Zel'dovich (1957) who argued that the ratio of the rms field strength
$B_{\rm rms}$ (principally tangled fields) to the large-scale field
strength $\overline B$ is given by $\overline B/B_{\rm rms}\simeq
R_M^{1/2}$.  This result is valid for a two-dimensional turbulent flow
and can be understood as follows (Parker 1979; Field 1995): Consider a
flux tube with cross-sectional width $\delta$ so that the timescale
for Ohmic diffusion is $\tau_D\simeq\delta^2/\eta\simeq \left
(\delta^2 R_M/v_TL\right )$.  Suppose that initially there is a
regular field $\overline{B}$ with coherence length $L$.  In a
turbulent flow and in the absence of backreaction, the length $l$ of a
flux tube grows as $l\simeq L\exp\left (t/2\tau_T\right )$ where
$\tau_T\simeq L/v_T$ is the characteristic turnover time for the
turbulence.  The field strength grows by the same factor (see
Eq.\,\EC{Solution} with the assumption that the flow is
incompressible) so that $B_{\rm rms}/\overline{B} \simeq\left
(l/L\right )\simeq\exp{(t/2\tau_T)}$.  Since the flux through the tube
is constant, the cross-sectional area must decrease by the
corresponding amount.  In a two-dimensional turbulent flow, the tube
stretches into a thin ribbon: One dimension shrinks exponentially
while the other dimension remains approximately constant.  The
diffusion time, which is governed by the small dimension of the tube,
therefore drops exponentially: $\tau_D\simeq \left ( L R_M/v_T\right
)\exp{(-t/\tau_T)} \simeq \tau_T R_M\left (\overline{B}/B_{\rm
rms}\right )^2$.  Again, if backreaction is ignored, the system
rapidly approaches the (Ohmic) diffusive limit when
$\tau_D\simeq\tau_T$ or equivalently $\overline{B}/B_{\rm rms}\simeq
R_M^{1/2}$.

Of course, once the field reaches its equipartition value $B_{\rm eq}
\equiv \left (4\pi\rho v_T^2\right )^{1/2}$, backreaction effects
become important.  For $\overline{B}\ga\beq$, $\tau_D$ is
approximately equal to the standard value for Ohmic diffusion,
$\tau_D\simeq L^2/\eta\simeq\tau_T/R_M$.  If $\beq>\overline{B}>
R_M^{-1/2}\beq$, the cascade to small scales is blocked by
backreaction.  In this case the diffusion time (assumed to be
equivalent to the dissipation time $\tau_D$) is modified by a factor
of $\beq^2/R_M \overline{B}^2<1$ (Cattaneo \& Vainshtein 1991).
Conversely, for $\overline{B}\la R_M^{-1/2}\beq$, the kinematic
assumption applies and $\tau_D\simeq\tau_T$.  These results are
summarized by the phenomenological formula

\be{taud}
\tau_D = \tau_T\left (1+\frac{R_M}
{\left (\beq/\overline{B}\right )^2+1}\right )~.
\ee

\noindent 
Two-dimensional simulations by Cattaneo \& Vainshtein (1991) appear to
support these conclusions.  The implication is that the $\beta$
coefficient in the mean-field dynamo equation is reduced relative to
the canonical turbulent diffusion value, $\beta_T\simeq vL/3$,
whenever the large-scale field is greater than $R_M^{-1/2}\beq$:

\be{beta_eff}
\beta_{\rm eff}=\beta_T\left
(1+\frac{R_M}{\left (\beq/\overline{B}\right )^2+1}\right )^{-1}~.
\ee

\noindent 
Vainshtein \& Cattaneo (1992) argued that $\alpha$ is suppressed in a
similar manner.  In present-day galaxies, $\overline{B}\simeq \beq$ and
$R_M\simeq 10^{20}$ implying that $\beta$ and $\alpha$ are reduced by
enormous factors.

Can the critisms described above be reconciled with the standard
dynamo hypothesis?  A somewhat extreme viewpoint, expressed by Kulsrud
\& Anderson (1992), is that the physical basis for the mean-field
dynamo theory is invalid and therefore the galactic magnetic field is
of primordial origin.  While this represents a departure from
mainstream thought on the problem of galactic magnetic fields, it
deserves careful consideration, especially in light of observations of
microgauss fields in high redshift galaxies and on supercluster scales
in the low redshift Universe.

A second possibility is that the arguments presented against the mean
field dynamo contain fundamental flaws.  Blackman \& Field (1999) 
pointed out that some of the analytic approaches (e.g., Gruzinov \& 
Diamond 1994) do not distinguish between turbulent quantities of
the ``zeroth-order'' state (i.e., the state with no large-scale
field) and higher order quantities.  The former are homogeneous,
isotropic, and stationary.  The same cannot be said of the latter.
Blackman \& Field (1999) are quick to point out that they do not
prove that a mean-field dynamo evades the backreaction problem,
but only that some of the objections can be challenged.

In a subsequent paper, Blackman \& Field (2000) argued that the
$\alpha$-suppression seen in numerical simulations by Cattaneo \&
Vainshtein (1991), Vainshtein \& Cattaneo (1992), and Cattaneo \&
Hughes (1996) are due to imposed (periodic) boundary conditions.
Indeed, Blackman \& Field (2000) noted that no such suppression is
seen in the simulations of Brandenburg \& Donner (1997) where dynamo
action in an accretion disk (and without periodic boundary conditions
imposed) is considered (see below).

Subramanian (1998) proposed a different resolution to the backreaction
controversy in suggesting that the fields generated by small-scale
dynamo action do not fill space but rather are concentrated into
intermittent rope-like structures.  In a weakly ionized gas, the
small-scale dynamo saturates when the field in the flux tube reaches
equipartition strength.  However, because of the low filling factor,
the average energy density of the field is still well below that of
the turbulent gas and therefore the gas is unaffected.  Fields do tend
to straighten out on small scales.  Lorentz forces act on the charged
component of the gas which leads to slippage between the ions and
magnetic field on one hand, and the neutral gas component on the
other.  This process is known as ambipolar diffusion and has been
considered by various authors in the context of galactic dynamo theory
(e.g., Zweibel 1988; Brandenburg \& Zweibel 1994).

A final possibility is that formulation of the mean-field dynamo in
terms of a turbulent eddy diffusion parameter $\beta$ and turbulent
$\alpha$-effect is in fact wrong but that some other mechanism
provides $\alpha$ and $\beta$-like effects which enter the mean field
dynamo equation more or less as in Eq.\,\EC{dynamo_eq}.  This point of
view was advocated by Parker (1992) who proposed a dynamo based on
buoyancy of magnetic flux tubes and neutral point reconnection.
Parker's model is discussed below.

One final note: Even if the problems at small scales are resolved,
backreaction of the field on the fluid certainly becomes important
once the energy density in the total magnetic field becomes comparable
to the kinetic energy associated with the turbulent eddies.  In this
case, a quasisteady-state ``equipartition'' system should emerge.  The
apparent approximate equality of magnetic field energy and ``turbulent
fluid energy'' in the Galaxy suggests that mature spiral galaxies have
already reached equipartition.  Observations of microgauss fields in
galaxies at cosmological redshifts lend further support to this
conjecture.  Quenching of the dynamo at equipartition field strengths
has been discussed by a number of authors including Moffatt (1978) and
Krause and R\"{a}dler (1980).  A phenomenological approach is to
replace $\alpha$ in the dynamo equation with

\be{quench}
\alpha_B = \alpha_T\left
(1+\frac{B^2}{\beq^2}\right )^{-1}~.
\ee

\noindent
Quenching, via Eq.\,\EC{quench}, can be easily incorporated into
numerical simulations of mean-field dynamo action (see below).

\subsection{Numerical Simulations of Disk Dynamos}

Numerical simulations provide an approach to the study of magnetic
dynamos that is complementary to the eigenvalue analysis described
above.  Fewer assumptions are required and one can include additional
effects not easily treated analytically.  For example:

\begin{itemize}

\item The separation-of-variables ansatz, Eq.\,\EC{ansatz}, which
follows from the thin-disk approximation, is no longer necessary.
Simulations allow one to explore a family of models that interpolate
between disklike and spherical systems.

\item The full mean-field dynamo equation can be used.  In particular,
the $\alpha$-term in the toroidal equation is retained allowing for
the possibility of a so-called $\alpha^2$-dynamo where toroidal as
well as poloidal fields are generated by small-scale turbulent
motions.

\item Nonlinear feedback of the field on the fluid can be included,
albeit in an {\it ad hoc} manner, by requiring that $\alpha$ and
$\beta$ approach zero as $B$ rises above the equipartition strength.
Feedback of this nature can be implemented by replacing $\alpha$ with
the form given in Eq.\,\EC{quench}.

\item The assumption that turbulence is isotropic can be dropped.  In
particular, $\alpha$ and $\beta$ can assume forms more general than
those given in Eq.\,\EC{alpha} and Eq.\,\EC{beta}.

\item Additional physics, such as the influence density waves and
feedback from star formation, can be included.

\item Simulations can be tailored to individual galaxies and can be 
used to generate synthetic observations such as radio continuum, 
polarization, and RM maps.

\item Ultimately, we may be able to bypass the mean-field
approximation by simulating the evolution of small-scale magnetic and
velocity fields explicitly.  

\end{itemize}

To be sure, many of the effects listed above have been addressed
through analysis.  For example, Mestel \& Subramanian (1991)
investigated the influence of spiral density waves via a
non-axisymmetric $\alpha$ tensor in an attempt to explain the
existence of bisymmetric magnetic fields in at least some spiral
galaxies.  Likewise, nonlinear effects have been treated analytically
by various authors such as Belyanin, Sokoloff, \& Shukurov (1994).

Over the years, galactic dynamo simulations have improved steadily
both in terms of the physics that is included and the dynamic range
achieved.  In early investigations (e.g., Elstner, Meinel, \&
R\"{u}diger 1990) $\alpha$, $\beta$, and $\omega$ were assumed to be
axisymmetric.  Moreover, quenching of $\alpha$ and $\beta$ due to
backreaction of the field on the fluid was ignored.  With these
assumptions, the different terms in the series expansion given by
Eq.\,\EC{fourier} decouple.  Their evolution can be studied
numerically by solving the finite-difference analog of
Eqs.\,\EC{dynamo_r}-\EC{dynamo_z} on a two-dimensional Eulerian mesh.
Simulations of this type have been carried out by Elstner, Meinel, \&
Beck (1992) who considered a suite of disklike galaxy models
characterized by an angular velocity profile $\omega(R)$ and analytic
forms for the $\alpha$ and $\beta$-effect tensors.  In almost all
cases, the preferred configuration was S0.  Only when the
$\alpha$-effect was confined to the inner part of the disk was the A0
configuration dominant.  Furthermore, no examples were found where a
bisymmetric mode was preferred, though in some cases a bisymmetric
mode could be excited.

Brandenburg et al.\,(1992) used two-dimensional calculations to
explore the difference between spherical and disk dynamo models.  In
particular, they considered models in which a disk is embedded in a
gaseous halo.  Not surprisingly, they found that for certain regions of
parameter space, mixed parity modes could be excited.  The relevance of
these results to galactic magnetic fields will be discussed below.

Only with 3D simulations have the true benefits of numerical
simulations been realized.  Deviations from azimuthal symmetry (for
example, due to spiral density waves) can be introduced by allowing
$\omega$, $\alpha$, and $\beta$ to depend on $\phi$.  In 3D
simulations, one can study mode coupling and in particular, the
effects of quenching via Eq.\,\EC{quench}.  Recently simulations have
included the effects of gas and stars using N-body and hydrodynamic
techniques.  These simulations are able to treat self-consistently the
interplay between spiral density waves and magnetic spiral structures
(see, for example, Elstner et al.\,2000).  Present day simulations
reveal a wide range of magnetic configurations, a reassuring result
given the rich variety of magnetic structures observed in actual
galaxies (see, for example, Panesar \& Nelson 1992; Rohde \& Elstner
1998).

In the simulations discussed so far, the effects of turbulence are
treated via the mean field quantities $\alpha$ and $\beta$.  High
performance computers make it possible to study dynamo processes from
first principles.  The challenge, of course, is to achieve the dynamic
range necessary to follow small-scale fluctuations in the velocity and
magnetic fields.  Here, we mention several techniques used to simulate
MHD phenomena that may soon yield useful results in the study of
dynamo theory.  Eulerian finite-difference codes such as ZEUS (Stone
\& Norman 1992a, 1992b) are designed to solve the equations of ideal
MHD.  A key feature of these codes is that $\bfn\cdot{\bf B}=0$ is
guaranteed explicitly at each timestep.  Recently, Roettighr, Stone,
\& Burns (1999) used ZEUS to study the evolution of magnetic fields in
merging galaxy clusters.  They found that the field strength and
structure are dramatically altered during the merger.  In the early
stages of the merger, the field is stretched by bulk flows and
compressed by shocks.  However, significant amplification of the field
does not occur until the later stages of the merger when gas motions
become turbulent.

Korpi et al.\,(1999) developed a numerical model for the ISM that
includes the effects of rotation, supernova heating, and nonideal MHD
phenomena.  In principle, simulations carried out within the context
of this model should be able to follow magnetic field amplification by
dynamo action.  However, the calculations performed thusfar have not
been run over an adequate time span to determine whether or not
magnetic field generation by dynamo action has occurred (see, also,
Shukurov (1999)).

Today, most cosmological simulations use a Lagrangian or particle
description for matter.  Stars and dark matter are modeled as
collisionless particles while gas is treated using smooth particle
hydrodynamics (SPH; Monaghan 1992 and references therein).  In SPH,
any local physical quantity $f({\bf r})$ (e.g., density, pressure)
can be approximated by taking a weighted average of the form

\be{SPH}
f({\bf r})\simeq \sum_i
\frac{m_i}{\rho({\bf r}_i)}
W\left ({\bf r}-{\bf r}_i,\,h\right )
\ee

\noindent 
where $m_i$ is the mass of the i'th particle, 

\be{sph_density}
\rho({\bf r})\simeq \sum_i
m_i W\left ({\bf r}-{\bf r}_i,\,h\right )
\ee

\noindent
is the density, and $W\left ({\bf r}-{\bf r}_i,\,h\right )$ is a
user-supplied window function whose characteristic width is $h$.

The SPH prescription can be used to model magnetic fields (see, for
example, Monaghan 1992).  Each ``gas particle'' carries with it an
additional physical quantity, a ${\bf B}$-vector, whose evolution is
followed by solving the SPH analog of the induction equation.  In
general, particle simulations such as those that use SPH have more
dynamic range than simulations done on a mesh.  One potentially
serious problem with using SPH to study MHD phenomena is that
$\bfn\cdot{\bf B}=0$ is not constrained to vanish.

Dolag, Bartelmann, \& Lesch (1999) used SPH to study the evolution of
magnetic fields in a cosmological setting.  In particular, they
followed the formation of a galaxy cluster from the linear regime
(redshift $z_i\simeq 15$) to the present epoch assuming an initial
magnetic field $B_i\simeq 10^{-9}\,{\rm G}$.  The magnetic field is
amplified by a factor $\sim 1000$, roughly an order of magnitude more
than what would be expected from simple collapse calculations.  The
simulations were used to generate synthetic RM maps for the simulated
clusters which are in good agreement with those observed by Kim et
al.\,(1990) and Kim, Kronberg, \& Tribble (1991).

Mean-field dynamo theory is based on the assumption that the coherent
component of a magnetic field fills space and described properly
by field equations.  (This assumption is the basis for both the
Eulerian and SPH methods described above.)  An alternative picture,
supported to some extent by both simulations and observations, is that
the magnetic field is confined to flux tubes.  For example, Roettighr,
Stone, \& Burns (1999) found filamentary structures in their cluster
simulations.  Moreover, it may be that magnetic fields in
extragalactic objects are filamentary but appear smooth because of
lack of resolution (Hanasz \& Lesch 1993).  In fact, filamentary
structures have been observed in the halo of M82 (Reuter et
al.\,1992).  The radio spurs seen in NGC 4631 (Golla \& Hummel 1994)
and NGC 5775 (T\"{u}llman (2000) may also be indicative of individual
flux tubes.  In addition, the magnetic field in the ISM appears to be
highly nonuniform, again suggesting that the magnetic field in the ISM
may be ``ropy'' (Heiles 1987).

The flux tube picture suggests an alternative treatment of
astrophysical magnetic fields based on the thin-tube approximation
(Spruit 1981).  The working assumption is that the flux tube radius is
much smaller than the characteristic scales over which physical
quantities (e.g., gravitational field, gas density) vary.  In this
limit all variables, such as the field strength and gas density inside
the tube, can be treated as functions of their position along the
tube.  These quantities evolve according to a set of (Lagrangian)
equations of motion.  Eqs.\,\EC{continuity} and \EC{MHD4} are examples
of this, applicable in the kinematic regime.  Thus, the illustrative
examples in Figures 8-10 may be regarded as (kinematic) flux tube
simulations.  A recent example of a flux-tube simulation can be found
in Vainshtein et al.\,(1996) where the so-called stretch-twist-fold
dynamo (Vainshtein \& Zel'dovich 1972) was investigated using
precisely these equations.

While the flux-tube model has been used extensively to study solar
magnetic fields, its application to the problem of galactic and
extragalactic magnetic fields has been rather limited.  One example is
the model of Hanasz \& Lesch (1993, 1997, 1998) who considered it in
the context of the Parker instability (see below).

\subsection{Diversity in Galactic Magnetic Fields}

The variety of field configurations observed in spiral galaxies
presents a challenge to the dynamo hypothesis.  In a thin-disk
$\aod$-dynamo, the fastest growing mode has S0 symmetry and therefore,
except in cases of very special initial conditions, this mode
dominates at late times.  However, there are a number of edge-on
spiral galaxies in which significant poloidal fields have been
observed.  For example, polarization data for NGC 4631 reveals strong,
highly ordered, poloidal magnetic fields that extend several
kiloparsecs above the galactic plane (Hummel et al 1991; Golla \&
Hummel 1994; Beck 2000).  These properties are by no means ubiquitous.
In NGC 4565, the magnetic fields are strongly confined to the disk
plane (Sukumar \& Allen 1991).  NGC 891 appears to be an intermediate
case.  Fields extend above the galactic plane, but they are not as
highly ordered as those in NGC 4631 and the toroidal components are
comparable to the poloidal ones (Hummel, Beck, \& Dahlem \,1991;
Sukumar \& Allen 1991).

Recall that in S0 configurations, the toroidal field dominates (the
poloidal field vanishes in the symmetry plane).  Thus the presence of
a vertically averaged poloidal field that is strong relative to the
azimuthal field may be indicative of an A0-type configuration.
Sokoloff \& Shukurov (1990) suggested that A0-type fields are
generated by dynamos that operate in the gaseous halos of spiral
galaxies.  In disklike systems, there is a high gradient energy cost
associated with A0-type field configurations.  In spheroidal systems,
this cost is reduced and therefore A0 modes can grow and even
dominate.  Sokoloff \& Shukurov (1990) proposed that two essentially
independent dynamos operate in spiral galaxies, one in the disk and
the other in the halo.  Their order of magnitude estimates
demonstrated that the model can explain the observations of NGC 4631
and NGC 891.  Brandenburg et al.\,(1992) studied disk-halo dynamos in
detail using numerical simulations.  Their results confirmed, at a
qualitative level, that a mean-field dynamo can operate in a
turbulent, differentially rotating, spherical halo.  One interesting
property of the halo fields is that they are oscillatory (not unlike
the solar cycle) with a period of oscillations comparable to the age
of the galaxy.  The ratio of odd to even parity fields therefore
varies with time and is also sensitive to initial conditions.
However, they found that it is generally difficult to construct models
in which poloidal fields dominate to the extent seen in NGC 4631.
Brandenburg et al.\,(1993) considered an alternative model in which
outflows from the galactic disk due to galactic fountains, chimneys,
and the Parker instability carry magnetic field from the disk to the
halo.  These fields are generally poloidal and might explain the
magnetic structures observed in NGC 4631.  They argued that by
contrast, the halo dynamo model works well for the galaxies NGC 891
and NGC 4565.

As discussed above, spiral galaxies are found with axisymmetric,
bisymmetric, and mixed symmetry field configurations (Beck et
al.\,1996).  In axisymmetric models, the $m=0$ mode is always
preferred which suggests that some form of symmetry breaking is
required in order to explain the observations.

Even in axisymmetric models, growing mode solutions with $m>0$ are
still possible.  The generation of axisymmetric and bisymmetric fields
was studied by Baryshnikova, Shukurov, Ruzmaikin, \& Sokoloff (1987)
and Krasheninnikova, Ruzmaikin, Sokoloff, \& Shukurov (1989).  They
solved the three-dimensional dynamo equations for a thin disk by
constructed a perturbation series where the small parameter is
essentially the scale height of the disk divided by its scale radius.
The growth rates for the two modes depend on the rotation curve one
assumes for the disk and this result may explain the variations in
field configurations observed in real galaxies.  Results assuming a
rotation curve appropriate to the galaxy M51 show that the
characteristic growth timescale for the $m=1$ bisymmetric mode can be
as short as $2\times 10^8\,{\rm yrs}$.  This timescale is still a
factor of four smaller than the growth rate for the axisymmetric mode.
In addition, the field appears to be confined to a relatively narrow
range in radius.

Mestel and Subramanian (1991) considered a modified dynamo with a
non-axisymmetric $\alpha$-effect.  Specifically, they imposed a
$\phi$-dependence for the $\alpha$-effect in the form of a uniformly
rotating spiral structure.  The motivation is the conjecture that the
$\alpha$-effect is enhanced along spiral arms.  In the spiral density
wave model a shock develops in the interstellar gas along the spiral
arms.  The jump in vorticity across the shock can lead to an
enhancement of the $\alpha$-effect which is constant in time but
azimuth-dependent.  As expected, an $\alpha$-effect with a
$\phi$-dependence of this type naturally excites bisymmetric fields.
The analysis of Mestel \& Subramanian (1991) made use of the thin disk
approximation and found rapidly growing bisymmetric magnetic fields
that can extend over a sizable range in radius.

\subsection{Variations on the Dynamo Theme}

The dynamo hypothesis for the origin of galactic magnetic fields faces
serious challenges beyond those raised above.  In particular,
observations have pushed back the epoch at which galactic-scale fields
are known to exist thus shortening the time dynamo processes have to
amplify small-amplitude seed fields.  These difficulties have led to a
number of variants of the standard $\aod$-dynamo.  Here we
discuss just a few.

\subsubsection{Parker Instability}

Parker (1992) proposed an alternative model that may alleviate the
problems mentioned above while retaining the attractive features of
the $\aod$-dynamo.  In his model, the combination of a hydromagnetic
instability and reconnection of oppositely-oriented magnetic field
lines (neutral point reconnection) provide the diffusion and
$\alpha$-effect necessary for the dynamo.  The gas-magnetic
field-cosmic ray system in galaxies is unstable (Parker 1979 and
references therein).  Consider a magnetic field that is initially
purely toroidal.  If any vertical waviness develops in the field, gas
will slide downward toward the galactic plane making the outward
bulges more buoyant.  Relativistic cosmic rays produced in OB
associations and supernovae can also force loops of magnetic field
into the galactic halo.  The result is a pattern of close-packed
magnetic lobes perpendicular to the disk plane (Figure 11(a)).
Magnetic reconnection can free these loops from the original toroidal
field (Figure 11(b)).  The tension along magnetic field lines is
reduced thereby enabling differential rotation, the Coriolis effect,
and further instabilities to act upon both the poloidal loops and
toroidal field.


\begin{figure}
\centerline{\psfig{file=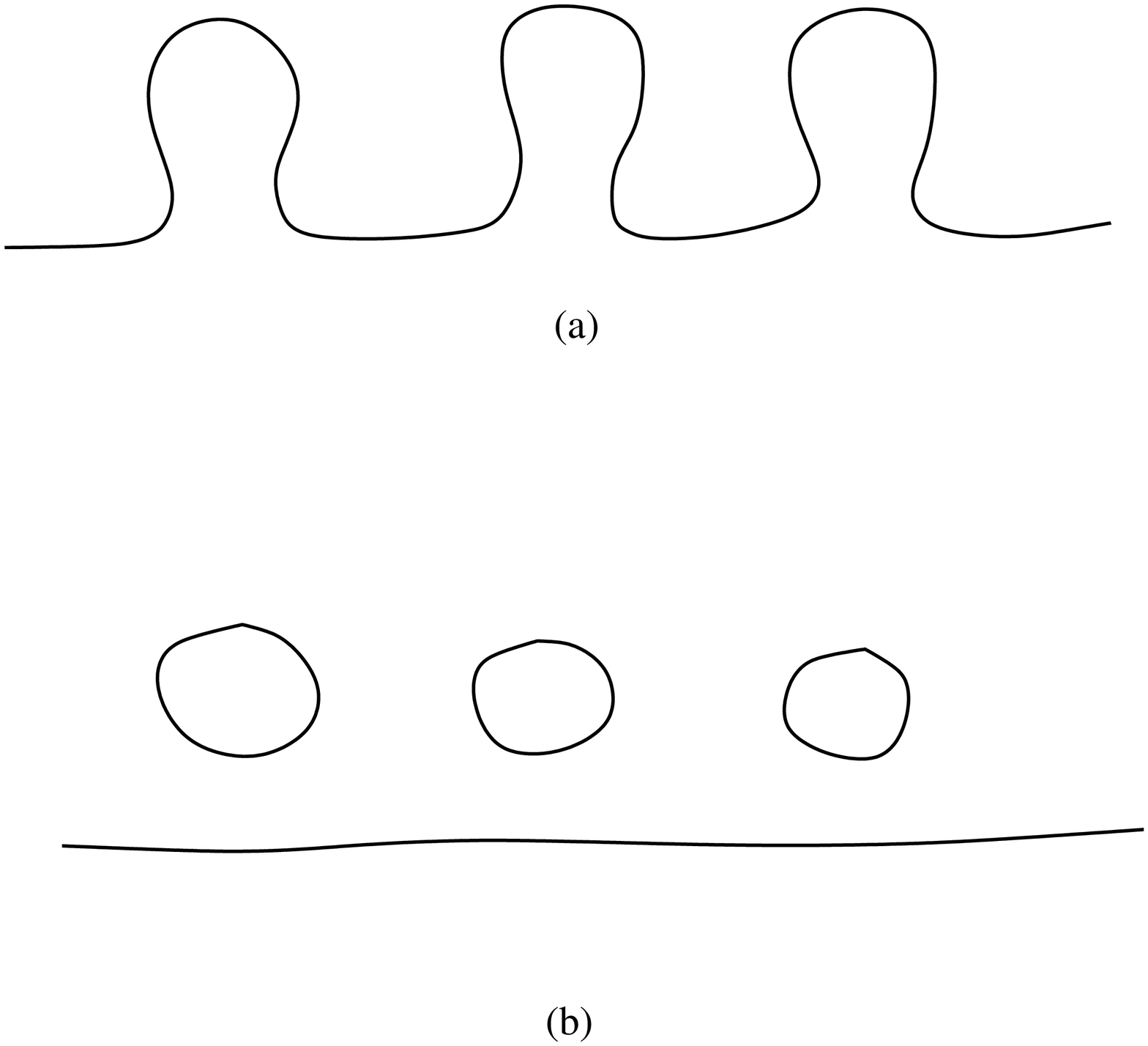,width=12.5cm,angle=-90}}

\caption{Schematic illustration of Parker's (1992) galactic dynamo.
(a) A magnetic field line is distorted by the Parker instability.
(b) Loops of magnetic field free themselves through reconnection.}

\label{parker}
\end{figure}


In classical mean-field dynamo theory, turbulence is presumed to arise
from nonmagnetic phenomena.  Since backreaction of the field on the
turbulence is assumed to be unimportant (at least until equipartition
strengths are reached) properties of the turbulence, as expressed
through the $\alpha$ and $\beta$ tensors, are independent of the
magnetic field.  A dynamo based on the Parker instability represents a
departure from this scenario in that the magnetic field (specifically,
buoyancy of magnetic flux tubes) drives the $\alpha$-effect.  But
while the underlying physics of Parker's model is quite different from
that of the classical dynamo, the effective equations are essentially
the same.  The generation of poloidal field from toroidal field as
well as the elimination of `unwanted' flux can be described by
$\alpha$ and $\beta$-like terms in an equation for the large-scale
field.  Moreover the timescale for these processes is similar to what
is obtained in the standard scenario.

Moss, Shukurov \& Sokoloff (1999) considered a mean-field disk dynamo
in which the $\alpha$-effect is driven by the Parker instability.  In
particular, they incorporated two modifications into the standard
dynamo equation (Eq.\,\EC{dynamo_eq}).  First, $\alpha$ is taken to be
an increasing function of $|{\bf B}|/B_{\rm eq}$ where $B_{\rm eq}$ is
the equipartition field strength.  Thus, the $\alpha$-effect becomes
stronger as the field reaches equipartition, i.e., as the buoyancy of
the flux tubes increases.  Second, the rotational velocity is
supplemented by a term corresponding to motions away from the disk,
again due to buoyancy, i.e., $\overline{\bf V}=r\omega(r)
\hat{\mbox{\boldmath ${\phi}$}} + V_B{\bf\hat{z}}$ $V_B$ also depends
on $|{\bf B}|/B_{\rm eq}$.  Solutions obtained from the full
three-dimensional dynamo equations were found to resemble those
obtained in conventional models in which the $\alpha$-effect comes
from cyclonic turbulence.  Some important differences did emerge.  In
particular, it was found that with a buoyancy-driven $\alpha$-effect,
the azimuthal field was no longer confined to a thin region near the
equatorial plane but rather extended into the halo region.

A variant of the Parker's model was considered in a series of papers
by Hanasz \& Lesch (1993, 1997, 1998).  In particular, they studied
the dynamics of a individual flux tubes under the influence of
gravity, pressure from the ambient medium, and buoyancy due to both
the magnetic field and cosmic rays pointing out that cosmic ray
pressure inside a flux tube will make it buoyant even if the magnetic
field is weak.  In fact, with a weak field, magnetic tension, which
tends to limit the Parker instability, will be unimportant.  The
hypothesis is then that star formation in an early phase of a galaxy
leads to an excess of cosmic ray pressure over magnetic pressure.  It
is the enhanced buoyancy of the flux tubes due to the cosmic rays that
drives and Parker-type fast dynamo, thereby converting cosmic ray
energy to magnetic field energy.

\subsubsection{Magnetorotational Instability}

Magnetic fields may play a more direct role in the onset of
turbulence.  Balbus \& Hawley (1991) have shown that a powerful,
local, shear instability occurs in weakly magnetized differentially
rotating disks.  Almost any small seed field, in combination with an
angular velocity profile that decreases with radius, will lead to
dynamical instability.  While the instability has been known for
some time (Chandrasekhar 1960) it was Balbus \& Hawley (1991) who
recognized its importance for astrophysical disks (see Balbus \&
Hawley 1998 for a review).

To understand the origin of the instability, consider an axisymmetric
rotating disk with angular velocity $\omega=\omega(R)$ and magnetic
field parallel to the spin axis.  Focus on two fluid elements along
the same field line and consider what happens when they are displaced,
one in the direction of rotation and the other in the opposite
direction.  Magnetic tension causes the leading element to slow and
move inward and the trailing element to speed up and move outward.  If
there is differential rotation with $d\omega/dR<0$ the fluid element
that flows inward will enter a more rapidly rotating region of the
disk and its angular velocity increases stretching the field line is
stretched even further.  This positive feedback loop is the essence of
the instability (Balbus \& Hawley 1991).

The instability can act as one component of a magnetohydrodynamic
dynamo (e.g., Brandenburg et al.\,1995; Hawley, Gammie, \& Balbus
1996).  The flows due to the instability regenerate magnetic fields
which in turn reinforce the turbulence.  A surprising feature of this
system is that the magnetic energy is somewhat higher than the kinetic
energy of the gas, i.e., super-equipartition field strengths are
achieved.

It is not clear whether the Balbus-Hawley instability is relevant to
galactic magnetic fields.  The underlying cause of the instability
comes from the tendency of a weak magnetic field to enforce
corotation.  Since the inner regions of a spiral galaxy are typically
in approximate solid body rotation a necessary ingredient for the
instability is all but absent.  However, the Balbus-Hawley instability
may be important in the creation of the magnetic fields that seed the
dynamo.  It is possible, for example, that seed fields are generated
in active galactic nuclei (AGN) and then dispersed through the ISM by
jets with the Balbus-Hawley instability playing a central role in the
generation of these AGN-fields.

\subsubsection{Cross-Helicity Dynamo}

The cross-helicity dynamo is a variant of the standard mean-field
dynamo which allows for more rapid amplification of a seed field at
early times.  The effect was first proposed in a general context by
Yoshizawa (1990) and later applied to the problem of galactic magnetic
fields by Yokoi (1996) and Brandenburg \& Urpin (1998).  The model
assumes that a cross-correlation exists between the fluctuating
components of the velocity and magnetic field, i.e., $\langle {\bf
v}\cdot {\bf b}\rangle\ne 0$.  The mean-field MHD equation is then
supplemented by a term that is proportional to the product of $\langle
{\bf v}\cdot {\bf b}\rangle$ and the large-scale vorticity field,
$\bzeta\equiv\bfn\times{\bf V}$.  Since this term is not proportional
to the large-scale magnetic field, it leads to linear rather than
exponential growth of ${\bf B}$.  Under certain assumptions, the
cross-helicity dynamo can dominate at early times speeding up the
amplification process (Brandenburg and Urpin 1998).

It is interesting to compare the cross-helicity effect with the
standard $\alpha$ mechanism.  In the cross-helicity dynamo, the mean
vorticity distorts ${\bf V}$ through the inertial term in the Euler
equation (Eq.\,\EC{euler})

\be{inertial}
\frac{\partial {\bf v}}{\partial t} = 
	-\left ({\bf v}\cdot \bfn\right )\overline{\bf V}
+\dots~,
\ee

\noindent and distorts ${\bf b}$ through the stretching term 
in the MHD equation (Eq.\,\EC{MHD3})

\be{stretch}
\frac{\partial{\bf b}}{\partial t} = 
	\left ({\bf b}\cdot \bfn\right )\overline{\bf V}
+\dots~.
\ee

\noindent 
By combining Eqs.\,\EC{inertial} and \EC{stretch} we find

\be{ucrossb}
\frac{\partial}{\partial t}\left ({\bf v}\times {\bf b}\right ) = 
	{\bf b}\times\left ({\bf v}\cdot \bfn\right )
	\overline{\bf V}+
	{\bf v}\times\left ({\bf b}\cdot \bfn\right )
	\overline{\bf V}
+\dots~.
\ee

\noindent 
If the fluctuation fields are isotropic, $\langle v_ib_j\rangle
=\frac{1}{3}\delta_{ij}\langle {\bf v}\cdot {\bf b}\rangle$.  The
cross-helicity effect therefore leads to an electromotive force of the
form $\bemf_{\rm CH}=\lambda \bzeta+=\dots$ where
$\lambda\equiv\frac{2}{3}\tau\langle {\bf v}\cdot {\bf b}\rangle$ and
$\tau$ is the characteristic correlation or turnover time of the
turbulence.  When this term is included, the mean-field dynamo equation
takes the form

\be{ch-dynamo}
\frac{\partial{\bf{\overline{B}}}}{\partial t} = 
\bfn\times\left (\overline{\bf V}\times\overline{\bf B}\right )
+\bfn\times\left (\alpha \overline{\bf B} -\beta\bfn\times 
\overline{\bf B} +\bemf_{\rm CH}\right ) ~.
\ee

\noindent
Thus, there is a source term ${\bf S}\equiv\bfn\times\bemf_{\rm CH}$ in
the induction equation which is independent of the large-scale
magnetic field.  The effective current associated with this source is
parallel to the mean vorticity which, in a disk galaxy, is parallel to
the spin axis.  Only toroidal fields are generated.  By contrast, the
$\alpha$-effect generates a current which is parallel to the magnetic
field and hence poloidal fields are generated from toroidal ones.

Brandenburg \& Urpin (1998) considered the cross-helicity dynamo in
the context of galactic magnetic fields.  In order to approximate the
importance of the effect, they analyzed different data sets from MHD
turbulence simulations and concluded that the relative
cross-correlation parameter, $\epsilon\equiv \langle {\bf v}\cdot {\bf
b}\rangle/\left (\langle {\rm v}\rangle\langle {\rm b}\rangle\right )$
is in the range $3\times 10^{-2}-3\times 10^{-4}$.  Using values
characteristic of spiral galaxies $S$ can be parametrized as follows:

\bea{characteristicS}
S &\simeq &
\frac{\lambda\zeta}{L}\nonumber \\
&\simeq &
10^{-7}{\rm G\,Gyr^{-1}}
\left (\frac{\epsilon}{0.03}\right )
\left (\frac{\tau}{10^7\,{\rm yr}}\right )
\left (\frac{v}{10\,{\rm km\,s^{-1}}}\right )
\left (\frac{b}{10^{-5}\,{\rm G}}\right )\\
&&~~~~~~~~~~~~~~\times
\left (\frac{\zeta}{30\,{\rm km\,s^{-1}\,kpc^{-1}}}\right )
\left (\frac{10\,{\rm kpc}}{L}\right )~.
\eea

\noindent
The cross-helicity dynamo dominates over the standard $\aod$-dynamo
for values of the field strength $B\la S\Gamma^{-1}$ where, as before,
$\Gamma$ is the growth rate of the fastest growing dynamo mode.  For
example, if $\Gamma\simeq 0.5\,{\rm Gyr^{-1}}$, the transition from
cross-helicity to $\aod$-dynamo occurs at a field strength $B\simeq
5\times 10^{-8}-5\times 10^{-10}$ (depending on the value used for
$\epsilon$) and at a time $t\simeq 1\,{\rm Gyr}$).  The cross-helicity
effect is potentially important in cases where the initial seed field
is extremely weak.  In the standard dynamo model, a seed field of
strength $10^{-18}\,{\rm G}$ is amplified to a strength of
$10^{-8}\,{\rm G}$ in $\simeq 4.5\,{\rm Gyr}$.  In a
cross-helicity/$\aod$-hybrid dynamo, the same field strength can be
reached in $\simeq 1\,{\rm Gyr}$
(Brandenburg \& Urpin 1998).

\subsection{Dynamos in Irregular and Elliptical Galaxies and Galaxy Clusters}

The existence of microgauss fields in elliptical galaxies and galaxy
clusters presents distinct challenges to models for the origin of cosmic
magnetic fields.  Rotation plays a central role in the mean-field
dynamo models devised to explain magnetic fields in disk galaxies,
stars, and planets.  Rotation provides a resevoir of energy for field
amplification through both $\alpha$ and $\omega$ effects.  In
particular, the $\alpha$-effect requires net helicity, or
equivalently, mirror symmetry violation in the turbulence.  In rapidly
rotating systems mirror symmetry violation occurs because of the
Coriolis effect.  This effect is weak, if not absent, in slow rotating
systems such as ellipticals and clusters.

A turbulent dynamo may well operate in ellipticals and clusters,
though without rapid rotation the coherence length of the fields is
limited by the characteristic scale of the turbulence.  In this
section, we discuss the necessity of dynamo action in ellipticals and
clusters, sources of energy for the generation of turbulence, and the
interplay between cooling flows and magnetic fields.

\subsubsection{Elliptical Galaxies}

Moss \& Shukurov (1996) identified two potential sources for seed
fields in elliptical galaxies: stellar magnetic fields ejected into
the ISM by supernavae and stellar winds and magnetic remnants that
arise if ellipticals formed from mergers of spiral galaxies.  The
former can lead to fields of about $10^{-16}\,{\rm G}$ (see below)
while the latter can lead to fields or about $10^{-8}\,{\rm G}$.  In
either case, Moss \& Shukurov (1996) concluded that the observed
microgauss fields in ellipticals require further amplification.
Moreover, simple stretching of fields lines by plasma motions without
dynamo action is probably unrealistic since Ohmic dissipation will
lead to loss of field energy.  They therefore concluded that magnetic
fields are amplified by dynamo processes.  However, the conditions in
elliptical galaxies do not appear to support a mean-field dynamo for
the reasons discussed above.  (See, however, Lesch \& Bender (1990)
where it is argued that a mean-field dynamo can operate in elliptical
galaxies.)  Instead, Moss \& Shukurov (1996) proposed that fields in
ellipticals are amplified by a so-called fluctuation dynamo
(Kazantsev, Ruzmaikin, \& Sokoloff 1985) in which turbulent motions
lead to rapid growth of the rms magnetic field.

Moss \& Shukurov (1996) identified on two potential sources of
turbulence in ellipticals, ejecta from supernovae and stellar winds
and the stirring of the interstellar gas by the random motions of
stars.  The former gives rise to acoustic turbulence --- essentially
the random superposition of sound waves --- with a characteristic
scale $\sim 1\,{\rm kpc}$.  The latter can drive vortical turbulence
with a much smaller characteristic scale ($\sim 3\,{\rm pc}$).  Moss
\& Shukurov (1996) suggested that amplification of magnetic fields in
ellipticals takes place in two stages.  First, vortical turbulence,
acting as a dynamo, amplifies a seed field to equipartition strengths
on $2-3\,{\rm pc}$ scales within a very short ($\la 10^5\,{\rm yr})$
timescale.  Acoustic turbulence amplifies the field and also leads to
an increase in the coherence length.  The model predicts that the
fields in the inner regions of ellipticals will have strengths that
range from $14\,\mu{\rm G}$ near the center to $1\,\mu{\rm G}$ at
$r=10\,{\rm kpc}$ and have a characteristic scale of $100-200\,{\rm
pc}$.

The effects of cooling flows as well as a time-dependent supernovae
rate were considered by Mathews \& Brighenti (1997).  Moreover,
they found that compression of the magnetic field in cooling flows
could lead to an additional amplification by a factor of $10^3$.
Moreover, the exact value of the field in ellipticals is sensitive to
interstellar turbulence in the distant past.

\subsubsection{Clusters}

The origin of magnetic fields in clusters is, at present, uncertain.
Cluster galaxies are an obvious source of magnetic fields.  Material
originally associated with individual galaxies is spread throughout
the intracluster medium by tidal stripping and galactic outflows.
Magnetic fields, tied to the fluid, are likewise dispersed throughout
the cluster, albeit with a smaller field strength due to the usual
${\cal V}^{2/3}$ dilution factor.  The observation that fields in
clusters are comparable in strength to those in galaxies implies that
field amplification is occurring in the intracluster medium.

Early attempts to understand the amplification of magnetic fields in
clusters focused on dynamo action in galactic wakes (Jaffe 1980;
Roland 1981; Ruzmaikin, Sokoloff, \& Shukurov 1989).  As galaxies move
through the intracluster medium they generate a turbulent wake which
can presumably support a dynamo.  However, more detailed calculations
have shown that it is very difficult to produce fields above $\sim
10^{-7}\,{\rm G}$ by this mechanism (Goldman \& Rephaeli (1991) and De
Young (1992)).

A more promising scenario relies on mergers to drive cluster-scale
dynamos (Tribble 1993).  Clusters are relatively young systems and
most have undergone at least one merger event during the last Hubble
time.  The amount of energy released during a major merger is
comparable to the gravitational and thermal energies of the system and
larger than the magnetic energy by a factor $\sim 10^2-10^3$.
Therefore only a small fraction of the available energy is required to
explain cluster fields.  Mergers lead naturally to turbulence and
shocks in the intracluster medium.  The former is necessary for a
dynamo while the latter can accelerate particles which can then
produce the synchrotron emission.  The lifetime of magnetic fields in
the intracluster medium can be relatively long (Soker \& Sarazin 1990;
Tribble 1993).  This timescale is set by the rate of magnetic
reconnection, which in turn is limited by the Alfv\'{e}n speed.  For a
$B\simeq 1\,\mu{\rm G}$ field and electron number density $n_e\simeq
2\times 10^{-3}{\rm cm}^{-3}$ one finds an Alfv\'{e}n speed $v_A\simeq
30\,{\rm km s}^{-1}$ so that the reconnection time for $10\,{\rm kpc}$
magnetic structures is $\ga 3\times 10^9\,{\rm yr}$.  (Note, however
that the timescale for the field to decay is shorter if the field is
concentrated in thin flux tubes (Subramanian 1998; Ruzmaikin et
al.~1989).)  By contrast, the lifetime of synchrotron emitting
electrons is short.  Relativistic electrons lose energy by inverse
Compton scattering off CMB photons.  The lifetime at the present epoch
is $\sim 10^8\,{\rm yr}$ and decreases rapidly with redshift (Tribble
1993).  Thus, radio halos fade quickly while cluster magnetic fields
survive for periods comparable to the Hubble time.  This simple
conclusion may explain the rarity of radio halos and the ubiquity of
cluster magnetic fields (Tribble 1993).

As in elliptical galaxies, cooling flows can have a significant 
effect on the magnetic fields in clusters.
Magnetic fields in cooling flows are amplified by radial infall and
shear (Soker \& Sarazin 1990) and can be expected to reach
equipartition with the gas.  A direct observable effect of these
fields is likely to be very strong Faraday rotation that increases
rapidly toward the cluster center.

\section{Seed Fields}

Though magnetic fields could have been a feature of the initial
conditions of the Universe a more appealing hypothesis is that they
are created by physical processes operating after the Big Bang.  These
first fields can be extremely small since subsequent dynamo action can
amplify them by many orders of magnitude.  However, even small fields
require an explanation (Zel'dovich \& Novikov 1983; Rees 1987;
Kronberg 1994; Beck et al.\,1996; Kulsrud 1999).

The list of proposals for the origin of seed fields is now long and
diverse.  However no single compelling model has emerged and thus the
following questions remain:

\begin{itemize}

\item When did magnetic fields first appear?  Were they present
during big bang nucleosynthesis? recombination? galaxy formation?

\item What was the spectrum (strength vs.\,coherence length) of the
first magnetic fields?  Was the galactic dynamo seeded by subgalactic,
galactic, or supergalactic-scale fields?  In the language of the old
debate on cosmological structure formation, are galactic magnetic
fields a top-down or bottom-up phenomena?

\item Is there a connection between the creation of the first fields
and the formation of large-scale structure?

\item Evidently, a dynamo is necessary to maintian galactic magnetic
fields.  Is dynamo action also necessary for amplification of an
initially small seed field?

\end{itemize}

Scenarios for seed fields fall into two broad categories, those that
rely on ordinary astrophysical processes and those that rely on new
and exotic physics.  By and large, astrophysical mechanisms exploit
the difference in mobility between electrons and ions.  This
difference can lead to electric currents and hence magnetic fields.
Mechanisms of this type can operate during galaxy formation or
alternatively they can operate in other systems such as stars and
active galactic nuclei.  In the latter case, the question is to
explain how magnetic fields in those systems make their way into the
interstellar or protogalactic medium.

Exotic processes in the very early Universe can also create magnetic
fields.  There is strong circumstantial evidence that the Universe has
undergone a series of phase transitions since the Big Bang including
an episode of inflation and the electroweak and quark-hadron
transitions.  Magnetic fields of interesting strength can arise during
these events though their connection to galactic and extragalactic
magnetic fields remains unclear.

\subsection{Minimum Seed Field for the Galactic Dynamo}

What is the minimum seed field required to explain the observed
galactic and extragalactic magnetic fields?  Uncertainties in our
understanding of the dynamo, galaxy formation, and cosmological
parameters make this quantity difficult to pin down.  In general,
the dynamo amplification factor ${\cal A}$ can be written

\be{growth}
{\cal A}\equiv \frac{B_f}{B_i}=e^{\Gamma\left (t_f-t_i\right )}
\ee

\noindent 
where $\Gamma$ is the growth rate for the dominant mode of the dynamo,
$B_i$ is the field strength at the time $t_i$ when the dynamo begins
to operate, and $t_f$ is the time when the fields reach the observed
value $B_f$.  

Irrespective of dynamo action, a seed field that is created prior to
galaxy formation is amplified as the protogalactic gas collapses to
form a disk.  This point was stressed by Lesch \& Chiba (1995) who
estimated the amount of pre-dynamo amplification that occurs during
galaxy formation.  Their analysis was based on a simple and well-worn
model for the formation of a disk galaxy proposed by Fall \&
Efstathiou (1980) (see, also White \& Rees 1978).  According to Fall
\& Efstathiou (1980), the first stage of galaxy formation is the
development of an extended virialized halo of gas and dark matter where
the gas and dark matter are assumed to have the same specific angular
momentum.  Gas elements cool and lose energy but, by assumption,
conserve angular momentum.  The end result is a disk which, for simple
and reasonable choices of initial conditions (e.g., a gas-dark matter
halo in solid-body rotation), resembles the disks of present-day
galaxies.

In the hierarchical clustering scenario, halos form around peaks in
the primordial fluctuation distribution.  Small-scale objects collapse
first and coalesce to form systems of increasing size.  In general,
the inner regions of halos form first.  The spherical infall model
(Gunn \& Gott 1972) provides a useful, albeit highly idealized,
picture of halo formation.  In this model, the mass distribution
around a given peak is assumed to be spherically symmetric with a
density profile that decreases monotonically with distance from the
peak center.  The evolution of a protohalo can therefore be described
in terms of the dynamics of spherical shells.  Initially, the shells
expand with the Hubble flow, though eventually they reach a maximum
radius known as the turnaround radius $R_{\rm ta}$, break away from
the general expansion and collapse.  After a shell collapses, it
virializes with the other shells (though this process is not
particularly well understood).  The model captures certain features of
the hierarchical clustering scenario (e.g., inside-out halo formation)
while ignoring the complexities of true hierarchical structure
formation.  A number of results obtained from this model (e.g.,
characteristic density, halo formation time) are in good agreement
with those obtained from N-body simulations of hierarchical structure
formations.  Simple arguments suggest that the characteristic radius
of a virialized shell is roughly one half of the turnaround radius,
$R_{\rm vir}\simeq R_{\rm ta}/2$.  For a typical spiral galaxy,
$R_{\rm vir}\simeq 150\,{\rm kpc}$.  A seed magnetic field created
prior to halo formation is therefore amplified by a factor $\left
(R_{\rm ta}/R_{\rm vir}\right )^2\simeq 4$.

A crucial assumption for what follows is that the gas-dark matter halo
acquire angular momentum as it forms.  In a hierarchical clustering
scenario, protogalaxies gain angular momentum through tidal
interactions with neighboring protogalaxies (Hoyle 1949; Peebles 1969;
White 1984).  (Tidal torque theory, of course, requires a departure
from spherical symmetry since tidal fields do not couple to spherical
shells.)  It is common practice to characterize the angular momentum
of a system in terms of the dimensionless spin parameter $\lambda
\equiv J|E|^{1/2}/GM^{5/2}$ where $J$ is the total angular momentum of
the system and $E$ is its total energy.  Roughly speaking, $\lambda$
is the square root of the ratio of the centripetal acceleration to the
gravitational acceleration.  Theoretical analysis and numerical
simulations suggest that, for a typical halo, $\lambda\simeq 0.07$
(Peebles 1969; Barnes \& Efstathiou 1987).

As described above the second stage in the formation of a disk galaxy,
the gas dissipates energy and sinks to the center of the halo's
gravitational potential well where it forms a rotationally supported
disk (i.e., a system in which $\lambda\simeq 1$).  If the protogalaxy
were composed entirely of gas, the radius of the disk would be $R_{\rm
disk}\simeq \lambda^2 R_{\rm vir}\la 1\,{\rm kpc}$.  Not only is this
value too small (the typical scale radius of a spiral galaxy is
$\sim 10\,{\rm kpc}$) but the time required for such a the disk to
form is greater than the age of the Universe (Peebles 1993).  In a CDM
model, 90\% of a protogalaxy is in the form of collisionless dark
matter which remains in an extended halo.  In this case, the disk
radius is $R_{\rm disk}\simeq \lambda R_{\rm vir}\simeq 10\,{\rm
kpc}$.

Magnetic fields, frozen into the gas, are amplified by a factor 
$\lambda^{-2}$.  Additional amplification occurs due to the collapse
along the spin axis as the gas forms a thin disk.  The net
pre-dynamo amplification factor is therefore

\be{amp_fac}
{\cal A}'\simeq 8\times 10^3
\left (\frac{R_{\rm ta}/R_{\rm vir}}{2}\right )^2  
\left (\frac{0.07}{\lambda}\right )^2  
\left (\frac{R_{\rm disk}/R_H}{10}\right )
\ee

\noindent
where $R_H$ is the disk scale height (Lesch \& Chiba 1995).
  
The quantity $\Gamma$ is model-dependent and not very well known.
In Section IV.E, we found

\be{gamma_max}
\Gamma < \gamma < c\omega_0\left (\frac{\alpha_0}{h\omega_0}
\right )^{1/2}
\ee

\noindent 
where $c=0.6-0.8$ depending on the functional form of $\alpha(z)$.
Dimensional arguments suggest that $\alpha_0 = L^2\omega/h$ where $L$
is the size of the largest eddies and $h$ is the scale height of the
disk (Ruzmaikin, Shukurov, and Sokoloff 1988a).  With $h\simeq
500\,{\rm pc}$, $\omega\simeq 29\,{\rm km\,s^{-1}\,kpc}$, and $L\simeq
100\,{\rm pc}$, we find $\alpha_0\simeq 6\times 10^4 \cms$ and $\Gamma
\la 0.7-0.9\,{\rm Gyr}^{-1}$.  Ferri\`{e}re (1992, 1993) derives the
somewhat smaller value $\alpha \simeq 2\times 10^4 {\rm cm\,s^{-1}}$
for the supernovae-driven model of turbulence and detailed models of a
galactic dynamo based on these results give $\Gamma\simeq 0.45{\rm
Gyr}^{-1}$ (Ferri\`{e}re and Schmidt 2000).  However, as discussed
above, their calculation may underestimate the strength of the
$\alpha$-effect from supernovae.

Observations of CMB angular anisotropy, high-redshift supernovae, and
large-scale structure indicate that the Universe is spatially flat
with $\Omega_m\simeq 0.15-0.40$ and $\Omega_\Lambda\simeq 1-\Omega_m$.
This conclusion implies an older Universe as compared with one in
which $\Omega_m=1$.  The lower bound on the required seed field is
therefore relaxed since there is more time for the dynamo to operate.
In a spatially flat, Robertson-Walker Universe,

\bea{time}
t_f-t_i &=&\frac{1}{H_0}\int_{a_i}^{a_f}
\frac{da}{\left (\Omega_m a^{-1} + \Omega_\Lambda a^2\right
)^{1/2}} \nonumber \\
& = & \frac{2}{3H_0\Omega_\Lambda^{1/2}}
\ln\left (\frac{\lambda_f + \left (\lambda_f^2+1\right )^{1/2}}
{\lambda_i + \left (\lambda_i^2+1\right )^{1/2}}\right )
\eea	 

\noindent 
where, as before, $a=\left (1+z\right )^{-1}$ is the scale factor at
time $t$ with $a(t_0)\equiv 1$ and $\lambda\equiv\left
(\Omega_\Lambda/ \Omega_m\right )^{1/2}\left (1+z\right )^{-1}$. An
upper bound on $z_i$ is given by the redshift at which a protogalaxy
separates from the Hubble flow, collapses, and virializes.  Simple
estimates based on the spherical infall model suggest a value
$z_i\simeq 50$.  However, disks are almost certainly assembled at a
much later epoch.  Observerations such as the Hubble Space Telescope
Deep Field Survey, for example, indicate that disk galaxies at $z\sim
3$ are highly irregular and still in the process of being formed
(see, for example, Lowenthal et al.\,1997)

A plot of $t_f-t_i$ vs.~$\Omega_m$ for $H_0=70\,{\rm km\, s^{-1}
Mpc}$, $t_f=t_0$, and various choices of $z_i$ is given in Figure
12(a).  The amplification factor ${\cal A}$, assuming a growth rate
$\Gamma = 2\,{\rm Gyr}^{-1}$, is shown on the right-hand vertical
axis.  Note that $\log{\cal A}$ scales linearly with $\Gamma/H_0$.
Evidently, ${\cal A}$ varies by several orders of magnitude depending
on $z_i$ and $\Omega_m$.  For $z_i=10$ and $\Omega_m=0.2$, the maximum
amplification factor is ${\cal A}\simeq 10^{14}$ and therefore a
present-day microgauss requires a seed field with strength $B_i\simeq
10^{-20}\,{\rm G}$.  (We have not included the amplification ${\cal
A}'$ which can occur during galaxy formation.)


\begin{figure}
\begin{center}
\begin{tabular}{cc}
\psfig{file=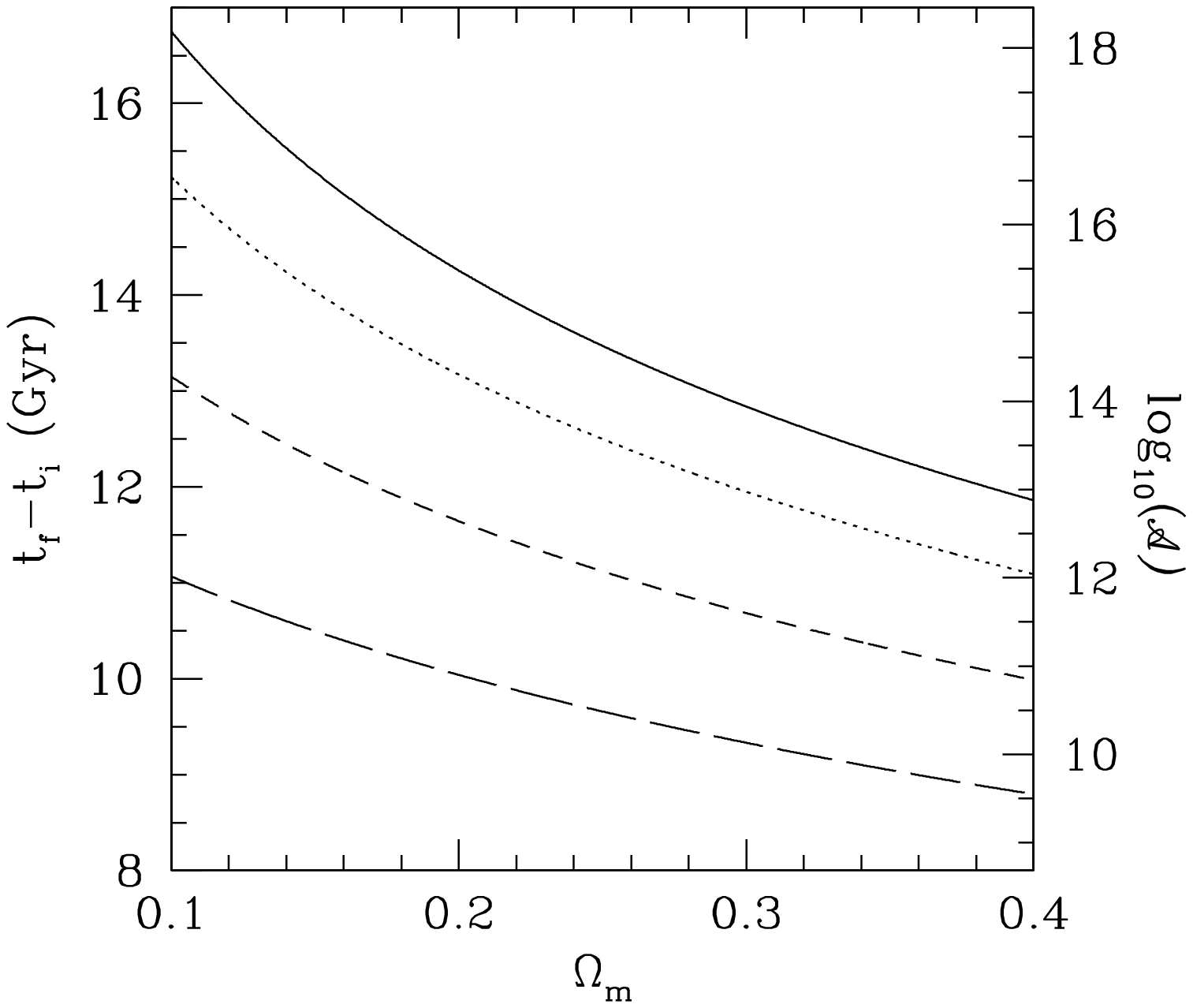,width=9.5cm,height=9.5cm} &
\psfig{file=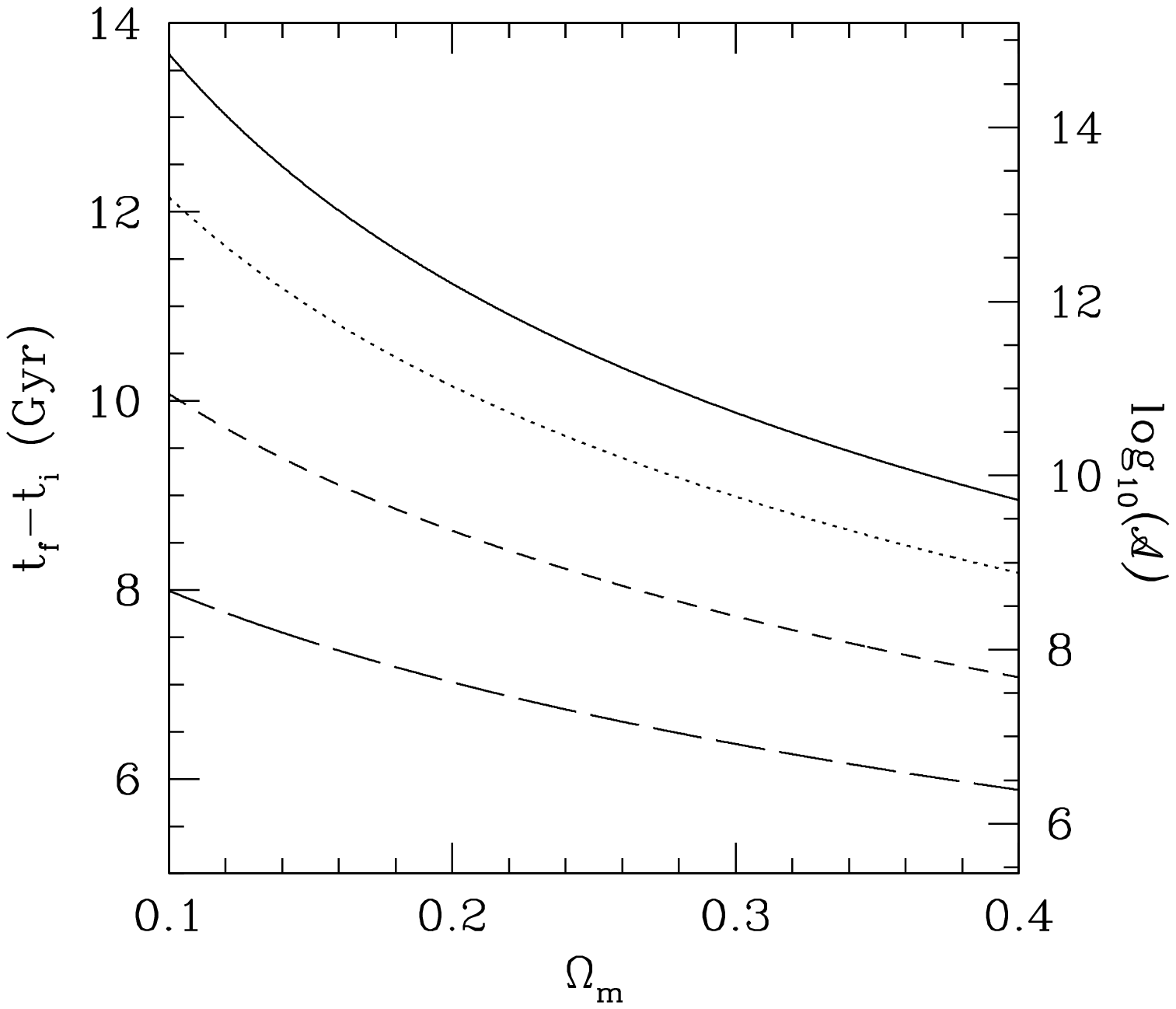,width=9.5cm,height=9.5cm} \\
\psfig{file=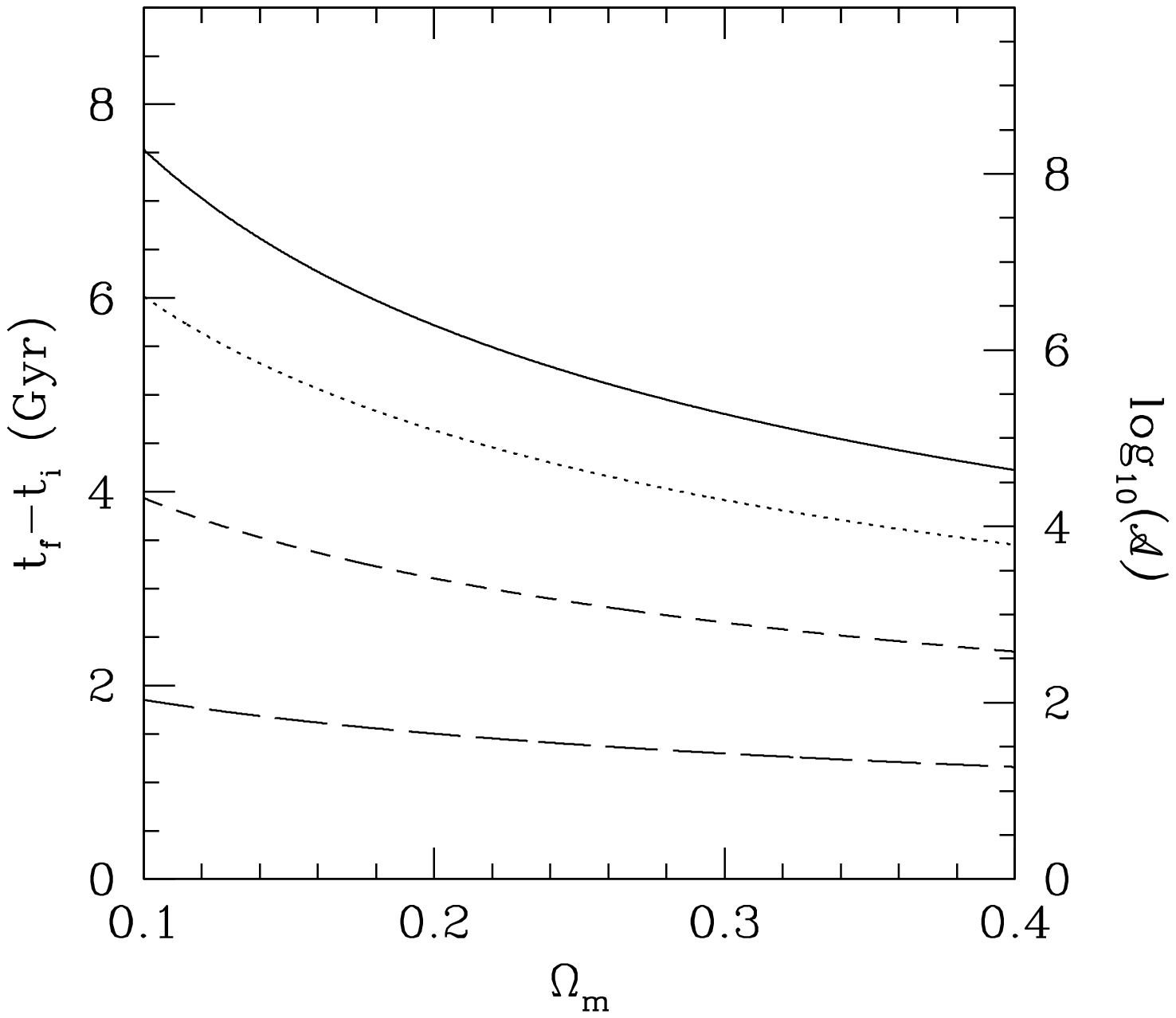,width=9.5cm,height=9.5cm} &
\\ 
\end{tabular}
\end{center}

\caption{Available time for dynamo action as a function of $\Omega_m$
in spatially flat cosmologies ($\Omega_\Lambda =1-\Omega_m$).
Right-hand vertical axis gives the amplification factor assuming a
growth rate $\Gamma = 2\,{\rm Gyr}^{-1}$.  We assume $H_0=70\, {\rm
km\,s^{-1}\,Mpc}$.  The four curves in each of the plots are for
different choices of $t_i$, or equivalently, $z_i$: $z_i=25$ --- solid
curve; $z_i=10$ --- dotted curve; $z_i=5$ --- dashed curve; $z_i=3$
--- long-dashed curve.  The three plots assume different values for
$t_f$: (a) $z_f=0$; (b) $z_f=0.4$; (c) $z_f=2$.}
 
\label{limits}
\end{figure}


Davis, Lilley, \& T\"{o}rnquist (1999) argued that the lower bound on
the strength of the seed field is $10^{-30}\,{\rm G}$ or less
depending on the cosmological model.  Their estimate assumes a lower
value for the Hubble constant ($H_0=50-65\,{\rm km\,s^{-1}\,{\rm
Mpc}}$), values for the dynamo growth rate as high $\Gamma=3.3\,{\rm
Gyr}^{-1}$, and an earlier choice for $t_i$.  The latter value is taken
to be the time when a region that corresponds in size to the largest
turbulent eddy ($\sim 100\,{\rm pc}$) breaks away from the Hubble flow
and collapses.  The implicit assumption is that dynamo processes begin
to operate on subgalactic scales well before the disk forms.  While
this assumption is reasonable, it requires further study.  In
particular, the estimates for $\Gamma$, derived for a disk dynamo, do
not necessarily apply to subgalactic objects in an evolving hierarchy
of structure.

Observations of microgauss fields in galaxies at moderate redshifts
tighten severely the lower bound on seed fields.  The results of
Kronberg, Perry, and Zukowski (1992) imply $z_f\ga 0.4$ while those of
Wolfe, Lanzetta, and Oren (1992) suggest a more stringent albeit
controversial choice $z_f\ga 2$.  $z_f\ga 2$ is also implied by the
observations of radio galaxies by Athreya et al.\,(1998).  Figures
12(b) and 12(c) show $t_f-t_i$ for $z_f=0.4$ and $2$ respectively.
With $z_f=0.4$, $z_i=10$, and $\Omega_m=0.2$, the limit on $B_i$ is
tightened to $10^{-16}\,{\rm G}$.  More severe is the case $z_f=2$
where the time available for the dynamo to operate is shortened to a
few billion years or less and a seed field with $B_i\simeq
10^{-10}\,{\rm G}$ is required.

\subsection{Astrophysical Mechanisms}

The difference in mobility between electrons and ions in an ionized
plasma leads to charge separation effects and a breakdown of the MHD
approximation.  Consider a multicomponent fluid composed of electrons,
protons, and photons.  (The discussion can be extended easily to
include heavier ions and neutral atoms).  The momentum equations for
the electrons and protons can be written:

\be{electron_eom}
\frac{d{\bf V}_e}{dt}
	~=~-\frac{\bfn p_e}{\rho_e} ~-~\frac{e}{m_e}\left (
	{\bf E}~+~\frac{{\bf V}_e\times {\bf B}}{c}\right )
	-\bfn\phi ~+~ \frac{{\bf K}_{ep}}{m_e}
\ee

\be{proton_eom}
\frac{d{\bf V}_p}{dt}
	~=~-\frac{\bfn p_p}{\rho_p} ~+~\frac{e}{m_p}\left (
	{\bf E}~+~\frac{{\bf V}_p\times {\bf B}}{c}\right )
	~-~\bfn\phi ~-~ \frac{{\bf K}_{ep}}{m_p}
\ee

\noindent where $p_i$, $m_i$, $\rho_i=m_i n_i$, $n_i$, and ${\bf v}_i$
are the partial pressure, mass, mass density, number density, and
velocity of the electrons $(i=e)$ and protons $(i=p)$ (e.g., Spitzer
1962; Sturrock 1994).  The rate of momentum transfer from protons to
electrons is ${\bf K}_{ep}=m_e\left ({\bf V}_p-{\bf V}_e\right )/\tau$
where $\tau=m_e\sigma/n_e e^2$ is the characteristic timescale of
electron-proton collisions.  The current density is
${\bf J}=e\left (n_p{\bf V}_p-n_e{\bf V}_e\right )$ and therefore, if
approximate local charge neutrality holds (i.e., $n_e\simeq n_p$),
${\bf K}_{ep}\simeq e{\bf J}/\sigma$.

Also important for some scenarios is the coupling of electrons and
photons due to Thomson scattering.  In the strong coupling limit,
photons behave like an ideal fluid.  Thomson scattering leads to an
additional momentum transfer term in Eq.\,\EC{electron_eom} of the
form ${\bf K}_{e\gamma}=4c\sigma_T\rho_\gamma\left ({\bf V}_\gamma
-{\bf V}_e\right )/3$ where $\sigma_T=\frac{8}{3}\pi \left
(e^2/m_ec^2\right )^2=6.65\times 10^{-25}\,{\rm cm^2}$ is the Thomson
cross section and $\rho_\gamma c^2$, $p_\gamma$, and ${\bf V}_\gamma$
are respectively the energy density, pressure, and (fluid) velocity of
the photon fluid.  The momentum equation for the photon fluid is given
by

\be{photon_eom}
\frac{4}{3}\rho_\gamma\frac{d{\bf V}_\gamma}{dt}=
-\bfn p_\gamma-
\frac{4}{3}\rho_\gamma\bfn\phi -n_e{\bf K}_{e\gamma}
\ee

\noindent (see, for example, Peebles 1980).

\subsubsection{Seed Fields from Radiation-Era Vorticity}

An early attempt to explain the origin of seed fields is due to
Harrison (1970, 1973) who considered the evolution of a rotating
protogalaxy prior to decoupling.  During this epoch, electrons and
photons are tightly coupled so that, for most purposes, they can be
treated as a single fluid.  The coupling between electrons and protons
is somewhat weaker allowing currents and hence magnetic fields to
develop.

Harrison considered a homogeneous, spherical protogalaxy in solid body
rotation with radius $R$.  As the protogalaxy expands, $\rho_p R^3$
and $\rho_\gamma R^4$ remain constant.  In the limit that the
electron-proton coupling is ignored the angular momenta of the
electron-photon and proton fluids are separately conserved, i.e.,
$\rho_p \omega_p^2 R^5$ and $\rho_\gamma\omega_\gamma^2 R^5$ are
constant.  This implies that $\omega_p\propto R^{-2}$ and
$\omega_\gamma\propto R^{-1}$.  The ions ``spin down'' faster than the
electrons and photons and a current $J\sim e\omega_\gamma R /m_pc$ and
hence a magnetic field $B\sim m_pc\omega_\gamma/e \simeq 10^{-4}\,{\rm
G} \left (\omega_\gamma/s^{-1}\right )$ develop.

A more formal derivation of this result, including the electron-proton
coupling, follows from Eqs.\,\EC{electron_eom}--\EC{photon_eom}.  The
velocity field for Harrison's spherical protogalaxy can be decomposed
into a homogeneous expansion and a circulation flow, i.e., ${\bf V}_p =
\left (\dot a/a\right ){\bf r}+{\bf u}_p$ where $\bzeta_p\equiv
\bfn\times {\bf V}_p=\bfn\times {\bf u}_p$ is the vorticity of the
proton fluid.  Combining the curl of Eq.\,\EC{proton_eom} with
Maxwell's equations gives

\be{proton_curl}
\frac{d}{dt}\left (a^2\left (\bzeta_p+\frac{e}{m_H c}{\bf B}\right )
\right )=\frac{e}{4\pi\sigma m_H}a^2\nabla^2{\bf B}
\ee

\noindent 
where, because of the assumed homogeneity, the pressure gradient term
in Eq.\,\EC{proton_eom} is dropped.  In the highly conducting
protogalactic medium, the diffusion term is negligible thus yielding
the conservation law

\be{conservation} 
\bzeta_p(t)+\frac{e}{m_Hc}{\bf B}(t) = \left
(\frac{a(t_i)}{a(t)}\right )^2\bzeta_p(t_i) 
\ee

\noindent where $t_i$ is some initial time characterized by zero
magnetic field.

The dynamics of the electrons is determined primarily by the photons.
The inertial and gravitational terms in Eq.\EC{electron_eom} are
therefore neglected leaving what is essentially a constraint
equation for the electron velocity field:

\be{electron_eom2}
{\bf E}+\frac{{\bf V}_e\times {\bf B}}{c}
=\frac{e{\bf J}}{\sigma}+\frac{{\bf K}_{e\gamma}}{e}~.
\ee

\noindent The curl of this equation, together with Maxwell's equations,
gives

\be{electron_curl}
\frac{1}{a^2}\frac{d}{dt}\left (a^2 {\bf B}\right ) = 
-\frac{1}{e}\bfn\times {\bf K}_{e\gamma}
\ee

\noindent while the curl of Eq.\,\EC{photon_eom} gives

\be{photon_curl}
\frac{1}{a}\frac{d}{dt}\left (a \bzeta_\gamma\right ) = 
-\frac{3\rho_p}{4\rho_\gamma m_p}\bfn\times {\bf K}_{e\gamma}~.
\ee

\noindent 
Eqs.\,\EC{proton_curl}, \EC{electron_curl}, and \EC{photon_curl} can
be combined to eliminate ${\bf K}_{e\gamma}$ and $\bzeta_p(t_i)$ in
favor of ${\bf B}$:

\be{harrison_b}
{\bf B}(t) = -\frac{m_p c}{e}\left (1-\frac{a(t_i)}{a(t)}\right )
\bzeta_p (t)~.
\ee

\noindent 
In deriving this equation we have used the fact that
$\rho_\gamma\propto a^{-4}$ and $\rho_p\propto a^{-3}$.  In addition,
to leading order, we set ${\bf V}_\gamma= {\bf V}_e={\bf V}_p$ and
$\bzeta_\gamma=\bzeta_p$, i.e., differences between the electron,
proton, and photon velocities enter the calculation only through the
collision terms. The desired result, $B\simeq m_pc\zeta/e$, follows
immediately if we take $a(t)\gg a(t_i)$.  By way of example, we note
that the solar-neighborhood value $\zeta\simeq 29\,\kms\,{\rm
kpc}^{-1}$ yields a seed magnetic field $B\simeq 10^{-19}\,{\rm G}$.

Harrison's scenario for the origin of magnetic fields has a number of
attractive features which are echoed in subsequent proposals.  First,
seed fields are generated as part of the galaxy formation process and
therefore naturally have a coherence length comparable to the size of
the galaxy.  Second, the seed fields have a strength that is set by
the vorticity of the protogalaxy which can be related to the
present-day vorticity in galactic disks, an observable quantity.
These fields, though weak, are sufficient to explain present-day
galactic fields, provided an efficient dynamo develops to amplify
them.  The most severe criticism of the model is that, prior to
structure formation, vorticity decays rapidly due to the expansion of
the Universe (Rees 1987).  The implication is that vorticity in disk
galaxies in not primordial but rather generated during structure
formation.  Harrison's scenario does suggest that there may be a
connection between the origin of galactic vorticity and the origin of
seed fields.

\subsubsection{Biermann Battery Effect}

In the hierarchical clustering scenario, protogalaxies acquire
rotational angular momentum from the tidal torques produced by their
neighbors (Hoyle 1949; Peebles 1969; White 1984).  However,
gravitational forces alone cannot generate vorticity and therefore its
existence in galactic disks must be due to gasdynamical processes such
as those that occur in oblique shocks.  In an ionized plasma, these
same processes produce magnetic fields (Pudritz \& Silk 1989; Kulsrud,
Cen, Ostriker, \& Ryu 1997; Davies \& Widrow 2000).

We consider vorticity generation first because it is conceptually
simpler.  In the absence of electromagnetic effects and viscosity, the
evolution of a collisional fluid is given by Eq.\,\EC{euler} with
${\bf J}=0$ and $\eta=0$.  Taking the curl yields the following
equation for the vorticity:

\be{vorticity}
\frac{\partial\bzeta}{\partial t}~-~
\bfn\times\left ({\bf V}\times \bzeta\right )~=~
\frac{{\bfn\rho}\times{\bfn p}}{\rho^2}~.
\ee

\noindent
The source term on the right-hand side comes from gasdynamical
effects, namely pressure and density gradients that are not collinear.
(The gravitational force term does not appear since the curl of
$\bfn\psi$ is identically zero.)  Vorticity generation is illustrated
in Figure 13 where we consider an ideal single component fluid
consisting of a particles of mass $m$.  The gravitational force $m{\bf
g}$ is only partially compensated by the pressure gradient force
$-\bfn p/n$.  The acceleration rate for high density regions is
greater than the rate from low density regions and therefore the
velocity field downstream of the pressure gradient has a nonzero
vorticity and shear ($\partial v_y/\partial x\ne 0$).


\begin{figure}
\centerline{\psfig{file=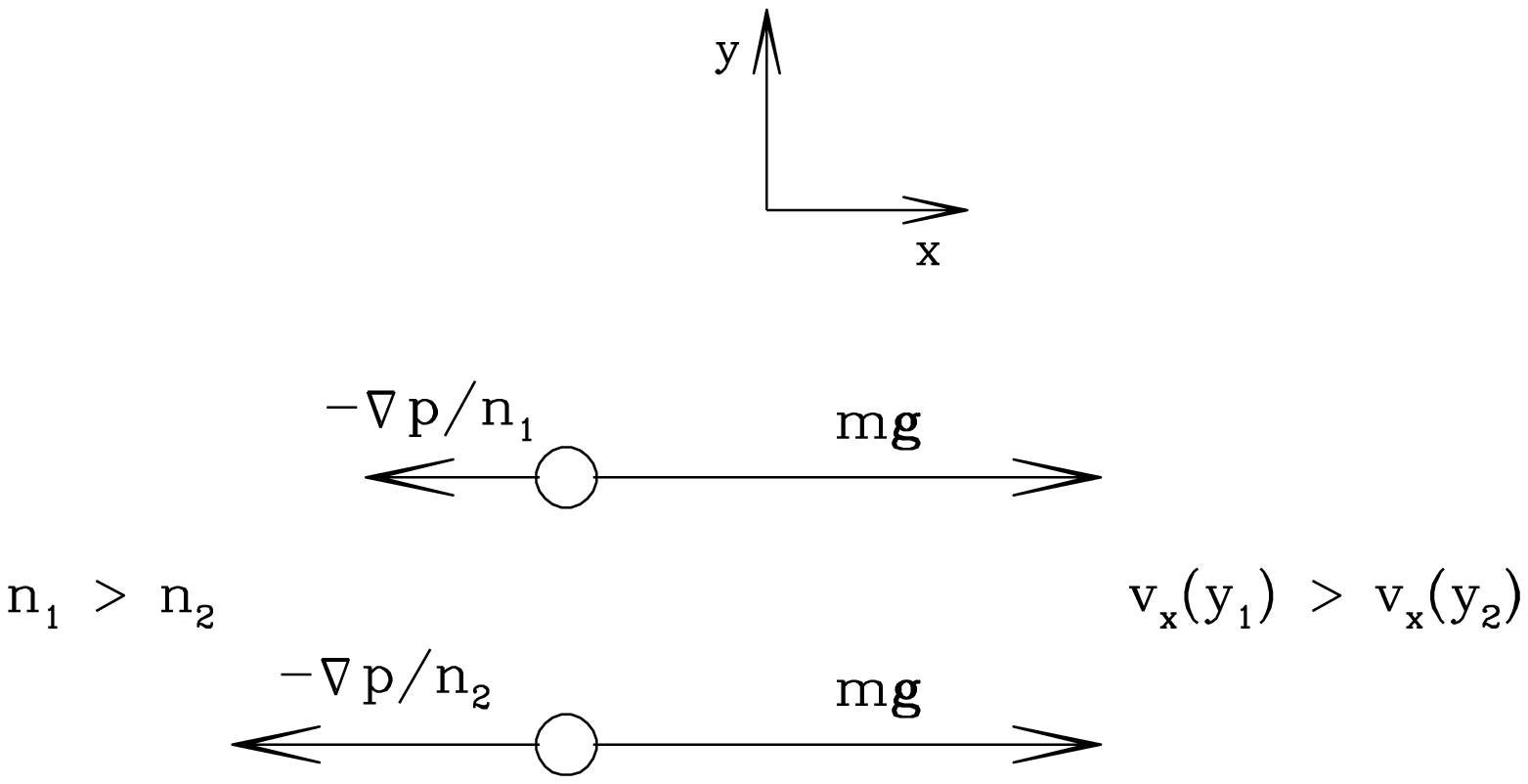,width=10cm}}

\caption{Schematic diagram illustrating the generation of vorticity in
a region of crossed pressure and density gradients.  The open circles
represent particles of equal mass separated in $y$.  The gravitational
force on the particles, $m{\bf g}$ is the same.  Particle 1 sits in a
higher density region and therefore experiences a smaller pressure
gradient force.  A velocity field with shear and vorticity develops.}

\label{vorticity}
\end{figure}


In an ionized plasma, similar inertial effects lead to electric
currents and magnetic fields.  This mechanism, known as the Biermann
battery effect (Biermann 1950; Roxburgh 1966) can be derived by
combining the generalized Ohm's law with Maxwell's equations.  The
former, found by taking the difference of Eqs.\,\EC{electron_eom} and
\EC{proton_eom} can be written

\bea{ohm1}
\frac{m_e}{e^2}
\frac{\partial}{\partial t}\left (\frac{{\bf J}}{n_e}\right )
&=&
\frac{m_e}{e}\frac{d\left ({\bf V}_p-{\bf V}_e\right )}{dt}
\nonumber \\
&=&\frac{\bfn p_e}{en_e} + {\bf E} +
\frac{{\bf J}\times {\bf B}}{cn_e} + \frac{{\bf V}_p\times {\bf B}}{c}
-\frac{{\bf J}}{\sigma}
\eea

\noindent 
where terms of order $m_e/m_p$, as well as terms quadratic in the
velocities, have been neglected (Spitzer 1962; Sturrock 1994).  In
addition, we have assumed that local charge neutrality ($n_e = n_p$)
holds and that the electron and proton partial pressures are equal.
The ${\bf J}\times {\bf B}$ term describes the backreaction of the
magnetic field on the fluid.  Since, in the present discussion, we are
interested in the creation of (small) seed fields, this term can be
ignored.  In ideal MHD, the left-hand side, which described charge
separation effects, is zero.  In addition, the pressure gradient term
is ignored.  We therefore recover the simple form of Ohm's law: ${\bf
J}=\sigma\left ({\bf E}+\left ({\bf V}_p\times {\bf B}\right )/c
\right )$ where ${\bf V}_p$ is essentially the fluid velocity.  In the
present discussion, we focus on charge separation effects and assume
that the conductivity is high and that ${\bf B}$ and ${\bf V}_p/c$ are
small.  The result is a form of Ohm's law which is relevant to the
Biermann effect:

\be{ohm2}
\frac{m_e}{e}\frac{\partial}{\partial t}\left (\frac{\bf J}{n_e}\right )
=\frac{\bfn p}{n_e} + e{\bf E}~.
\ee

\noindent
The curl of this equation,

\be{curl_ohm}
\frac{m_e}{e}\bfn\times\frac{\partial}{\partial t}
\left (\frac{\bf J}{n_e}\right )
=-\frac{\bfn n_e\times\bfn p}{n_e^2} + e\bfn\times {\bf E}~,
\ee

\noindent
together with Maxwell's equations, gives

\be{biermann}
\frac{\partial{\bf B}}{\partial t} -
\bfn\times\left ({\bf V}\times {\bf B}\right ) =
\frac{m_e c}{e}
\frac{\bfn p_e\times\bfn \rho_e}{\rho_e^2}~.
\ee

The Biermann effect can be illustrated by the following simple
example.  (For a similar pedagogical discussion of the Biermann effect
in stars, see see Kemp (1982)).  Consider, first, the plane-symmetric
flow in Figure 14(a).  Since the gravitational force on the protons is
much greater than that on the electrons, electrostatic equilibrium
($d{\bf V}_p/dt=d{\bf V}_e/dt$) requires an electric field ${\bf
E}\simeq -\bfn p/en_e\simeq m{\bf g}'/2$ where ${\bf g}'\equiv {\bf
g}-d{\bf V}_p/dt$ is the effective gravitational force in the rest
frame of the fluid.  An electrostatic force of this type exists in
stars (Rosseland 1924) but is extremely small and inconsequential to
understanding stellar structure.


\begin{figure}
\begin{center}
\begin{tabular}{c}
\psfig{file=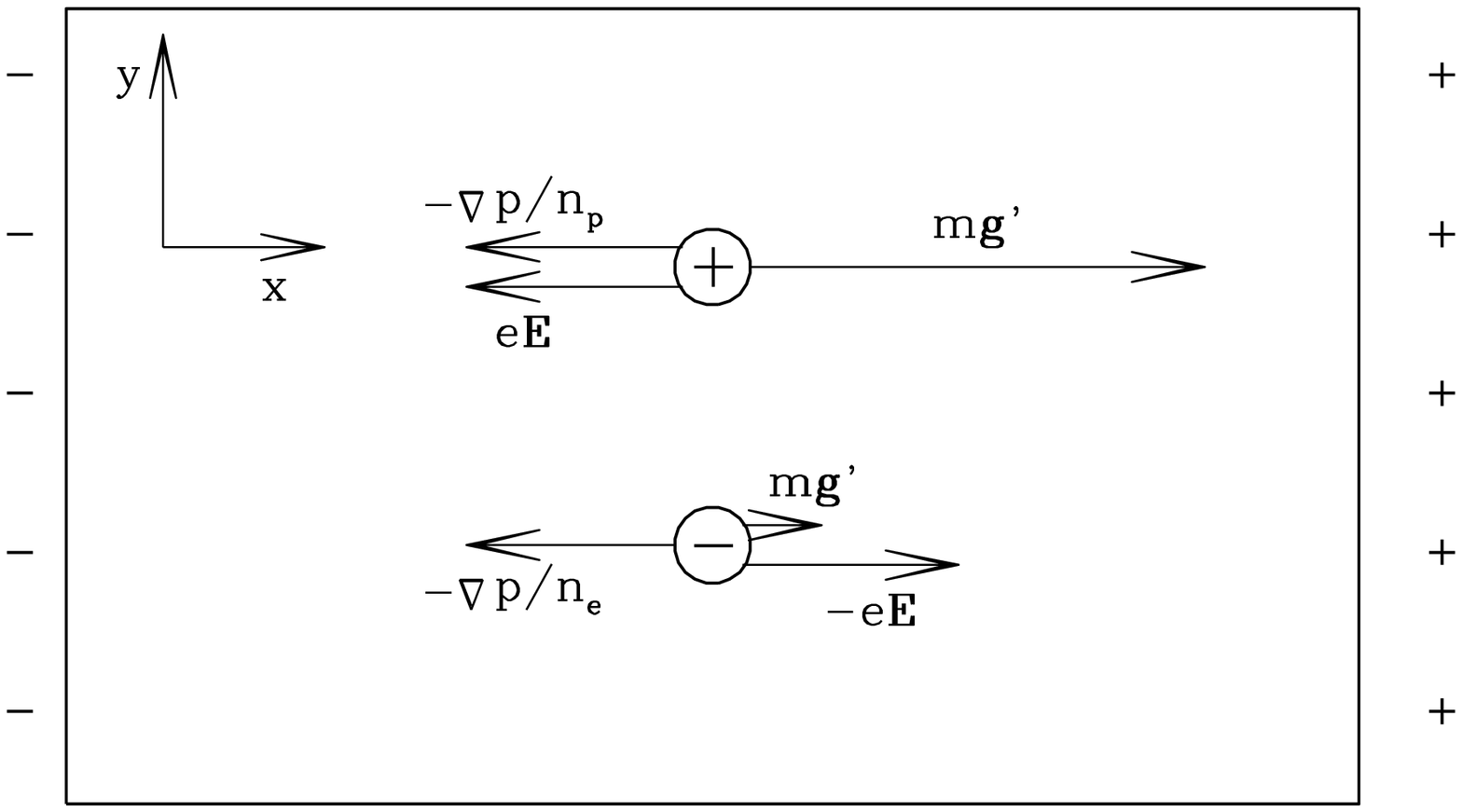,width=9.5cm,height=9.5cm} \\
\psfig{file=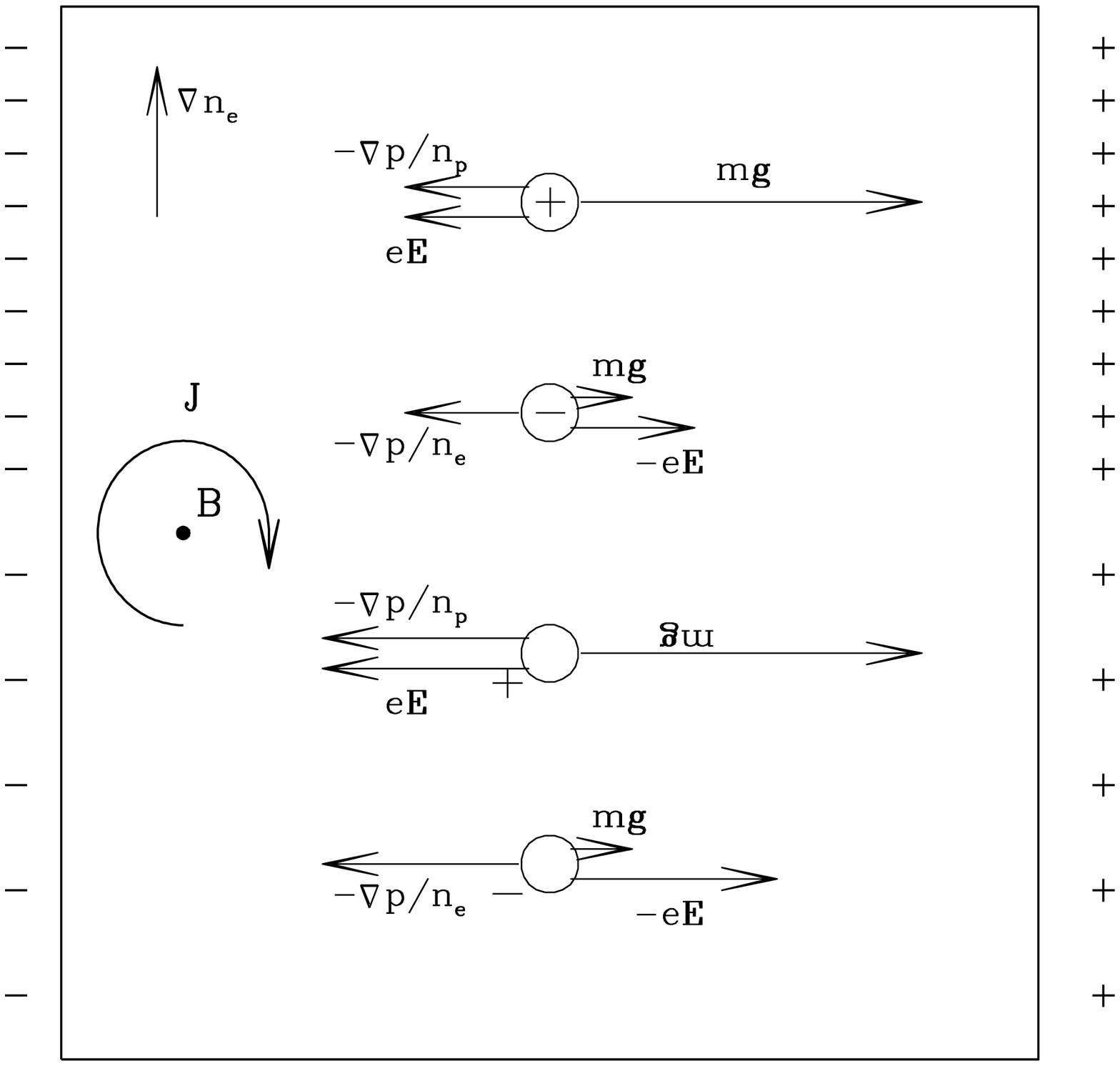,width=9.5cm,height=9.5cm} \\ 
\end{tabular}
\end{center}

\caption{Schematic diagram illustrating the Biermann battery effect.
Open cirles represent ions and electrons (plus and minus symbols,
respectively).  In (a), physical quantities do not depend on $y$ an
electric field develops in the $x$ direction, and electrostatic
equilibrium is achieved.  In (b), the density increases in the $y$
direction and electrostatic equilibrium is no longer possible.
Instead, this situation leads to a time-dependent magnetic field into
the page and a clockwise current as shown.}

\label{biermann}
\end{figure}


Electrostatic equilibrium is no longer possible when gradients in
theremodynamic quantities such as the density and temperature are not
parallel to the pressure gradient.  This situation is illustrated in
Figure 14(b) where the density increases in the $+y$-direction.  The
second term in Eq.\,\EC{curl_ohm} is a nonzero vector in the $+z$-direction
and therefore at least one of the other terms in this equation is
nonzero.  In fact, since a nonzero $\bfn\times {\bf E}$ implies a
time-dependent magnetic field, which in turn implies time-dependent
currents, both terms are nonzero.  However, the term on the left-hand
side is negligible:

\bea{compare}
\frac{|\frac{m_e}{e}\bfn\times\frac{\partial}{\partial t}
\left ({\bf J}/n_e\right )|}
{|e\bfn\times {\bf E}|}&\sim&
\frac{m_e c^2}{4\pi e^2 n_e L^2}\nonumber \\
&\simeq & 10^{-26}
\left (\frac{10^{-4}{\rm cm}^{-3}}{n_e}\right )
\left (\frac{100\,{\rm pc}}{L}\right )^2
\eea

\noindent
where $L$ is the typical length scale for the system.  We therefore
find a time-dependent magnetic field, in agreement with
Eq.\,\EC{biermann}.  For the situation shown in Figure 14(b), the
magnetic field is in the $+z$-direction.

Approximate local charge neutrality implies that $n_e\simeq n_p \equiv
\chi\rho/m_p$ where $\chi$ is the ionization fraction.  In addition,
since the electrons and protons are expected to be in approximate
thermal equilibrium, $p_e\simeq pn_e/\left (n_e+n_p\right )=p\chi/\left
(1+\chi\right )$ where $p$ is the total gas pressure.  The source term
on the right-hand side of Eq.\,\EC{biermann} can therefore be written

\be{source}
\mbox{\boldmath $\Gamma$}\simeq\frac{\alpha}{1+\chi}
\frac{\bfn\rho\times\bfn p}{\rho^2}~.
\ee

\noindent
and is evidently proportional to the source term for the vorticity.

It is interesting to note that the Biermann effect has been observed
in lasar-generated plasmas (Stamper \& Ripin 1975; also see Loeb \&
Eliezer 1986).  The schematic diagram of a typical experiment is shown
in Figure 15.  The plasma has strong temperature and density gradients
that are nearly perpendicular leading to a source term for the
magnetic field.  These magnetic fields, typically megaguass in
strength, were studied through Faraday rotation induced on a second
laser beam that was shined through the plasma.


\begin{figure}
\centerline{\psfig{file=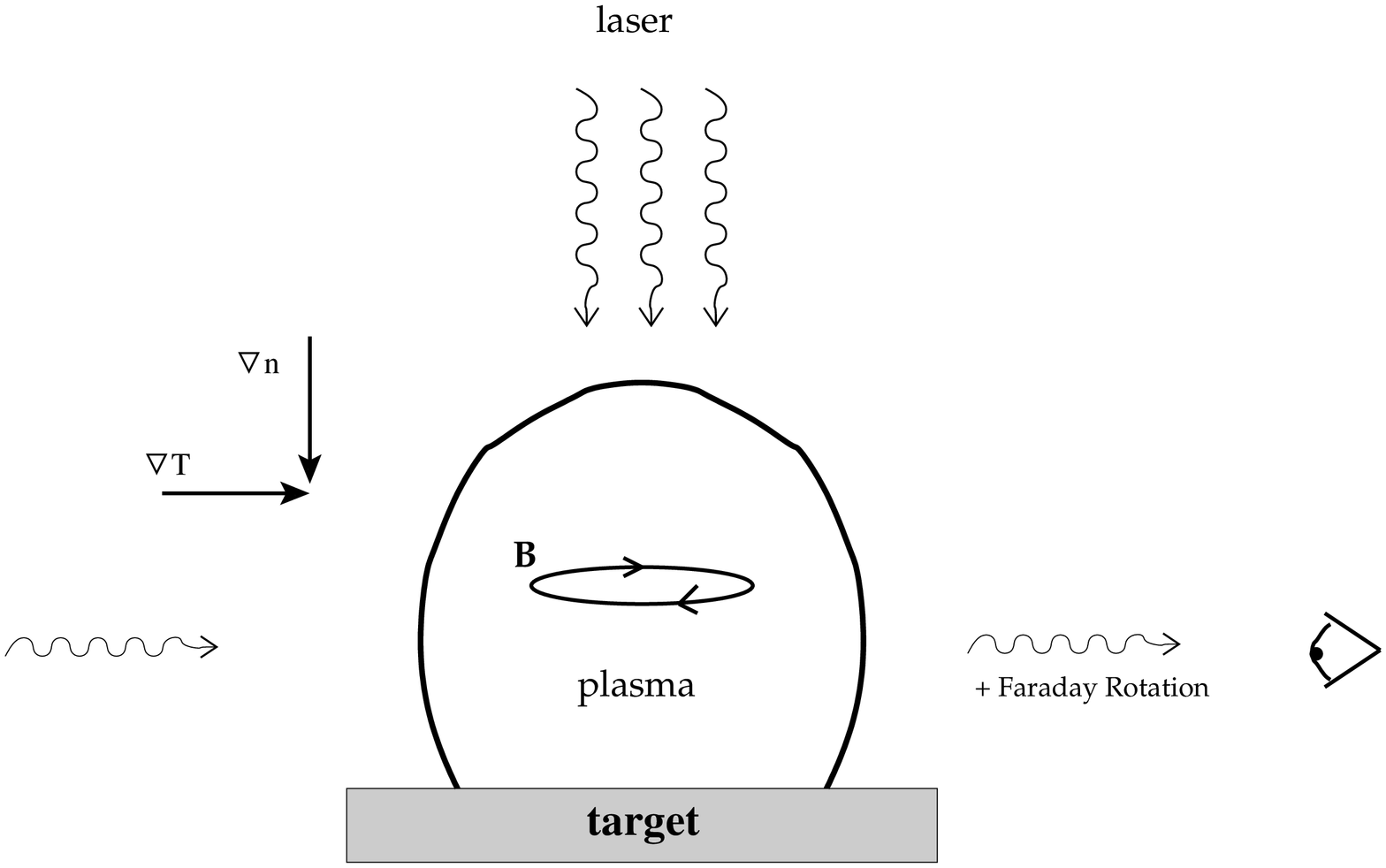,width=12.5cm,angle=-90}}

\caption{Experimental set-up showing how the Biermann effect can be
produced in the laboratory.  Radiation from a laser is incident on the
target and produces a plasma.  This plasma is characterized by density
and pressure gradients which are not parallel leading to a
time-dependent magnetic field.  The field produces Faraday rotation in
radiation that traverses the plasma from left to right. }

\label{exp_biermann}
\end{figure}


Kulsrud et al.\,(1997) simulated the creation of magnetic fields via
the Biermann battery effect.  They used a cosmological hydrodynamic
code that was designed to handle shocks as well as gravitational
collapse (Ryu et al.\,1993).  Since the fields are extremely weak,
their backreaction on the fluid could be ignored.  The MHD equation,
supplemented by the Biermann term (Eq.\,\EC{biermann}) was solved with
${\bf V}$ obtained from the hydrodynamic code.  Kulsrud et al.\,(1997)
assumed parameters appropriate to the CDM model that was popular at
the time of their work ($h=0.5,~\Omega_B=0.06$ and $\Omega_0=1$) and
found that galactic-scale fields are created with strengths of order
$10^{-21}\,{\rm G}$.  However, the spatial resolution of their
simulations is $\sim {\rm Mpc}$, comparable to, if not somewhat larger
than, the scales of interest.

Davies \& Widrow (2000) took a complementary approach by focusing on
the collapse of an isolated galaxy-sized object in an otherwise
homogeneous and isotropic universe.  Their investigation is in the
spirit of semianalytic models of disk galaxy formation by Mestel
(1963), Fall \& Efstathiou (1980), and Dalcanton, Spergel, \& Summers
(1997) as well as numerical simulations by Katz \& Gunn (1991).
However, those models assume that angular momentum and vorticity are
present {\it ab initio} and therefore they are unable to shed light on
voriticity and hence magnetic field generation.  In Davies \& Widrow
(2000) vorticity generation is followed explicitly.  The system
considered consists of collisionless dark matter and collisional gas
with an initial perturbation whose density distribution is assumed to
be nearly spherical, smooth (i.e., no small-scale perturbations) and
monotonic, specifically, that of an axisymmetric, nonrotating prolate
protogalaxy.  As discussed above, each fluid element expands to a
maximum or turnaround radius before falling in toward the center of
the protogalaxy.  During the early stages of collapse, an outward
moving shock develops and as the infalling gas crosses the shock, it
is heated rapidly and decelerated (Bertschinger 1985).  In addition,
the velocity changes direction at the shock.  This is where both
magnetic fields and vorticity are generated.

Analytic calculations based on the thin-shock approximation (Landau \&
Lifshitz 1987) and numerical simulations using SPH follow explicitly
the generation of vorticity at the shock that forms.  In this simple
model, the vorticity and hence magnetic field are in the azimuthal
direction and are antisymmetric about the equatorial plane so that the
total vorticity of the system is zero.  This of course reflects that
fact that angular momentum has not been included.  Rotation about a
short axis of the protogalaxy will ``shear'' the vorticity and
magnetic field lines into a dipole-like configuration.

The results of this analysis indicate that by a redshift $z\simeq 8$,
a $10^{-20}\, {\rm G}$ field, coherent on $10-20\,{\rm kpc}$ scales,
is generated.  This field is amplified by a factor $10^{2}-10^3$ as
the protogalaxy collapses (Lesch \& Chiba 1995) leading to a
$10^{-17}\,{\rm G}$ field in the fully formed disk galaxy.

In a hierarchical scenario, the Biermann effect also operates in
subgalactic objects suggesting an alternate route to galactic-scale
seed fields.  Consider a region that contains a mass $M\simeq
10^6\,{\rm M}_\odot$ in baryons.  This mass corresponds to the Jeans
mass at decoupling and represents a lower bound on the mass of the
first generation of gravitationally bound objects.  Moreover, gas
clouds of this size are thought to be the first sites of
star-formation.  The Biermann effect leads to seed fields of order
$10^{-18}\,{\rm G}$ by $z\simeq 40$ (Pudritz \& Silk 1989; Davies \&
Widrow 2000).  The dynamical time for these objects is relatively
short ($\la 10^8\,{\rm yr}$) and therefore amplification by a dynamo
can be very fast.  The field reach equipartion (microgauss) strength
on these subgalactic scales.  As discussed below, small-scales fields
of this type can act as a seed for a galactic-scale dynamo.  Moreover,
the existence of magnetic fields in star-forming clouds may resolve
the angular momentum problem for the first generation of stars
(Pudritz \& Silk 1987).

Lazarian (1992) considered a variant of the Biermann battery in
which electron diffusion plays the key role in establishing electric
currents and hence magnetic fields.  This effect relies on the
observation that the ISM is nonuniform, multiphase, and clumpy.
Diffusion of electrons from high to low density regions results in an
electric field whose curl is given by

\be{lazarian}
\bfn\times{\bf E}=-\frac{4k_B}{\pi e n_e}\bfn T_e\times \bfn n_e~.
\ee

\noindent
Lazarian (1992) estimated that magnetic fields as high as $3\times
10^{-17}\,{\rm G}$ can be generated on large scales in the ISM.

A battery of this type may have operated prior to the epoch of galaxy
formation (Subramanian, Narasimha, \& Chitre 1994).  At decoupling
($z_d\simeq 1100$), protons and electrons combine to form neutral
hydrogen.  However, the absence of \lya~absorption in quasar spectra
(above and beyond the features in the \lya~forest) severely limit the
amount of smoothly distributed HI in the intergalactic medium.  The
implication is that hydrogen has been reionized almost completely at
some epoch between decoupling and $z\simeq 5$ (Gunn \& Peterson 1965).
Subramanian, Narasimha, \& Chitre (1994) assumed that reionization is
characterized by ionization fronts propagating through the
intergalactic medium.  Electric currents and magnetic fields are
produced when these fronts encounter density inhomogeneities.  Gnedin,
Ferrara, \& Zweibel (2000) performed numerical simulations detailing
this mechanism.  These simulations followed the reionization of the
Universe by stars in protogalaxies.  Ionization fronts formed in high
density protogalaxies propagate through the Universe.  As they cross
filamentary structures in the so-called cosmic web, they drive
currents which in turn give rise to magnetic fields.  The fields
produced are highly ordered on a megaparsec scale and have a strength
of $10^{-19}\,{\rm G}$.  Furthermore, the fields are stronger inside
dense protogalaxies though a detailed analysis on these scales was
limited by resolution of the simulations.

\subsubsection{Galactic Magnetic Fields from Stars}

The first generation of stars can be a source of galactic-scale seed
fields.  Even if a star is born without a magnetic field, a Biermann
battery will generate a weak field which can be amplified rapidly by a
stellar dynamo.  If the star explodes or undergoes significant mass
loss, magnetized material will be spread throughout the ISM.  The
following simple argument illustrates why this proposal is attractive
(Syrovatskii 1970).  Over the lifetime of the Galaxy, there have been
$\sim 3\times 10^8$ supernova events, roughly one for each $\left
(10\,{\rm pc}\right )^3$ volume element.  The magnetic field in the
Crab nebula, taken as the prototypical supernova remnant, is
$300\,\mu{\rm G}$ over a region $1\,{\rm pc}$ in size.  A galaxy
filled with (ancient) Crab-like nebulae would therefore have an
average field with strength $\simeq 3\,\mu {\rm G}$.  Similarly, gas
lost from stars via a wind will carry stellar magnetic fields into the
ISM (Michel \& Yahil 1973).

The magnetic fields produced by stars will have a tangled component
which is orders of magnitude larger than the coherent component.  By
contrast, the random and regular components of a typical spiral galaxy
are nearly equal.  Therefore, stellar magnetic fields, in and of
themselves, cannot explain magnetic fields in disk galaxies.
Nevertheless, the (rms) large-scale component due to an ensemble of
small-scale stellar fields can act as a seed for a galactic dynamo.

A naive model for the large-scale structure of stellar-produced
magnetic fields divides the galactic disk into a large number of random
cells of size $R_{\rm cell}$.  The magnetic field is assumed to be
uniform within each cell but uncorrelated from one cell to the next.
The rms flux through a surface of scale $L$ is $F\sim B_{\rm cell}
N^{1/2}\sim  B_{\rm cell}\left (L/R_{\rm cell}\right )$ where $N$
is the number of cells that intersect the surface.
The rms field on the scale $L$ is therefore

\be{rmsfield}
B_{0} \simeq 
B_{\rm cell}\left (\frac{R_{\rm cell}}{L}\right )
\ee

\noindent
where $B_{\rm cell}$ is the average field in an individual cell.

A more careful treatment takes into account the expected topology of
the magnetic field.  In particular, magnetic field lines within an
individual cell should close.  This point was stressed by Hogan (1983)
in his discussion of the magnetic fields produced in a cosmological
phase transitions (see, also Ruzmaikin, Sokoloff \& Shukurov 1988a,
1988b).  Consider an idealized model in which the field within a given
cell has a dipole structure.  The orientation and strength of the
dipoles are assumed to be uncorrelated from one cell to the next.  Let
us estimate the magnetic flux through a surface $S$ enclosed by a
circular contour $C$ of radius $L\gg R_{\rm cell}$ as shown in Figure
16.  The contribution to the flux from cells ``inside'' the contour is
identically zero: the only nonzero contribution comes from cells along
$C$.  The number of such cells is $\sim L/R_{cell}$ and therefore the
flux through $C$ scales as $\left (L/R_{\rm cell}\right )^{1/2}$.
Hence the magnetic field, averaged over a length scale $L$ is


\begin{figure}
\centerline{\psfig{file=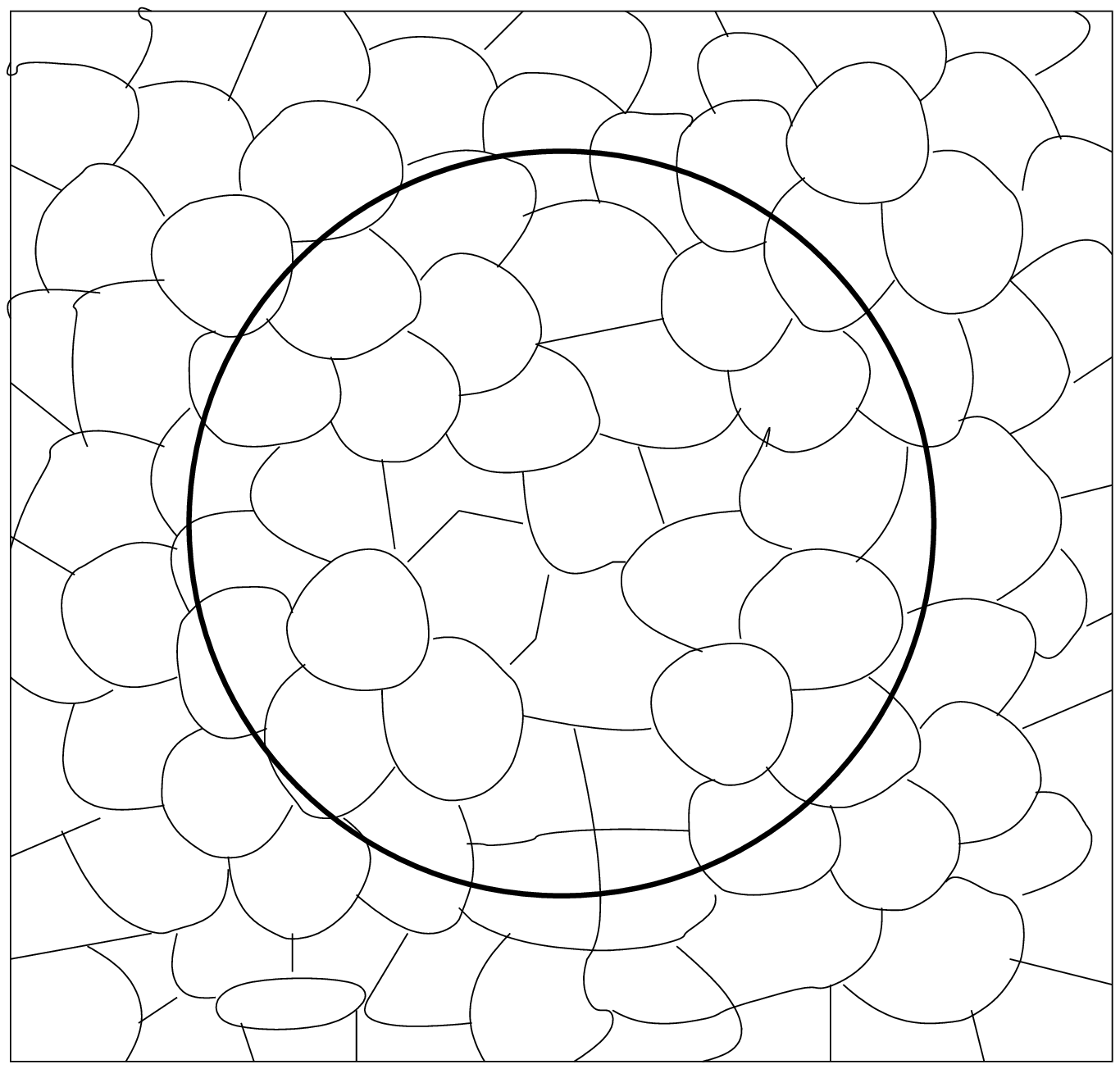,width=12.5cm,angle=-90}}

\caption{Schematic diagram showing a cross-section of a distribution of 
magnetic dipole cells in a galactic disk.  The circular contour C
has a radius $L$ and encloses $O(L/R_{cell})^2$ cells.  Only cells
on the border contribute to the magnetic flux through $C$.}

\label{stars}
\end{figure}


\bea{rmsfield2}
B_{0} &\simeq &
B_{\rm cell}\left (\frac{R_{\rm cell}}{L}\right )^{3/2}\nonumber \\
&\simeq& 10^{-11}\,{\rm G}
\left (\frac{B_{\rm cell}}{3\,\mu {\rm G}}\right )
\left (\frac{R_{\rm cell}}{100\,{\rm pc}}\right )^{3/2}
\left (\frac{10\,{\rm kpc}}{L}\right )^{3/2}
\eea

\noindent
Thus, a significant large-scale magnetic field results if the Galaxy
is filled with Crab-like regions.

The proposal that the galactic dynamo is seeded by magnetic fields
first created in stars has a potentially fatal flaw: Stars form in
protostellar gas clouds.  Since the presence of angular momentum in a
cloud halts its collapse, efficient angular momentum transport is
necessary if star formation is to occur.  Magnetic field can remove
angular momentum from protostellar clouds, but if the first fields
form in stars, then the first clouds will not have any fields.
Nevertheless, stellar field may be important in establishing the
galactic dynamo.  Bisnovatyi-Kogan, Ruzmaikin, \& Sunyaev (1973) (see,
also Pudritz \& Silk 1989) proposed a model in which fields are
generated in protostellar clouds by a Biermann-type mechanism and
dynamo action.  These fields facilitate star formation and the stars
that form then act as sites for rapid field amplification.  Finally,
supernova explosions and stellar winds disperse stellar fields into
the ISM thereby seeding the dynamo.

A common thread between the battery mechanism described in the
previous section and fields produced by stars is that small-scale
fields can provide the seed for a large-scale dynamo.  Indeed, the
fluctuation dynamo is very efficient at amplifying fields on scales up
to that of the largest turbulent eddies ($l\sim 100\,{\rm pc}$) to
equipartition $(1\,\mu{\rm G})$ strengths on a relatively short
timescale.  The projection of this highly tangled field onto the
dominant mode of an $\aod$ dynamo can act as the seed.  The amplitude
of this component is, in general, a factor of $100$ smaller than the
rms field thus giving a seed field of $B_s\simeq 10^{-8}\,{\rm G}$.
Simulations of an $\aod$ dynamo seeded by tangled field were carried
out by Poedz et al.\,(1993) and Beck, Poedz, Shukurov \& Sokoloff
(1994).  The initial field was chosen to be random on sub-kiloparsec
scales with an rms strength of $1\,\mu{\rm G}$.  At first, the
amplitude of the field declines sharply while its scale length
increases rapidly.  There follows a period of exponential growth to a
present day value for the regular field strength of $\sim 3\,\mu{\rm
G}$.

\subsubsection{Active Galactic Nuclei}

Active galactic nuclei (AGN) are promising sites for the production of
galactic and extragalactic magnetic fields (Hoyle 1969, Rees 1987,
Daly \& Loeb 1990; Chakrabarti, Rosner, \& Vainshtein 1994, Rees
1994).  AGN are powered by the release of gravitational potential
energy as material accretes onto a central compact object, presumably
a supermassive black hole.  There, dynamo processes can amplify
magnetic fields on relatively short timescales.  Moreover, even if the
central region of an AGN forms without a magnetic field, one will
develop quickly through Biermann battery-type mechanisms.  Finally,
well-collimated jets can transport magnetic field energy away from the
central object and into the protogalactic or intergalactic medium.

The following order of magnitude estimate, due to Hoyle (1969) gives
an indication of the field strengths possible in an AGN scenario.  The
rotational energy of a compact object and the material in its
immediate vicinity (total mass $M$) can be parametrized as $fMc^2$
where $f<1$.  Equipartition between the magnetic field and the
rotational energy in the fluid (achieved by differential rotation
and/or dynamo action) implies a magnetic field strength

\be{hoyle1}
B_c \simeq \left (\frac{8\pi fMc^2}{V_c}\right )^{1/2}
\ee

\noindent
where ${\cal V}_c$ is the volume of the central region.  If this 
field expands adiabatically to fill the volume ${\cal V}_g\simeq 
100\,{\rm kpc}^3$ of the Galaxy, a field with strength

\bea{hoyle2}
B_g &=& B_c\left (\frac{{\cal V}_c}{{\cal V}_g}\right )^{2/3}
\nonumber \\
    &=& \left (\frac{8\pi fMc^2}{{\cal V}_g}\right )^{2/3}B_c^{-1/3}
\eea

\noindent 
will result.  As an example, Hoyle (1969) considers the values
$M=10^9\,M_\odot$, $f=0.1$, ${\cal V}_g=10^{67}\,{\rm cm}^3$, and $B_c
= 10^9\,{\rm G}$ where one finds $B_g\simeq 10^{-5}\,{\rm G}$.

Our understanding of AGN has improved to the extent that detailed
discussions of these objects as potential sources of seed magnetic
fields is now possible.  Daly \& Loeb (1990) proposed that the
magnetic field in a particular galaxy originates during an AGN phase
of that galaxy.  This scenario presumes that all galaxies (or at
least, all galaxies with strong magnetic fields) have compact central
objects, a conjecture that is now supported by numerous observations
(see, for example, Magorrian et al.\,1998).  An AGN phase is
characterized by two oppositely-directed, well-collimated jets which
transport material into the ISM.  Since magnetic fields in the central
region are frozen into the jet material, they too are carried into the
ISM.

Daly \& Loeb (1990) described, in detail, the interaction of the jet
with the ISM (also see Daly 1990).  A shock forms where this material
collides with the ambient gas.  At the same time, a blast wave
develops propagating perpendicular to the jet axis and carrying
magnetized material into the so-called shock cocoon.  The timescale
for this process is short relative to the lifetime of the galaxy and
material from the central AGN engine can reach the outer part of a
protogalaxy in a time $\sim v_s/L_J$ where $v_s$ is the shock speed and
$L_J$ is the length of the jet.  An order of magnitude estimate
for the strength of the galactic field that results is obtained by
assuming that the total magnetic field energy in the jet resides
ultimately in the disk galaxy that develops, i.e., ${\cal V}_JB_J^2=
{\cal V}_D B_D^2$ where ${\cal V}_J$ and $B_J$ are the volume and
field strength for the jet and ${\cal V}_D$ and $B_D$ are the
corresponding quantities for the disk.  If the jet has a mean
cross-sectional area $\pi R_J^2$ and scale length $L_J$ we have

\be{AGN_field}
B_D\simeq 
\left (\frac{R_J^2 L_J}{R_D^2 h}\right )^{2/3}B_J
\ee

\noindent 
where $h$ and $R_D$ are the characteristic height and radius of the
disk.  Typically, $R_J\simeq h\simeq 0.5\,{\rm kpc}$, $R_D\simeq L_J
\simeq 10\,{\rm kpc}$ and $B_J\simeq 10\,\mu{\rm G}$.  With these
values, Daly \& Loeb (1990) estimated that the galactic scale would
have a strength $\simeq \mu{\rm G}$.

An essential feature of Daly and Loeb's model is that each galaxy
produces its own seed field.  An alternative is that a population of
AGN at high redshift `contaminates' the protogalactic medium with
magnetic field prior to the epoch when most galaxies form (Rees 1987,
1994; Furlanetto \& Loeb 2001).  The jets that originate in AGN often
end in giant radio lobes which are tens of kiloparsecs in size and
have magnetic fields $\ga 10\mu\,{\rm G}$.  The field in a typical
protogalaxy due to pre-galactic AGN is

\be{contaminate}
B_s \simeq 10^{-11}\,{\rm G}\left (\frac{n_{\rm AGN}}{{\rm Mpc^{-3}}}
\right )
\ee

\noindent 
where $n_{\rm AGN}$ is the number density of AGN.  Thus, a relatively
small number of AGN can seed the protogalactic medium with fields in
excess of those produced by battery-type mechanisms.

The amplification of magnetic fields in an accretion-disk environment
has been the subject of a number of studies.  Pudritz (1981), for
example, demonstrated that a mean-field dynamo can operate in a
turbulent accretion disk.  The Balbus-Hawley instability provides an
alternate route to strong fields since a weak magnetic field leads to
turbulence which in turn can rapidly amplify the field (Brandenburg et
al.\,1995; Hawley, Gammie, \& Balbus 1996).

Kronberg, Lesch, \& Hopp (1999), Birk, Wiechen, Lesch, \&
Kronberg (2000) considered a somewhat different scenario in which the
seed fields for spiral galaxies such as the Milky Way are created in
dwarf galaxies that form at a redshift $z\sim 10$.  In their model
starburst-driven superwinds rather than jets transport magnetic flux
into the IGM.  In the hierarchical clustering scenario, dwarf galaxies
are the first to form and hence the first to support active star
formation.  In addition, the consensus is that these systems suffer
substantial mass loss via supernovae and stellar winds.  The scenario
requires that strong magnetic fields exist at early times in dwarf
galaxies a reasonable assumption given that star-formation and outflow
activity increase the amplification rate in models such as Parker's
(1992) modified $\aod$-dynamo.

\subsection{Seed Fields from Early Universe Physics}

The very first magnetic fields may have been created during an early
universe phase transition.  These events typically involve fundamental
changes in the nature of particles and fields as well as a significant
release of free energy over a relatively short period of time, two
conditions that lead naturally to electric currents and hence magnetic
fields.  Often, the question is not whether magnetic fields are
created during an early universe phase transition, but whether these
fields are appropriate seeds for galactic dynamos.

Causality imposes a fundamental constraint on early universe scenarios
for the generation of seed fields.  The Hubble distance, $L_H(t)\equiv
c/H(t)$, sets an upper bound on the size of a region that can be
influenced by coherent physical processes.  (In a radiation or matter
dominated Universe, $L_H$ is, up to a constant of order unity, equal
to the causal horizon, i.e., the age of the Universe times the speed
of light.  Even if an inflationary epoch changes the causal structure
of the Universe, $L_H$ still sets an upper limit on the scale over
which physical processes can operate.)  In comoving coordinates, the
Hubble distance $\lambda_H\equiv L_H(t)/a(t)$ reaches the scale
$\lambda_H\simeq 100\,{\rm kpc}$ of galactic disks at a time $t\simeq
10^7\,{\rm s}$ after the Big Bang.  Contrast this with the age of the
Universe at the electroweak phase transition ($t_{EW}\simeq
10^{-12}\,{\rm s}$) or QCD phase transition ($t_{QCD}\simeq 10^{-4}
\,{\rm s}$) and the nature of the causality problem becomes clear:
Fields generated in the early Universe have a coherence length that is
much smaller than what is required of seed fields for a galactic
dynamo.

Many of the scenarios that operate after inflation rely on either
statistical fluctuations of strong small-scale fields to yield (weak)
large-scale ones or dynamical processes, such as an inverse cascade of
magnetic energy, to channel field energy from small to large scales.
Conversely, scenarios that operate during inflation produce fields on
scales up to the present-day Hubble radius.  However, these models
have their own set of difficulties.

\subsubsection{Post-inflation Scenarios}

We begin our discussion with the cosmological QCD phase transition
(e.g., Boyanovsky 2001).  At high temperatures, quarks and gluons are
weakly coupled and exist as nearly free particles in a plasma.  The
transition to the hadronic phase, in which quarks are bound into
mesons and baryons, occurs at a temperature $\tqcd\simeq 150\,{\rm
MeV}$.  The order of the quark-hadron phase transition is not known.
If it is second order, the transformation from quark-gluon plasma to
hadrons occurs adiabatically, i.e., approximate thermodynamic
equilibrium is maintained locally and at each instant in time.  A more
dramatic sequence of events occurs if the phase transition is first
order.  As the Universe cools below $\tqcd$, bubbles of hadronic phase
nucleate and grow.  Shocks develop at the bubble walls, latent heat is
released, and the Universe reheats back to $\tqcd$.  The two phases
now coexist with the hadronic regions growing at the expense of
regions still in the quark phase.

Hogan (1983) was the first to investigate the possibility that
magnetic fields could be generated in a cosmological first-order phase
transition.  His model assumes that battery and dynamo processes
create and amplify magnetic fields, $B_B$ that are concentrated in the
bubble walls.  When the walls collide the fields from each bubble are
``stitched'' together by magnetic reconnection.  In this way, magnetic
field lines, following random paths, can extend to scales much larger
than the characteristic scale, $L_B$, of the bubbles at the time of
the phase transition.  The situation is analogous to the one that
arises with seed fields from stellar outflows and supernovae.  The
component of the magnetic field that is coherent on scales $L\gg L_B$
has a typical strength $B_L\simeq B_B\left (L_B/L\right )^{3/2}$.

Detailed calculations of magnetic field generation during the
electroweak and QCD phase transitions, under the assumption that they
are first-order, have been carried out by numerous groups.  Quashnock,
Loeb, \& Spergel (1989), for example, demonstrated that a Biermann
battery operates during the QCD phase transition.  The baryon
asymmetry of the Universe implies that there are more quarks than
anti-quarks.  If the number densities of the light quarks up, down
and strange (charges $2/3, -1/3,$ and $-1/3$ respectively) were equal,
the quark-gluon plasma would be electrically neutral.  However, the
strange quark is heavier than the other two and therefore less
abundant so that there is a net positive charge for the quarks.  This
charge is compensated by an excess of negative charge in the lepton
sector.  Shocks that develop during the nucleation of hadronic bubbles
are characterized by strong pressure gradients which affect the quarks
and leptons differently.  Therefore, electric currents develop as
bubble walls sweep through the quark-gluon plasma.  Quashnock, Loeb,
\& Spergel (1989) estimated the strength of the electric fields to be
$E\simeq \left (\nabla p_e\right )/en_e \simeq \epsilon\delta kT_{\rm
QCD}/L_B$ where $\delta\equiv L_B\nabla p_e/p_e$ characterizes the
magnitude of the pressure gradient and $\epsilon$ is the fractional
difference between energy densities of the quark and lepton fluids.
$L_B$ depends on the details of the transition.  A reasonable estimate
is $L_B\simeq 100\,{\rm cm}$ which is an order of magnitude or two
smaller than the Hubble distance at the time of the phase transition.
With $\epsilon\simeq\delta\simeq 0.1$, Quashnock, Loeb, \& Spergel
found $B\simeq 5\,{\rm G}$.  $L_B$ sets the scale for the coherence
length of the field.  A distance $L_B\simeq 100\,{\rm cm}$ at $t_{\rm
QCD}$ corresponds to a present present-day distance $l\simeq 6\times
10^{13}\,{\rm cm}\simeq 4\,{\rm AU}$.  The field strength at
recombination on these scales is $\simeq 2\times 10^{-17}\,{\rm G}$.
Assuming $B\propto L^{-3/2}$ (Hogan 1983) this yields a galactic-scale
($L\simeq 100\,{\rm kpc}$) field with strength at recombination of
$B\simeq 6\times 10^{-32}\,{\rm G}$.

At the time of the QCD phase transition, the energy density of the
Universe is $\sim 1\,{\rm Gev\,fm^{-3}}$ corresponding to an
equipartition field strength of $\beq\simeq 10^{18}\,{\rm G}$.  Thus,
the battery mechanism proposed by Quashnock, Loeb, \& Spergel (1989)
taps into a tiny fraction of the available energy.  Cheng \&
Olinto (1994) and Sigl, Olinto, \& Jedamzik (1997) suggested that
stronger field-generating mechanisms operate during the coexistence
phase that follows bubble nucleation.  As hadronic regions grow, there
is a tendency for baryons to concentrate in the quark phase.
Essentially, the bubble walls act as `snowplows' sweeping up baryons.
In doing so, they create currents of order $J\sim en_+ v$ where $n_+$
is the number density of positive charge carriers in the quark phase
and $v$ is the typical velocity of the bubble walls.  The
corresponding magnetic field is $B\sim en_+ vr_d$ where $r_d$ is the
thickness of the charge layer.  For reasonable parameters, this leads
to an estimate $B_{\rm QCD}\simeq 10^6-10^8\,{\rm G}$ on scales of
$100\,{\rm cm}$ at $t_{\rm QCD}$.  If $B\propto L^{-3/2}$, a field
strength at recombination on $100\,{\rm kpc}$ scales (comoving
coordinates) will be $B\simeq 6\times 10^{-26}- 10^{-24}\,{\rm G}$.
 
A first-order electroweak phase transition can also generate
large-scale magnetic fields (Baym, B\"{o}deker, \& McLerran 1996;
Sigl, Olinto, \& Jedamzik 1997).  During the electroweak phase
transition, the gauge symmetry breaks from the electroweak group
$SU(2)_L\times U(1)_Y$ to the electromagnetism group $U(1)_{\rm EM}$.
The transition appears to be weakly first order or second order
depending on parameters such as the mass of the Higgs particle
(Boyanovsky 2001; Baym, B\"{o}deker, \& McLerran 1996 and references
therein).  If it is first order, the plasma supercools below the
electroweak temperature $\tew\simeq 100\,{\rm GeV}$.
Bubbles of broken symmetry phase nucleate and expand, eventually
filling the Universe.  As in the case of the QCD transition, we write
the bubble size $L_B=f_B L_H$ where $f_B\simeq 10^{-3}-10^{-2}$ and
$L_H\simeq 10\,{\rm cm}$.  The wall velocities are believed to be in
the range $v_{\rm wall}\simeq (0.05-0.9)c$.  Baym, B\"{o}deker, \&
McLerran 1996 discussed the structure of the bubble walls
and the associated shocks in detail.  The key observation is that the fluid
becomes turbulent where two walls collide.  Fully developed MHD
turbulence leads rapidly to equipartition of field energy up to the
scale of the largest eddies in the fluid, assumed to be comparable to
$L_B$.  Thus, the field strength on this scale is

\bea{B_EW}
B &\simeq &\left (4\pi\epsilon\right )^{1/2}
\left (\tew\right )\tew^2
\left (\frac{v_{\rm wall}}{c}\right )^2\nonumber \\
&\simeq & \left (7\times 10^{21}-2\times 10^{24}\right ){\rm G}
\eea

\noindent where $\epsilon=g_* a \tew^4/2\simeq 4\times 10^{11}{\rm
GeV\,fm^{-3}}$ is the energy density at the time of the electroweak
phase transition.

Magnetic fields can arise in cosmological phase transitions even if
they are second order (Vachaspati 1991).  In the standard model,
electroweak symmetry breaking occurs when the Higgs field $\phi$
acquires a vacuum expectation value (VEV): $\langle\phi\rangle =\eta$.
Interactions between $\phi$ and the gauge fields $A_\mu$ are
described by the kinetic energy term in the Lagrangian which can be
written ${\cal L}_{\rm kin}=D_\mu\phi D^\mu\phi$ where $D_\mu\equiv
\partial_\mu -igA_\mu$ is the covariant derivative and $g$ is the
coupling constant.  (Gauge group indices have been suppressed.)  In
principle, $\partial_\mu\langle \phi\rangle$ and $A_\mu$ can conspire
to give $\langle D_\mu\phi\rangle =0$.  However, since all quantities
are uncorrelated over distances greater than the Hubble distance at
the time of the phase transition, $\langle D_\mu\phi\rangle$, in
general, does not vanish (Vachaspati 1991).  Dimensional arguments
imply $\langle D_\mu\phi\rangle\sim \eta/\xi$ where $\xi$ is the
correlation length for the field at the time of the phase transition.
The correlation length is actually set by the temperature and is much
smaller than the Hubble length: $\xi\sim \hbar c/k\tew\gg c/H_{{\rm
EW}}$.  We are interested in the electromagnetic field which is but
one component of the complete set of gauge fields $A_\mu$.  Vachaspati
(1991) finds $F_{\mu\nu}^{\rm em}\sim g^{-1}\eta^{-2} \partial_\mu
\phi\partial_\nu\phi$ which, using dimensional analysis, implies a
field strength $B\sim g^{-1}\xi^{-2}$.

To estimate the field strength on larger scales Vachaspati (1991)
assumed that $\phi$ executes a random walk on the vacuum manifold with
stepsize $\xi$.  Over a distance $L=N\xi$ where $N$ is a large number,
the field will change, on average, by $N^{1/2}\eta^{-1}$.  Thus, the
gradient in the Higgs field is $\partial\langle\phi\rangle\sim
\eta\,N^{1/2}\,L^{-1} =\eta N^{-1/2}\xi^{-1}$.  The magnetic field
strength, at the time of the phase transition, will be $B\sim B_{CD}
N^{-1}\propto L^{-1}$.  On a galactic scale ($L=100\,{\rm kpc}$ in
comoving coordinates; $N\simeq 10^{24}$) at decoupling, we find
$B\simeq 10^{-23}\,{\rm G}$.

A somewhat different analysis by Enqvist \& Olesen (1993) suggests
that the magnetic fields decrease with scale more slowly than
suggested by Vachaspati (1991).  They argued that the mean magnetic
field satisfies

\be{meanB}
\langle {\bf B}\rangle = 0~~~~~~~~~~\langle B^2\rangle^{1/2}\simeq 
\frac{B_{CD}}{N^{1/2}}
\ee

\noindent where $\langle\dots\rangle$ denotes an average over 
regions of size $N\xi$.  Thus, in their model, $B\propto L^{-1/2}$.

Dynamical mechanisms can also lead to an increase in the coherence
length of a magnetic field that is produced in an early Universe phase
transition.  Cornwall (1997), Son (1999) and Field \& Carroll (2000)
have considered the transfer of magnetic field energy from small to
large scales by a process known as inverse cascade.  This process
occurs when there is substantial magnetic helicity in a fluid
(Pouquet, Frisch, \& L\`{e}orat 1976).  Assume that small-scale
helicity is injected into the fluid in a short period of time.
Magnetic energy will shift from small to large scales as the system
attempts to equilibrate while conserving magnetic helicity and total
energy.  An inverse cascade in a cosmological context leads to the
following scaling laws for the field energy and coherence length:

\be{inverseB}
B_{\rm rms}(t) = \left (\frac{a(t_i)}{a_0}\right )^2
\left (\frac{t_i}{t_{\rm eq}}\right )^{1/6}
B_{\rm rms}(t_i)
\ee

\noindent and

\be{inverseL}
L(t) = \frac{a_0}{a(t_i)}
\left (\frac{t_{\rm eq}}{t_i}\right )^{1/3}
L(t_i)
\ee

\noindent where $t_i$ is the time when the fields are created.  In the
above analysis, it is assumed that the inverse cascade operates during
the radiation-dominated but not matter-dominated phase of the
Universe.  For illustrative purposes, Field \& Carroll (2000) considered
the evolution of fields created at the electroweak scale.  The
coherence scale and strength of these initial fields are written as
$L(t_i) = f_L c H_{ew}^{-1} \simeq 0.6 f_L{\rm cm}$ and $B_{\rm
rms}=f_B\sqrt{8\pi\epsilon_{\rm ew}} \simeq 8\times 10^{25}\,f_B\,{\rm
G}$ where $f_L$ and $f_B$ are dimensionless parameters.  Today, these
fields would have a coherence length and scale $L(t_0)=13f_L{\rm kpc}$
and $B_{\rm rms}(t_0)\simeq 5\times 10^{-10}f_B{\rm G}$.  Thus, if the
generation mechanism is efficient in producing horizon-sized helical
fields $(f_B\simeq f_L\simeq 1)$ then the result will be strong fields
on very large scales.  However, no compelling mechanism is known for
generating large net helicity in a cosmic fluid and therefore
scenarios based on the inverse cascade must be considered highly
speculative.

\subsubsection{Inflation-Produced Magnetic Fields}

The inflationary Universe paradigm provides both the kinematic and
dynamical means of producing a nearly scale-free spectrum of energy
density perturbations (e.g., Kolb \& Turner 1990 and references
therein).  This feature makes inflation an attractive candidate for
the production of magnetic fields (Turner \& Widrow 1988).  

In most models, inflation is driven by the dynamics of a weakly
coupled scalar field known as the inflaton.  During inflation, the
energy density of the Universe is dominated by the vacuum energy of
the inflaton (Guth 1981; Linde 1982; Albrecht \& Steinhardt 1982).
Since this energy density is approximately constant, the Universe is
in a nearly de Sitter phase with $H(t)\simeq {\rm constant}$ and
$a\simeq \exp{(Ht)}$.  Exponential growth of the scale factor has two
important consequences.  First, during inflation the physical length
$L$ corresponding to a fixed comoving scale $\lambda = L(t)/a(t)$
grows relative to the (nearly constant) Hubble distance $L_H$.  By
contrast, during the radiation and matter dominated epochs, $L$ grows
more slowly than $L_H$.  Therefore, a given mode of fixed comoving
size (e.g., a single Fourier component of a small-amplitude density
perturbation) starts out subhorizon-sized and crosses outside a Hubble
volume during inflation only to reenter the Hubble volume during
either radiation or matter dominated epochs (Figure 17).
Microphysical processes can influence this mode during the first
subhorizon-sized phase.  In particular, de Sitter space
quantum-mechanical fluctuations continuously excite all massless or
very light fields (i.e., all fields whose Compton wavelength $\hbar
c/m$ is greater than $L_H$).  Both the amplitude and coherence length
of these fluctuations are set by the Hubble parameter so that the
energy density per logarithmic wavenumber bin in these excitations at
the time when they are produced is $d\epsilon/d\ln{k}\simeq \hbar
H/L_H^3\simeq \hbar H^4/c^3$.  Therefore, to the extent that $H$ is
constant during inflation (or more generally, a power-law function of
time) the fluctuation spectrum is scale-free.


\begin{figure}
\centerline{\psfig{file=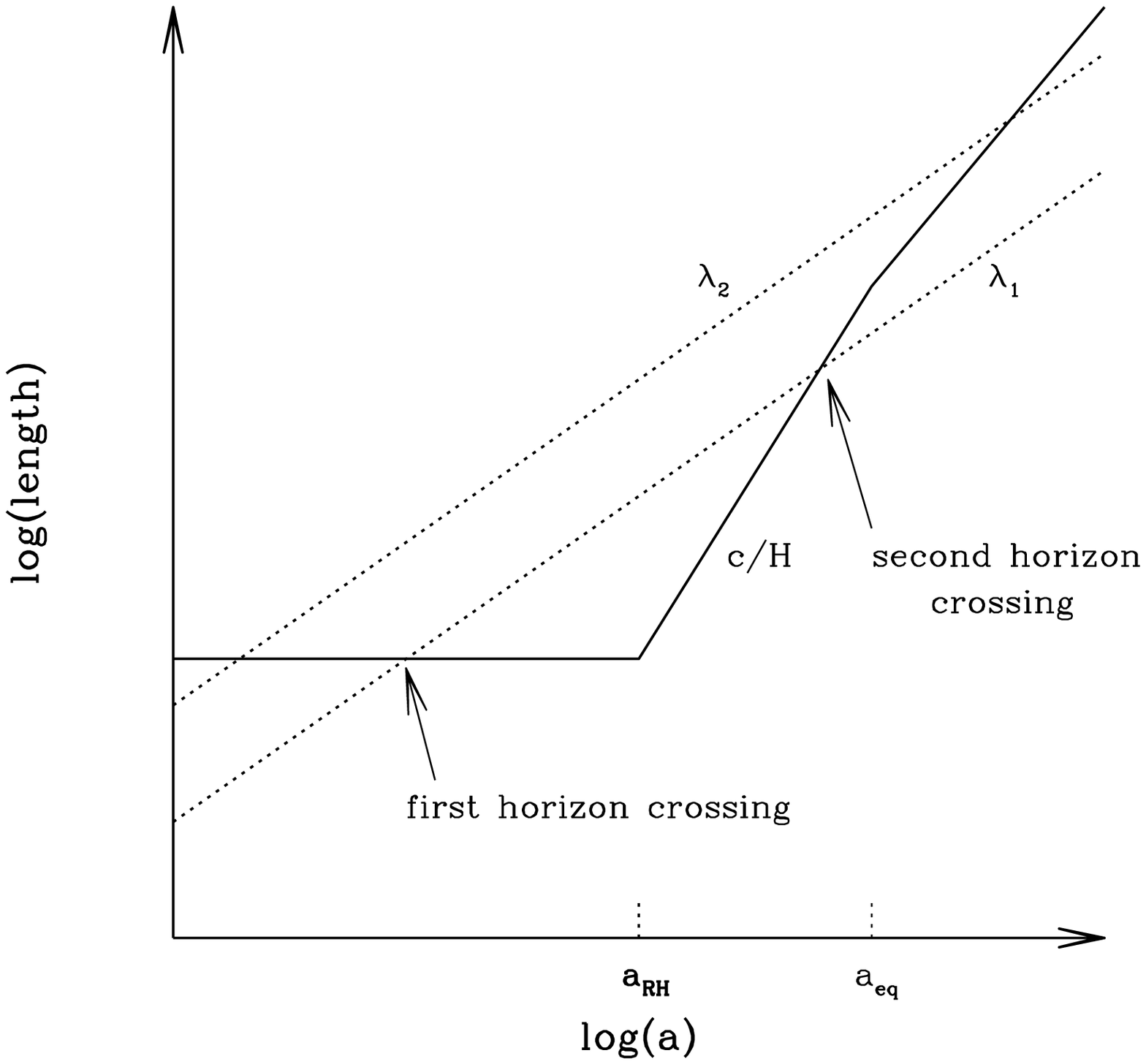,width=12.5cm}}

\caption{Evolution of the physical size for two comoving scales,
$\lambda_1$ and $\lambda_2$, and the Hubble radius, $c/H$ in an
inflationary cosmology.  During inflation, the Hubble radius is
approximately constant and modes cross outside the horizon.  (This
event for $\lambda_1$ is labelled `first horizon crossing'.  Modes
cross back inside the horizon during the post-inflationary epoch.}

\label{inflate}
\end{figure}


While all massless fields are excited during inflation, their
subsequent evolution varies dramatically depending on how they couple
to gravity.  Fluctuations of the inflaton give rise to an
energy-density perturbation spectrum at second horizon crossing that
is consistent with what is required to explain large-scale structure
in the Universe (Guth \& Pi 1982; Starobinskii 1982; Hawking 1982;
Bardeen, Steinhardt, \& Turner 1983).  In simple inflationary Universe
models, perturbations have an amplitude at second horizon crossing
that is approximately constant with scale.  This result is consistent
with what is called for in hierarchical clustering models and is in
agreement with measurements of the angular power spectrum of microwave
background anisotropies.

Quantum-produced de Sitter-space fluctuations in minimally coupled
scalar fields (i.e., fields that do not couple explicitly to the Ricci
scalar $R$) and in gravitons can also be significant (see, Kolb \&
Turner 1990 and references therein).  By contrast, since
electromagnetic fields are conformally coupled to gravity, the energy
density in a fluctuation decreases as $a^{-4}$ leading to amplitudes
that are uninterestingly small.  Conformal invariance must therefore
be broken in order to produce significant primeval magnetic flux.

In a conformally trivial theory, the field equations are invariant
under a rescaling of lengths at each location in space.  Likewise,
conformally flat spacetimes, such as de Sitter space and the radiation
and matter dominated Robertson-Walker models, can be written as
(time-dependent) rescalings of Minkowski space.  It follows that the
form of the field equations of a conformally invariant theory in a
conformally flat spacetime are time-independent.  For example, the
equations of motion for the magnetic field in a Robertson-Walker space
time are:

\be{eomB}
\left (\frac{\partial^2}{\partial\eta^2}-\nabla^2\right )
\left (a^2{\bf B}\right )=0
\ee

\noindent where $\eta$ is the conformal time related to the clock time
$t$ through the expression $d\eta = dt/a(t)$.  For massive fields, 
conformal invariance is broken by the introduction of a length scale,
namely, the Compton wavelength.  Conformal invariance can also be
broken through couplings to other fields.

Turner \& Widrow (1988) suggested a number of ways of breaking
conformal invariance in electromagnetism: (1) introduce the
gravitational couplings $RA^2$, $R_{\mu\nu}A^\mu A^{\nu}$,
$RF^{\mu\nu}F_{\mu\nu}$, etc; (2) couple the photon to a charged field
that is not conformally coupled to gravity; (3) couple the photon to
an axion-like field.  Only (1) was considered in any great detail.
The $RA^2$ terms are the least attractive possibility since they
explicitly break gauge invariance by giving the photon an effective
mass.  However, computationally, this case is the easiest to analyze
and for a wide range of parameters (the coupling constants for the
various terms) interesting large-scale magnetic fields can be
generated.  The $RF^2$ terms are theoretically more palpable but the
fields that result are very small.

Numerous authors have attempted to find more natural and effective
ways to break conformal invariance.  Ratra (1992) calculated the
spectrum of magnetic fields produced in a set of inflation models 
characterized by the inflaton potential $V(\phi)\propto \exp{(\phi)}$.
Potentials of this type can be motivated by superstring theory.
Ratra (1992) assumed a coupling of the inflaton to electromagnetism
through a term $\propto\exp{(\phi)}F_{\mu\nu}F^{\mu\nu}$ and found
that fields as large as $10^{-9}\,{\rm G}$ could be produced.

Magnetic fields due to a charged scalar field were considered by
Calzetta, Kandus, \& Mazzitelli (1998) and Kandus et al.\,(2000).
These authors found that charged domains form during inflation which
give rise to currents and hence magnetic fields during the
post-inflation era.  A different mechanism has been proposed by Davis
et al.\,(2001) who show that the backreaction of the scalar field
gives the gauge field an effective mass thus breaking conformal
invariance.  The mechanism is attractive, in part, because it operates
in the standard model.  Actually, it is the standard-model $Z$-boson
that is amplified through its coupling to the Higgs field.  As
inflation comes to a close, the fluctuations in the $Z$ field are
transferred to the hypercharge field (i.e., a linear combination of
the $Z$ and photon fields).  The $Z$ boson acquires a mass at the
electroweak scale leaving behind pure magnetic field.  Davis et
al.\,(2001) obtain a magnetic field strength of order $10^{-24}\,{\rm
G}$ on a scale of $100\,{\rm pc}$ provided certain conditions during
reheating are met.  

Garretson, Field, \& Carroll (1992) analyzed the amplification of
inflation-produced electromagnetic fluctuations by their coupling to a
PGB.  This coupling, which takes the form $\sim \phi {\bf E}\cdot {\bf
B}$, leads to exponential growth but only for modes whose wavelength
is smaller than the Hubble radius.  No amplification is found for
modes outside the horizon.  The net result is that large-scale
magnetic fields of an interesting strength are not produced by this
mechanism.

\section{Summary and Conclusions}

It was the late 1940's when the Galactic magnetic field was
independently proposed by theorists and detected by observers.  Since
then, galactic and extragalactic magnetic fields have been the subject
of intense and fruitful research.  Nevertheless, fundamental questions
concerning their origin, evolution, and nature remain unanswered.

Magnetic fields have been detected in over one hundred spiral
galaxies, in numerous elliptical and irregular galaxies, in galaxy
clusters, and in the Coma supercluster complex.  New instruments such
as the planned square kilometer array radio telescope will no doubt
reveal new magnetic structures.

It is of interest to note that at present, there is no example of a
meaningful null detection of magnetic field in a collapsing or
virialized system.  Conversely, only upper limits exist on the
strength of truly cosmological magnetic fields.  The fact that these
limits are several orders of magnitude lower than the strength of
galactic and cluster fields suggests that magnetic fields are
amplified, if not created, during structure formation and evolution.

The magnetic fields found in spiral galaxies are unusual in that the
strength of the large-scale component is comparable to that of the
tangled component.  By contrast, the fields in ellipticals are random
on $\la 100\,{\rm pc}$ scales.  Likewise, cluster fields are tangled
on the scale of the cluster itself.  The distinction no doubt reflects
a key difference between spiral galaxies on the one hand and
ellipticals and clusters on the other.  Namely, the stellar and
gaseous disks of spiral galaxies are dynamically ``cold'',
rotationally supported systems while ellipticals and clusters are
dynamically ``hot''.  Evidently, the scale of the largest component of
the magnetic field in any system is comparable to the scale of the
largest coherent bulk flows in that system.

The $\aod$-dynamo is the most widely accepted paradigm for the
amplification and maintenance of magnetic fields in spiral galaxies.
The hypothesis that magnetic fields are continuously regenerated by
the combined action of differential rotation and helical turbulence is
compelling especially in light of the observation that the magnetic
structures in disk galaxies are in general spiral.  One may think of
these structures as the MHD analogue of material spiral arms.  Spiral
structure is believed to be a wavelike phenomenon where the crests of
the waves are characterized by enhanced star-formation activity which,
in turn, is triggered by an increase in the local density.  Likewise,
magnetic spiral arms may reflect low-order eigenmodes in a disklike
magnetized system.  A more direct connection between material and
magnetic spiral structure is evident in certain galaxies where strong
magnetic fields appear in the regions between the material arms.  The
FIR-radio continuum correlation provides further evidence in support
of a connection between star formation and large-scale galactic
magnetic fields.

The magnetic fields in ellipticals and clusters require a different
explanation.  Mergers may play the central role in the establishment
of these fields since they are likely to be present in merger
remnants, typically tidally shredded spiral galaxies.  The magnetic
debris from merger events can act as seeds for subsequent dynamo
action.  In addition, the energy released during a merger event can
drive turbulence in the interstellar or intercluster medium.  It is
unlikely that either ellipticals or clusters will support an
$\aod$-dynamo since differential rotation in these systems is
relatively weak.  However, they may support fluctuation dynamos in
which turbulence amplifies magnetic fields on scales up to the size of
the largest eddies in the systems.

From a theoretical perspective, the greatest challenge in the study of
galactic magnetic fields comes from the tremendous dynamic range
involved.  The scale radius and height of a typical galactic disk are
of order $10\,{\rm kpc}$ and $1\,{\rm kpc}$, respectively, while
turbulent eddies in the ISM extend in size from subparsec to
$100\,{\rm pc}$ scales.  Naive arguments suggest that even if
initially, magnetic energy is concentrated at large scales, in a
turbulent medium, there is a rapid cascade of energy to small scales.
A mean-field approximation, where velocity and magnetic fields are
decomposed into large-scale and small-scale components, bypasses this
problem.  The equation for the large-scale magnetic field, known as
the dynamo equation, incorporates the effects of the small-scale
fields through the $\alpha$ and $\beta$ tensors which, in turn,
attempt to capture the gross properties of the turbulence (e.g.,
helicity, spatial anisotropy).

The dynamo equation for an axisymmetric thin disk can be solved by
means of a quasi-separation of variables which leads to eigenvalue
equations in $t$, $\phi$, $R$, and $z$.  The $t$-eigenvalue gives the
growth rate of the magnetic field while the $\phi$-eigenvalue
characterizes the symmetry of the field under rotations about the
spin-axis of the disk.  The $R$ equation is similar, in form, to the
Schr\"{o}dinger equation and its eigenvalue feeds back into into the
value of the growth rate.  The separation-of-variables analysis has
yielded a number of encouraging results.  Chief among these is a
demonstration of principle, namely, that disklike systems with a
rotation curve similar to that of a spiral galaxy, can support a
magnetic dynamo.  Numerical simulations provide the means to study
more realistic models.  In particular, the effects of a finite disk
thickness and deviations from axisymmetry can be explored.  These
investigations suggest ways in which bisymmetric and/or odd party
magnetic fields can be excited.

Unless one is willing to accept magnetic fields as a property of the
Big Bang, their existence today implies a violation of the MHD
approximation at some stage during the history of the Universe.  While
MHD processes can stretch, twist, and amplify magnetic field, by
definition, they cannot generate new field where none already exist.
Proposals for the origin of the first magnetic fields are as varied as
they are imaginative.  For example, interest in the exotic environment
of the very early Universe, and in particular cosmological phase
transitions, has spawned numerous ideas for the creation of seed
magnetic fields.  A perhaps more appealing set of proposals relies on
the ordinary astrophysical phenomena that occur during structure
formation.  Magnetic fields will develop in AGN, stars, and the shocks
that arise during gravitational collapse.  Indeed, rough estimates
suggest that AGN and/or an early generation of stars will yield fields
of strength $10^{-11}\,{\rm G}$ on galactic scales.  Dynamo action can
amplify a field of this strength to microgauss levels by a redshift
$z\simeq 2$, a result consistent with observations of magnetic fields
in high-redshift radio galaxies.

The astrophysical mechanisms mentioned above were proposed at a time
when our understanding of structure formation was relatively crude.
It is in part for this reason that the creation of seed fields
and the dynamo have been treated as separate and distinct processes.
Indeed, most studies of disk dynamos do not make specific references
to particular models for seed field production.  Likewise, few papers
on seed fields follow the resultant fields into the dynamo regime.

Today, semi-analytic models and numerical simulations enable us to
study galaxy formation in detail, taking into account hierarchical
clustering, tidal torques from nearby objects, gasdynamics, and
feedback from star formation.  Moreover, observations of high-redshift
supernovae, the CMB angular anisotropy spectrum, and large scale
structure have pinned down key cosmological parameters such as the
densities of baryons and dark matter and the Hubble constant.  In
light of these developments, it may be possible to achieve a more
complete description of the origin of galactic magnetic fields,
one that begins with the production of seed fields and follows
smoothly into the dynamo regime.

\acknowledgments{I am grateful to J. Irwin, D. McNeil, A. Olinto, and
S. Toews for carefully reading and commenting on early versions of the
manuscript and to R. Beck and A. Shukurov extremely detailed
and valuable suggestions.  I am also grateful to P. Kronberg and
R. Beck for providing several of the figures.  Finally, I would like
to thank O. Blaes, G. Davies, R. Henriksen, J. Irwin, P. Kronberg,
C. Thompson, and J. Weingartner for useful discussions.  This work is
supported in part by the Natural Sciences and Engineering Research
Council of Canada.}

\end{document}